\documentclass[twocolumn]{aastex63}
\usepackage{comment}
\usepackage[caption=false]{subfig}
\usepackage{graphicx}
\usepackage{amsmath}
\usepackage{gensymb}
\usepackage[section]{placeins}

\newcommand*{\rom}[1]{\expandafter\@slowromancap\romannumeral #1@}

\newcommand{\lya}{Ly$\alpha$}
\newcommand{\ha}{H$\alpha$}

\newcommand{\heii}{He {\sc II}}

\newcommand{\civ}{C {\sc IV}}

\newcommand{\oiii}{[O {\sc III}]}

\newcommand{\kms}{km\,s$^{-1}$} 

\newcommand{\ferg}{erg s$^{-1}$ cm$^{-2}$ }
\newcommand{\fergarc}{erg s$^{-1}$ cm$^{-2}$ arcsec$^{-2}$}

\newcommand{\ergs}{erg s$^{-1}$ }

\received{July 7th, 2025}
\revised{March 7th, 2026}
\accepted{March 20th, 2026}

\submitjournal{ApJ}

\shorttitle{CGM around z $\sim$2 radio-loud quasars}
\shortauthors{Sabhlok et al.}

\graphicspath{{./}{ }}

\begin{document}

\title{The Two Component Circumgalactic Medium Emission around z$\sim$2 Radio-loud Quasars}

\correspondingauthor{Sanchit Sabhlok}
\email{ssabhlok@arizona.edu}

\author[0000-0002-8780-8226]{Sanchit Sabhlok}
\affiliation{Department of Astronomy and Steward Observatory, University of Arizona, 933 N. Cherry Ave, Tucson, AZ 85721, USA}
\affiliation{Department of Physics, University of California San Diego, 
9500 Gilman Drive 
La Jolla, CA 92093 USA}

\author[0000-0003-1034-8054]{Shelley A. Wright}
\affiliation{Department of Astronomy and Astrophysics, University of California San Diego,
9500 Gilman Drive 
La Jolla, CA 92093 USA}
\affiliation{Department of Physics, University of California San Diego, 
9500 Gilman Drive 
La Jolla, CA 92093 USA}

\author[0000-0002-0710-3729]{Andrey Vayner}
\affiliation{Florida Gulf Coast University, 10501 FGCU Blvd. South, Fort Myers, 33965, FL, USA}
\affiliation{IPAC, California Institute of Technology, 1200 E. California Blvd., Pasadena, CA 91125 USA}

\author[0000-0002-8659-3729]{Norman Murray}
\affiliation{Canadian Institute for Theoretical Astrophysics, University of Toronto, 60 St. George Street, Toronto, ON M5S 3H8, Canada}

\author[0000-0003-3498-2973]{Lee Armus}
\affiliation{IPAC, California Institute of Technology, 1200 E. California Blvd., Pasadena, CA 91125 USA}

\author[0000-0002-2248-6107]{Maren Cosens}
\affiliation{The Observatories, Carnegie Science, 813 Santa Barbara Street, Pasadena, CA 91101}

\author[0000-0003-2687-9618]{James Wiley}
\affiliation{Department of Physics, University of California San Diego, 
9500 Gilman Drive 
La Jolla, CA 92093 USA}
\affiliation{Caltech Optical Observatories
MS 11-17
1200 E California Blvd.
Pasadena, CA. 91125}

\author[0000-0003-2687-9618]{Eileen Meyer}
\affiliation{Department of Physics, University of Maryland Baltimore County, 1000 Hilltop Circle, Baltimore, MD 21250, USA}

\author[0000-0003-2687-9618]{Karthik Reddy}
\affiliation{Division of Physics Mathematics and Astronomy, California Institute of Technology, 1200 E. California Blvd., Pasadena, CA 91125 USA}
\affiliation{School of Earth and Space and Exploration, Arizona State University, 1151 S. Forest Ave., Tempe, AZ 85281}

\author[0000-0003-2687-9618]{Marie Wingyee Lau}
\affiliation{Department of Physics and Astronomy, University of California, Riverside, CA 92521, USA}

\begin{abstract}
We present \lya, \heii, and \civ\, observations of 7 redshift $\sim$ 2 radio-loud quasars observed using the Keck Cosmic Web Imager (KCWI) and compare it to observed radio jet emission using archival VLA and ALMA radio observations. We detect 80-120 kpc diameter \lya\, and 10-40 kpc \heii\, and \civ emission around the targets. We find the \lya\, emission to be brighter in the inner 30 kpc by factors of 2-10 compared to other literature samples. We reproduce the trend for increased total luminosity for a larger area on sky, but find our targets tend to be brighter for a given area when compared to literature observations, even when adjusting for the observational sensitivity. We infer that the \heii\, and \civ\, is likely powered by quasar photoionization, with the ionizing radiation likely escaping along the radio jet axis which is aligned with the \heii\, and \civ\, emission. The observations agree with a two component model of the CGM where the inner CGM ($< 30$ kpc) is directly influenced by the host galaxies, whereas the gas motion in the outer CGM ($> 30$ kpc) is influenced by gas turbulence and the larger environment around the host galaxies. 
\end{abstract}

\keywords{galaxies: active, galaxies: high-redshift, galaxies: kinematics and dynamics ---  quasars: emission lines}

\section{Introduction} \label{sec:intro}
The ubiquity of Lyman-$\alpha$ nebulae is now an established paradigm in observational astronomy. Initially, the cosmic UV background was theorized to illuminate these sources \citep{Hogan1987,Gould1996}. It was soon realized that observing these nebulae in the vicinity of a quasar would be easier since the surface brightness of the nebulae would be boosted by the fluorescence emission due to recombination radiation \citep{Rees1988, Haiman2001, Cantalupo2005,Kollmeier2010}. Initially seen in narrowband images of high redshift quasars \citep{Heckman1991a} and quickly confirmed by spectroscopy \citep{Heckman1991b}. The following decade had studies that showed the first \lya\, nebulae revealed by Integral field Spectroscopy (IFS) on the Potsdam Multi Aperture Spectrophotometer \citep{Christensen2006}. While absorption line studies probing the Circumgalactic Medium (CGM) around foreground sources using background quasars were used succesfully to probe the equivalent width and column densities of \lya\, absorbers along a line of sight \citep{Hennawi2013}, the development and commissioning of new optical Integral Field Spectrographs (IFS) over the past decade have finally allowed astronomers the opportunity to directly observe these low surface brightness nebulae. The Keck Cosmic Web Imager (KCWI) \citep{Morrissey2018KCWI} and Multi Unit Spectroscopic Explorer (MUSE) \citep{Bacon2010MUSE} have been used to identify and study CGM emission. \lya\, nebulae have been reported around quasars at z$\sim$2 \citep{Cai2019, Vayner2023,Sabhlok2024a}, z$\sim3$ \citep{Borisova2016, AB2019}, faint quasars \citep{Mackenzie2021}, Type I and Type II quasars \citep{denBrok2020}, Extremely Red Quasar (ERQ) systems \citep{Lau2022, Gillette23}, radio-loud sources at $2.0 < z < 4.5$ \citep{Shukla2022}, including those comparing spatial \lya\, emission with jets \citep{Wang23,Vayner2023,Sabhlok2024a}, a hot dust obscured galaxy \citep{Ginolfi22} and quasars with $z > 5.5$ \citep{Farina2019, Drake2019}. 

The interpretation of the observed \lya\, emission profiles is difficult due to the resonant scattering of \lya\, photons. The observed profiles effectively only trace the last scattering of the photons, which can emerge either from the surface nearest to the observers due to scattering in physical space, or from within the bulk of the cloud due to a random walk in frequency space. Thus, \lya\, surface brightness maps do not necessarily trace gas motion in the CGM and require additional information to disentangle them from the effects of scattering. The addition of other rest-frame UV or optical lines is necessary to understand the dynamics and the photoionization conditions of the gas in the CGM \citep{Langen2023,Chen25,Vayner24}. \heii\,  emission coincident with \lya\, traces the velocity distribution of the underlying gas and is not susceptible to scattering as is \lya. \civ\, emission in the CGM traces the metallicity of the medium. Lastly, \civ\, and \heii\, are both high ionization lines which indicate regions of photoionization by the quasar, or depending on morphology of the observed emission, shocks in the CGM \citep{Sabhlok2024a}. While both are high ionization lines, their emission mechanisms are different, as \heii\, is a recombination line, similar to \ha\, emission for Hydrogen, whereas \civ\, is a resonant line and is an important coolant for the CGM. Thus, when observed together, one can deduce the extent of the cold gas halo in the CGM using \lya, gas kinematics (using \heii), trace metals (using \civ) and ionization state (using \heii\, and \civ) of the CGM. The gas kinematics and metallicity are then used as probes of tidal interactions and interactions with companion galaxies. Thus, in tandem, the emission lines provide significant details about the underlying structure of the CGM. 

Recent work has indicated the CGM structure consists of an inner ($< 30$ kpc) and outer ($> 30$ kpc). This two component model asserts that the kinematics of the inner CGM component interact more directly with host galaxy feedback, whereas the outer component interacts broadly with the larger intergalactic medium and any satellite galaxies. The gravitational motion of the gas is largely determined by the gravitational potential of the host galaxy dark matter halo. This has been observed in rest frame emission lines displaying coherent smooth velocity gradients in the inner region \citep{Lau2022} and in kinematics of high ionization emission lines \citep{Sabhlok2024a}. Line of sight absorption from foreground/background galaxy pairs show that Equivalent Width of \lya\, absorption $\mathrm{EW(Ly\alpha) \propto D_{tran}^{-1.1}}$ for $\mathrm{D_{tran} < 100 \, pkpc}$, where $\mathrm{D_{tran}}$ is the transverse on sky separation from the host galaxy of the background source \citep{Chen2020}. The observation of spatially resolved emission lines such as \heii\, and \civ\, specifically trace high ionization regions around quasars, and can probe the origins of such a dichotomy. The presence of radio jets provides another indicator, since the emission lines are sensitive to the radiation field of the quasar and the presence of companion galaxies, whereas radio jet emission traces changes in IGM density and pressure as seen in expanded lobes or bends in the jet. 

The Quasar hosts Unveiled by high Angular Resolution Techniques (QUART) sample is a multi-wavelength survey of radio-loud quasars at cosmic noon, with observations of spatially resolved host galaxy emission \citep{QUART1}, kinematics of the ISM \citep{QUART2}, ALMA observations of the molecular gas \citep{QUART_ALMA}, and observations of the larger environments indicating the ongoing formation of groups/clusters \citep{Vayner2023, Sabhlok2024a}. The purpose of this work is to study \lya, \heii, and \civ\, nebulae around radio-loud quasars at z$=$2 with spatially resolved radio jet emission and compare the CGM emission to the radio jet emission. Previous work in the literature has explored the radio loud/quiet dichotomy when comparing the extent of \lya\, emission, this work compares the morphology of CGM emission to radio jet emission. The resolved radio-loud compact jet selection offers key advantages in the interpretation of the observed CGM emission. The use of \civ\, and \heii\, in addition to \lya\, allows us to interpret the location and kinematics of the CGM gas around the host galaxy. Interpreting line of sight velocity relative to the host galaxy is challenging since the derived velocities can be infalling or outflowing depending on whether they are in the foreground or behind the galaxy along the line of sight. For example, foreground gas falling towards the galaxy and background gas outflowing from the host galaxy will both be redshifted, whereas the converse will always be blueshifted. The observation of radio jets has the potential to resolve this degeneracy since we know that radio jets with a brighter hotspot are pointed towards the observer along line-of-sight and the emission is Doppler boosted. Thus, if we can successfully associate the gas emission with the radio emission, it will allow us to disentangle whether the emission occurs in the foreground or the background of the galaxy, which then allows us to interpret whether the observed kinematics constitute gas motion towards or away from the galaxy.

This paper is organized as follows. In section \ref{sec:Observations}, we describe the observations and the data reduction techniques employed. In section \ref{sec:Results}, we look at the PSF subtracted Lyman-$\alpha$ distribution of gas around the quasars. In section \ref{sec:Discussion}, we compare our results to other quasar surveys of CGM at redshifts of 2 to 4 and compare the CGM emission to the radio jet emission from the quasars. We summarize our findings in section \ref{sec:Summary}. We assume $\Omega_M = 0.308$, $\Omega_{\Lambda} = 0.692$ and $H_0 = 67.8 \, \mathrm{km \, s}^{-1}\, \mathrm{Mpc}^{-1}$ for the cosmological parameters \citep{Planck2016}.

\section{Observations} \label{sec:Observations}
\subsection{Strategy}
The observations consist of 20-minute exposures on the source and 10-minute sky frames with an ABAAB pattern dithered around the source where A frame targets the quasars and the B frames are ``pure" sky frames,. We used the medium slicer with the BL grating and KBlue filter with the wavelength centered at 4500 $\AA$, except for the 4C 04.81 where the central wavelength was 5200 $\AA$. We observed standard stars at the beginning and end of the night selected from the ESO list for Spectrophotometric standards \footnote{https://www.eso.org/sci/observing/tools/standards/spectra/stanlis.html} for flux calibration. The dataset was gathered over an extended period, and all frames have seeing within 0.6-0.9 $\prime \prime$, which corresponds to 2-3 pixel FWHM on the KCWI detector.
\subsection{Data Reduction}
For the final reduced data, our wavelength coverage was $3532\AA$ to $5527 \AA$, at a spectral resolution of 1,800. The field of view for this KCWI configuration is $16.5 \textrm{\arcsec}\, \times 20.4 \textrm{\arcsec}$. We used half slicer increments to dither perpendicular to the slicer and random dither parallel to the slicer. Primary data reduction was carried out using the KCWI Data Reduction Pipeline v1.1. The data reduction follows the procedure described in Section 3.2 from \cite{Sabhlok2024a} and is summarized here briefly. We used the pipeline to perform cosmic ray removal, wavelength solution, flat-fielding and dark corrections, differential atmospheric refraction corrections, and flux calibration. We skipped the pipeline sky subtraction and used custom routines for this, scaling the sky in object frames to the average flux in a pure sky frame. The WCS coordinates in the header were corrected by 2D fitting a gaussian to the quasar continuum emission, far from any emission lines, which was then assigned the coordinates from Gaia \citep{GAIARef}. The reduced data cubes were then co-added using CWITools \citep{CWIToolsRef}, from which we subtracted the quasar PSF emission spectrum as detailed in \cite{QUART_OSIRIS}. The quasar PSF is reconstructed by assuming a continuum template and fitting this to the observed spectrum far from the wings of broad line emissions or any narrow lines. This is normalized against the peak flux and then subtracted for all wavelength channels. Good PSF subtraction is confirmed by observing residuals in the vicinity of broad line regions. The PSF subtraction leaves no significant residuals beyond 2 pixels from the quasar ($\sim 0.6^{ \prime \prime}$).  The co-added cubes are used to construct the moment maps using CWITools. After final coaddition, we get surface brightness residuals of $\sim 2.2 \times 10^{-19}$ \fergarc\, for the deepest exposure time of 560 minutes, compared to 6 $\times 10^{-19}$ \fergarc\, for the shallowest source with an exposure time of 100 minutes. The maps produced here are using the surface brightness map derivation in CWITools for Moment 0 maps, and the first and second moment calculations for the Moment 1 and 2 maps, which derive from equations 18--21 in \cite{CWIToolsRef}.

\begin{deluxetable*}{lcccccccc}
\tablenum{1}
\caption{Properties of the QUART sample \label{tab:QUART_Members}}

\tablehead{\colhead{Quasar}&
\colhead{RA}&
\colhead{Dec}&
\colhead{Redshift} \tablenotemark{a}&
\colhead{Exp. time} & 
\colhead{Q$_{bol}$ \tablenotemark{b}} &
\colhead{F$_{\mathrm{Ly\alpha}}$ \tablenotemark{c}}&
\colhead{F$_{\mathrm{He II}}$ \tablenotemark{c}}&
\colhead{F$_{\mathrm{C IV}}$ \tablenotemark{c}}
\\  
\colhead{}&
\colhead{HMS}&
\colhead{DMS}&
\colhead{}&
\colhead{minutes}&
\colhead{$\times$ 10$^{46}$ \ergs}&
\colhead{F$_{\lambda}$}&
\colhead{F$_{\lambda}$}&
\colhead{F$_{\lambda}$}
}

\startdata
3C 9       & 00:20:25  & $+$15:40:54 & 2.02 & 390 & 8.17 $\pm$ 0.31 & 1393.3 & 39.8 & 44.9 \\
4C 09.17   & 04:48:21 & $+$09:50:51 & 2.1083 & 320 & 2.88 $\pm$ 0.14 & 171.5 & 5.9 & 8.9\\
4C 11.45   & 13:21:18 & $+$11:06:50 & 2.178 & 100 & $-$  & 591.2 & 8.5 & 7.6 \\
7C $+$1354   & 13:57:06 & $+$25:37:24 & 2.032 & 240 & 2.75 $\pm$ 0.11 & 1083.9 & 39.4 & 60.0 \\
4C 57.29   & 16:59:46 & $+$57:31:32 & 2.173 & 160 & 2.1 $\pm$ 0.1 & 440.2 & 30.3 & 55.3 \\
4C 05.84   & 22:25:14 & $+$05:27:08 & 2.32 & 560 & 20.3 $\pm$ 1.0 & 300.9 & 1.8 & 9.8 \\
4C 04.81   & 23:40:58 & $+$04:31:16 & 2.589 & 340 & 0.62 $\pm$ 0.02  & 281.2 & $-$ \tablenotemark{d} & 8.2 \\
\enddata
\tablenotetext{a} {Redshift taken from SDSS.}
\tablenotetext{b} {Q$_{bol}$ taken from values in \cite{QUART1}}
\tablenotetext{c} {Integrated line fluxes for emission lines in units of $10^{-17} \, \mathrm{ergs \, cm}^{-2} \mathrm{\, s}^{-1}$}
\tablenotetext{d} {Flux not recovered due to sky line.}
\end{deluxetable*}

\begin{deluxetable*}{lccccc}
\tablenum{2}
\tablewidth{0.5 \linewidth}
\caption{Radio observations and W3 magnitudes around the QUART sample \label{tab:QUART_Radio}}
\tablehead{\colhead{Source Name} &
\colhead{Telescope} &
\colhead{Program} &
\colhead{Frequency (GHz)} &
\colhead{S$_{\mathrm{5GHz}}$ (mJy) \tablenotemark{a}} &
\colhead{W3 (mag)}
}

\startdata
3C 9 & VLA & AK307 & 8.44 & 443  & 15.40 \\
4C 09.17 & ALMA & 2017.1.01527.S & 141 & 443 & 15.54 \\
4C 11.45 & VLA & AB332 & 4.86 & 793 & 16.52 \\
7C +1354 & VLA & AB322 & 4.86 & 120 & 16.86 \\
4C 57.29 & VLA & AG220 & 1.48 & 114 & 15.87 \\
4C 05.84 & VLA & 2017.1.01527.S & 143.8 & 253 & 15.18 \\
4C 04.81 & ALMA & AL93 & 4.86 & 470 & 15.68 \\
\enddata
\tablenotetext{a} {Radio flux densities taken from  \cite{Barthel1988} and \cite{Lonsdale1993}}

\end{deluxetable*}

\section{Spatially mapping the CGM emission around radio-loud quasars} \label{sec:Results}

\begin{figure*}[!ht]
    \centering
    \includegraphics[width=\linewidth]{ 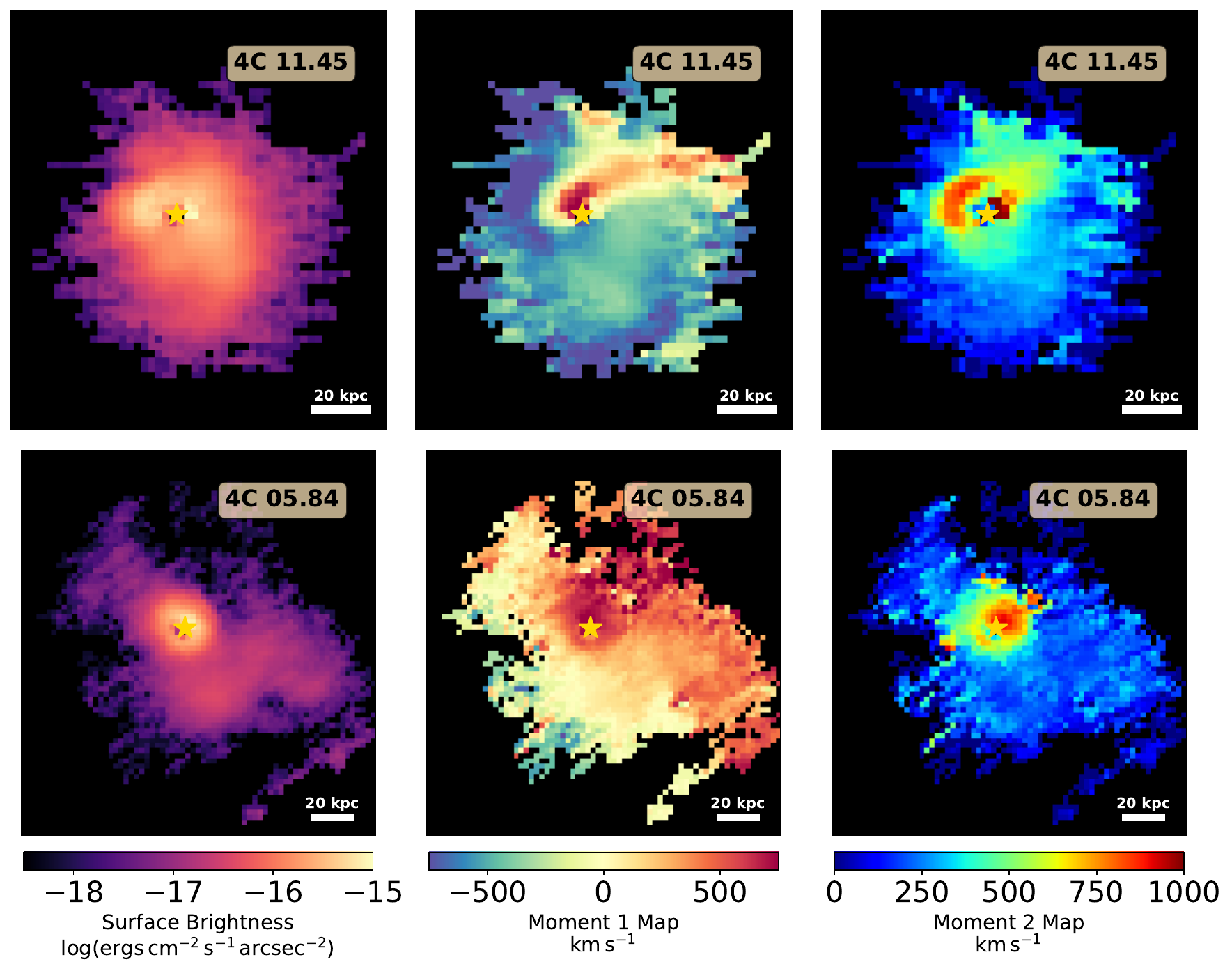}
    
    \caption{PSF-subtracted images of the \lya\, surface brightness, Moment 1 and Moment 2 maps for the two sources with the lowest (4C 11.45; 100 minutes) and highest (4C 05.84; 560 minutes) exposure times on source from the QUART sample. The exposure times for the full sample are reported in Table \ref{tab:QUART_Members}. All figures have an on-sky position angle of 0, North is up and East is to the left. The systemic redshift used to calculate the moment maps is used from OSIRIS measurements in \citet{QUART_OSIRIS} and is shown in Table \ref{tab:QUART_Members}. The location of the quasar is denoted with a gold star marker. The complete figure set (7 images) is available in the online journal. }
    \label{fig:KCWI_Lya}
\end{figure*}

\begin{figure*}[!ht]
    \centering
    \includegraphics[width=\linewidth]{ 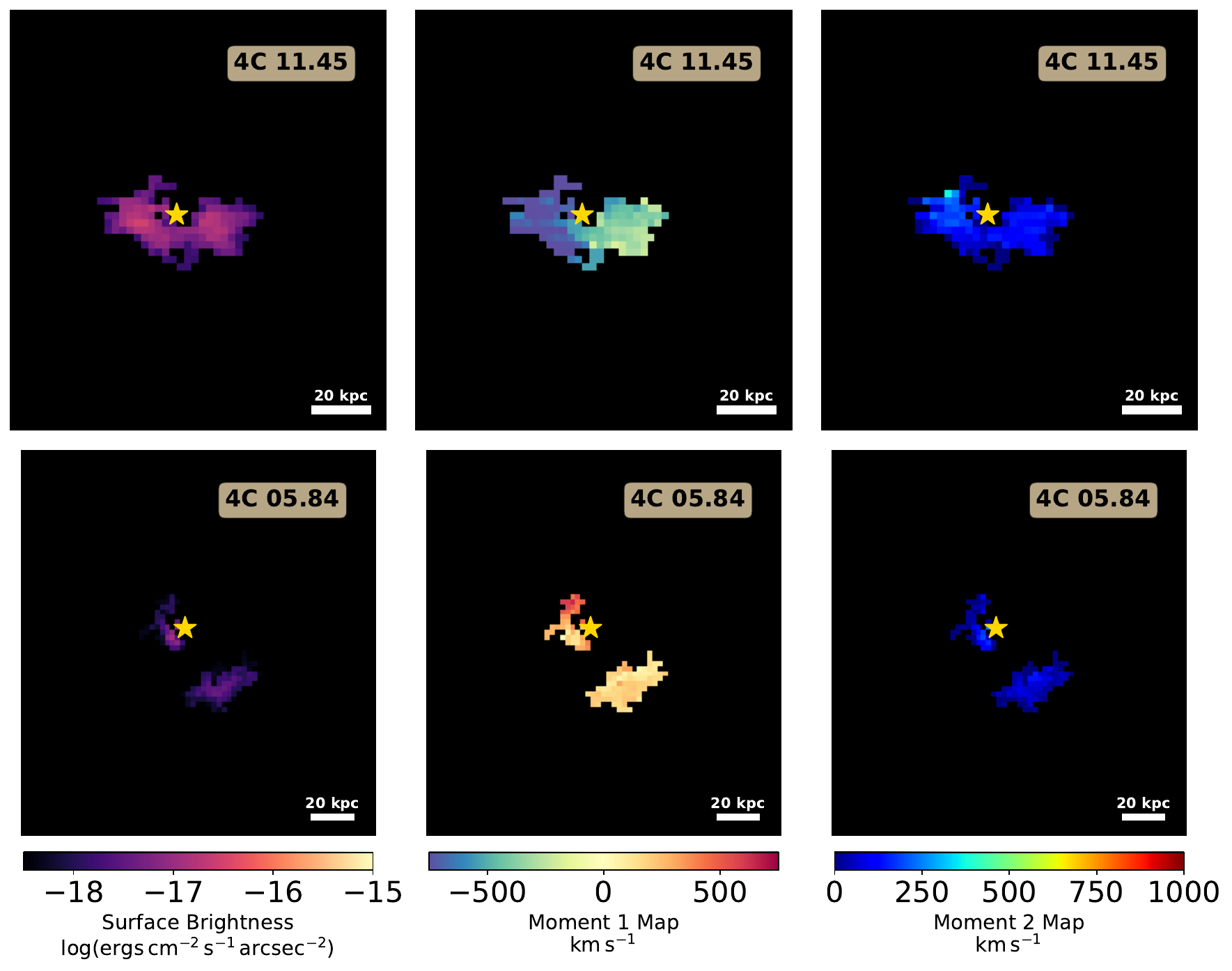}
    
    \caption{PSF-subtracted images of the \heii\, surface brightness, Moment 1 and Moment 2 maps for the two sources with the lowest (4C 11.45) and highest (4C 05.84) exposure times on source from the QUART sample. All figures have an on-sky position angle of 0. The systemic redshift used to calculate the moment maps is used from OSIRIS measurements in \citet{QUART_OSIRIS} and as shown in Table \ref{tab:QUART_Members}. The location of the quasar is denoted with a gold star marker. The complete figure set (6 images) is available in the online journal. }
    \label{fig:KCWI_HeII}
\end{figure*}

\begin{figure*}[!ht]
    \centering
    \includegraphics[width=\linewidth]{ 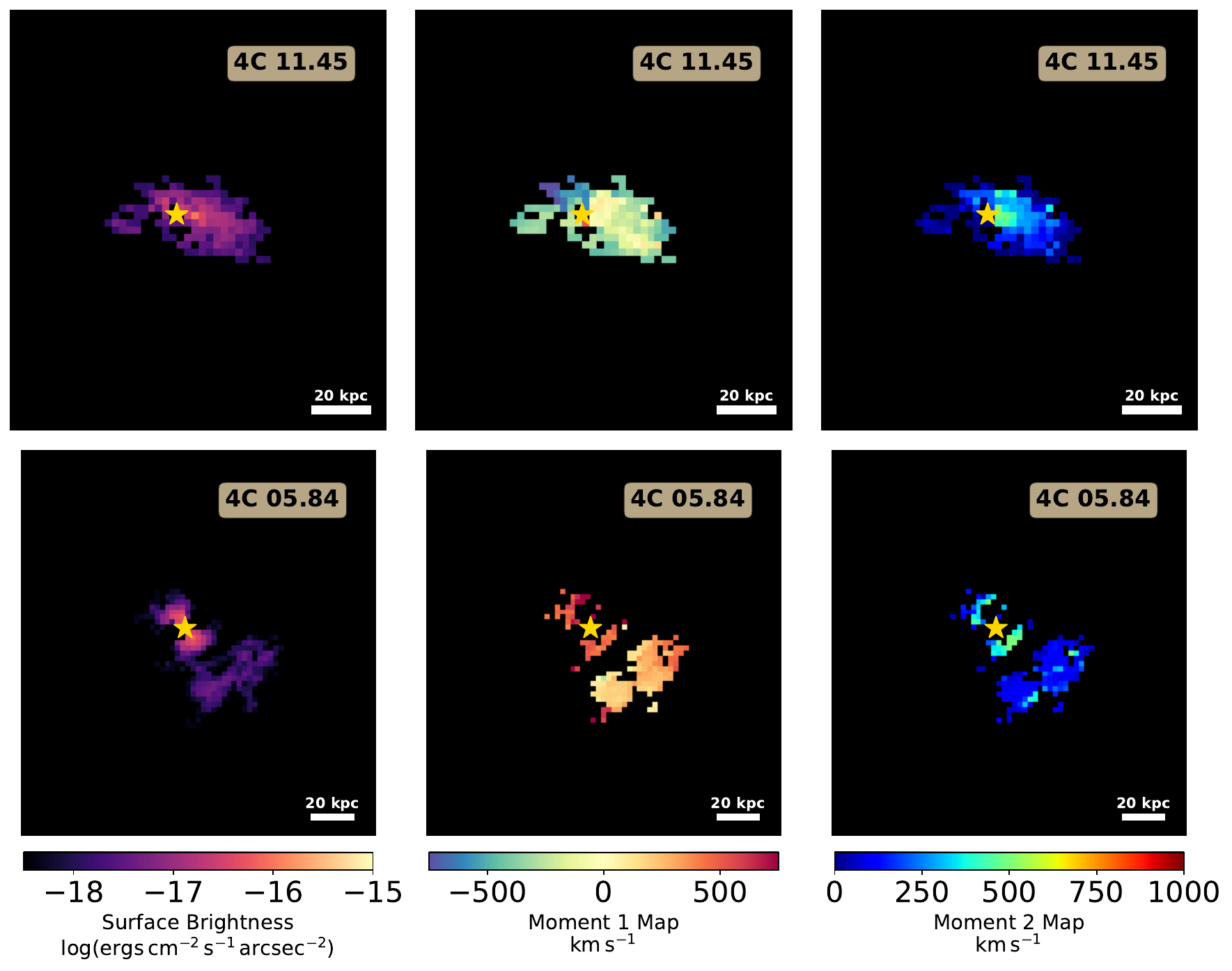}
    
    \caption{PSF-subtracted images of the \civ\, surface brightness, Moment 1 and Moment 2 maps for the two sources with the lowest (4C 11.45) and highest (4C 05.84) exposure times on source from the QUART sample. The systemic redshift used to calculate the moment maps is used from OSIRIS measurements in \citet{QUART_OSIRIS} and as shown in Table \ref{tab:QUART_Members}. The location of the quasar is denoted with a gold star marker. The complete figure set (7 images) is available in the online journal. }
    \label{fig:KCWI_CIV}
\end{figure*}

\subsection{Spatially mapping the emission around quasars}
The emission around the quasars 3C 9 and 4C 05.84 is described in \cite{Sabhlok2024a}, while that around 7C +1354 and 4C 09.17 is described in \cite{Vayner2023}. The emission around these sources is briefly summarized here, and subsequently the other three quasars are described. The \lya, \heii, and \civ\, moment maps correspond to Figures \ref{fig:KCWI_Lya}, \ref{fig:KCWI_HeII} and \ref{fig:KCWI_CIV} respectively. 

\subsubsection{3C 9 and 4C 05.84}
3C 9 and 4C 05.84 both exhbit \lya\, nebulae $\sim$ 100 kpc in diameter, with major axis from N to S for 3C 9 and NE to SW for 4C 05.84. Both sources show biconical morphology for \heii\, and \civ\, emission in the inner CGM ($<$ 15 kpc radius) and extended 20 kpc diameter \heii\, nebulae, kinematically distinct from the CGM around the quasar host galaxy which are CGM subhalos of companion galaxies as spectroscopically confirmed in \cite{Sabhlok2024a}. While the subhalo around the 3C 9 companion does not show any \civ\, emission and is redshifted with respect to the quasar by $\sim 900$ \kms, the 4C 05.84 companion subhalo does show \civ\, emission and is redshifted by $\sim 450$ \kms. 

\subsubsection{7C 1354+2552 and 4C 09.17}
7C+1354 and 4C 09.17 both have \lya\, emission with a maximum extent of $\sim 90$ kpc and \heii\, and \civ\, emission with a maximum extent of $\sim 30-50$ kpc. The radial velocity maps show velocity gradients which ranges from $-500$ \kms to $+500$ \kms. The \lya\, emission in 7C 1354+2552 connects a bridge between quasar host and three ALMA galaxies, whereas narrow gas streams associated with companion galaxies are detected in the 4C 09.17 system. The \heii\, nebulae in both systems are found to be associated with an overdensity of galaxies, based on observed kinematics and spatial correlation. The \heii\, emission encompasses the associated galaxies in the observed field of view of KCWI. 

\subsubsection{4C 57.29}
4C 57.29 displays a \lya\, nebula $\sim$ 100 kpc in diameter, with the major axis extending from NE to SW of the quasar. The \lya\, emission contains velocities ranging from -500 \kms to 600 \kms, but appears to be tracing the turbulent motion in the CGM, as it does not show any signs of distinct kinematic components or systematic trends associated with observed sources in the field of view. The \heii\, emission is detected very close to the quasar host galaxy and only a small fraction of detected emission is spatially resolved and free of any PSF subtraction residuals. The \heii\, velocity suggests a small blueshift with respect to the quasar, corresponding to a velocity of -200 \kms. 

\subsubsection{4C 04.81}
4C 04.81 displays a \lya\, nebula $\sim$ 80 kpc in diameter, with the majority of emission originating close to the quasar host galaxy. The emission close to the quasar host galaxy is redshifted, with velocities $<$ 500 \kms. The nebula shows kinematic components with velocity -300 \kms\, SE and North of the quasar, but these are low surface brightness features, so it is hard to state whether this is a distinct kinematic component or turbulent gas in the CGM. While we detect \heii\, the flux is mixed with the sky line so moment maps for \heii\, could not be extracted at the same level of confidence and have been omitted here. We detect an extended $\sim$ 25 kpc diameter \civ\, nebula around the host galaxy.

\subsubsection{4C 11.45}
4C 11.45 displays a \lya\, nebula $\sim$ 100 kpc in diameter. The nebula displays strong velocity gradients from -900 \kms\, to $\sim$ 500 \kms. 4C 11.45 also displays a distinct filament shaped kinematic component in \lya, extending NW from the quasar $\sim$ 40 kpc away, with a velocity that decreases from 400 \kms as the on sky separation from the quasar increases. This is also visible in the Moment 2 map, where the structure appears to create a high dispersion velocity, which is in fact two \lya\, kinematic components superimposed on each other, artificially boosting the dispersion values. 4C 11.45 also displays an extended \heii\, nebula about 40 kpc in diameter, showing a smooth velocity gradient from -700 \kms on the West to -300 \kms on the East of the quasar.

\subsection{Radial profiles}

\begin{figure*}
    \centering
    \includegraphics[width=\linewidth]{  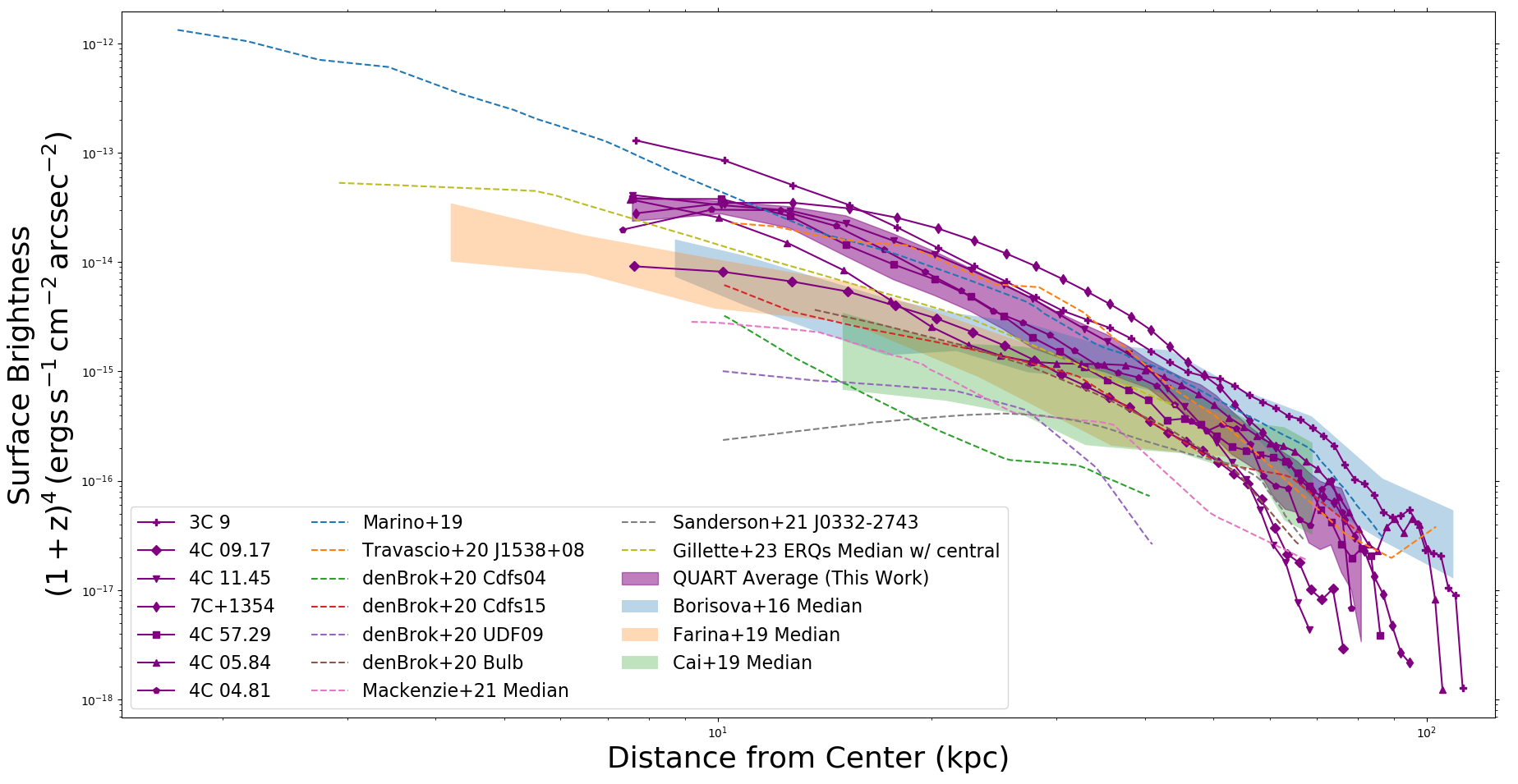}
    \caption{\lya\, surface brightness radial profiles extracted from Moment 0 maps for the QUART sample shown for individual sources (purple). The 25th and 75th percentile centered on the median are indicated with the purple fill. Radial profiles start at 8 kpc from the quasar, to remove any contamination due to PSF subtraction residuals. We find the QUART sample profiles to be brighter than other samples shown here from the literature for the inner 30 kpc. }
    \label{fig:Radial_Profiles}
\end{figure*}

Figure \ref{fig:Radial_Profiles} shows the radial surface brightness profiles around the QUART  sample in comparison to other \lya\, nebulae from the literature. Emission from the inner 8 kpc is not considered for the entire sample to avoid any issues due to residuals from PSF subtraction. However, we only indicate the percentile values indicating 25th and 75th percentile of the sample radial profiles out to $\sim$ 80 kpc, since half our sample have maximum extent up to this distance. 

The maximum variation observed surface brightness radial profiles is within a factor of 10. This is interesting since the environments around these quasars exhibit a range of substructures. The quasars 3C 9 and 4C 05.84 display evidence of companion galaxies in the CGM, confirmed with follow-up spectroscopy \citep{Sabhlok2024a} whereas 7C +1354 and 4C 09.17 show \lya\, substructure in absorption \citep{Vayner2023} as well as presence of companion galaxies in molecular emission \citep{QUART_ALMA}. The quasar 4C 11.45 displays a kinematic component suggestive of a cosmic filament in \lya. \heii\, mapping around the quasars, likely a tracer of strongly ionized gas, suggests ionization states ranging from extended nebulae around quasar host galaxies and companion galaxies to no detection around the quasar host. Thus, for radial profiles of \lya\, to follow a relatively small scatter indicates that a wide variety of CGM substructure can exhibit similar \lya\, emission.

The \lya\, emission in the inner 10--30 kpc is 2--10 times higher than values taken from the literature that are shown by the dashed lines in Figure 4. While this could be due to a selection bias, it can also originate due to enhanced emission in the inner CGM due to an increased contribution from scattering of \lya\, photons from the quasar. When considering the surface brightness at larger radii ($> 60$ kpc), due to the KCWI field of view, the observations are limited to within 8--10 \arcsec of the quasar in the EW direction, which translates to $\sim$ 68--85 kpc from the quasar at this redshift. As MUSE observations have a wider field of view, the higher brightness of $z\sim$ 3 quasars beyond the inner 60 kpc can be due to the KCWI FoV not covering the full extent of emission. This is supported by FoV filling \lya\, nebulae observed in 7C+1354, 3C 9 and 4C 05.84.  \\

\subsection{Redshift evolution}
\begin{figure*}
    \centering
    \includegraphics[width=\linewidth]{  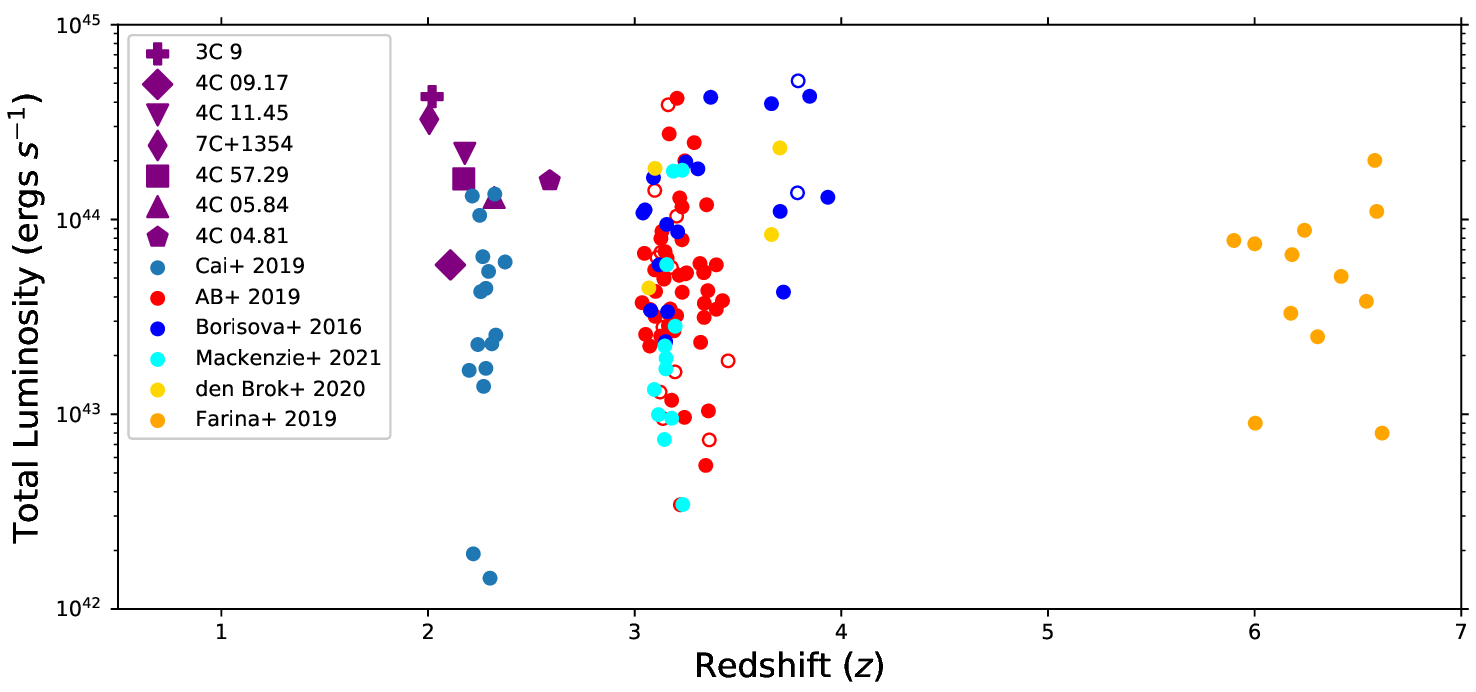}
    \caption{Redshift evolution of total \lya\, luminosity. Open circles denote radio-loud sources in the other samples where data is available. }
    \label{fig:Luminosity_Redshift}
\end{figure*}
Figure \ref{fig:Luminosity_Redshift} shows the total \lya\, luminosity as a function of redshift for the QUART sample and other \lya\, nebulae in literature \citep{Cai2019, AB2019, Borisova2016, Farina2019,  denBrok2020, Mackenzie2021}. Previous work in literature has only shown a weak redshift dependence on the total \lya\, luminosity between $2<z<4$ \citep{Borisova2016, AB2019, Cai2019}, whereas the total luminosity appears to be higher than the median luminosities at $z=6$ from \cite{Farina2019}.  While we observe an increase in luminosity from $z = 2.5$ to $z = 2$, this is statistically not a large enough sample to robustly identify evolution with redshift. However, we note that the QUART sample displays higher total \lya\, luminosity compared to other quasars at this redshift from \citep{Cai2019} and the median luminosity is comparable to the upper end of the range seen in $z > 3$ \lya\, nebulae. 

Particularly, we illustrate the radio-loud sources in \citep{AB2019} with open circles, and they display a larger scatter in luminosity than the QUART sample. Radio loud sources in \cite{AB2019} have an average (median) radio flux density values of 164 (27) mJy, which are higher than QUART average (median) values of 376 (443) mJy, but comparable at rest frame luminosity once the difference in redshift is taken into account. Assuming a spectral index of 0.7, all sources in \cite{AB2019} have intrinsic luminosity exceeding $10^{25}$ W Hz$^{-1}$ at 1.4 GHz, implying they are FR-II sources, same as the QUART sample. Similarly, the two radio sources in \cite{Borisova2016} have flux densities of 806 and 403 mJy, making the intrinsic radio luminosity brighter than the QUART sources. This indicates that radio-loudness may not be the only contributing factor in the enhanced luminosity of the QUART sample. 
\subsection{Asymmetry} \label{subsec:Asymmetry}
\begin{figure*}
    \centering
    \includegraphics[width=\linewidth]{  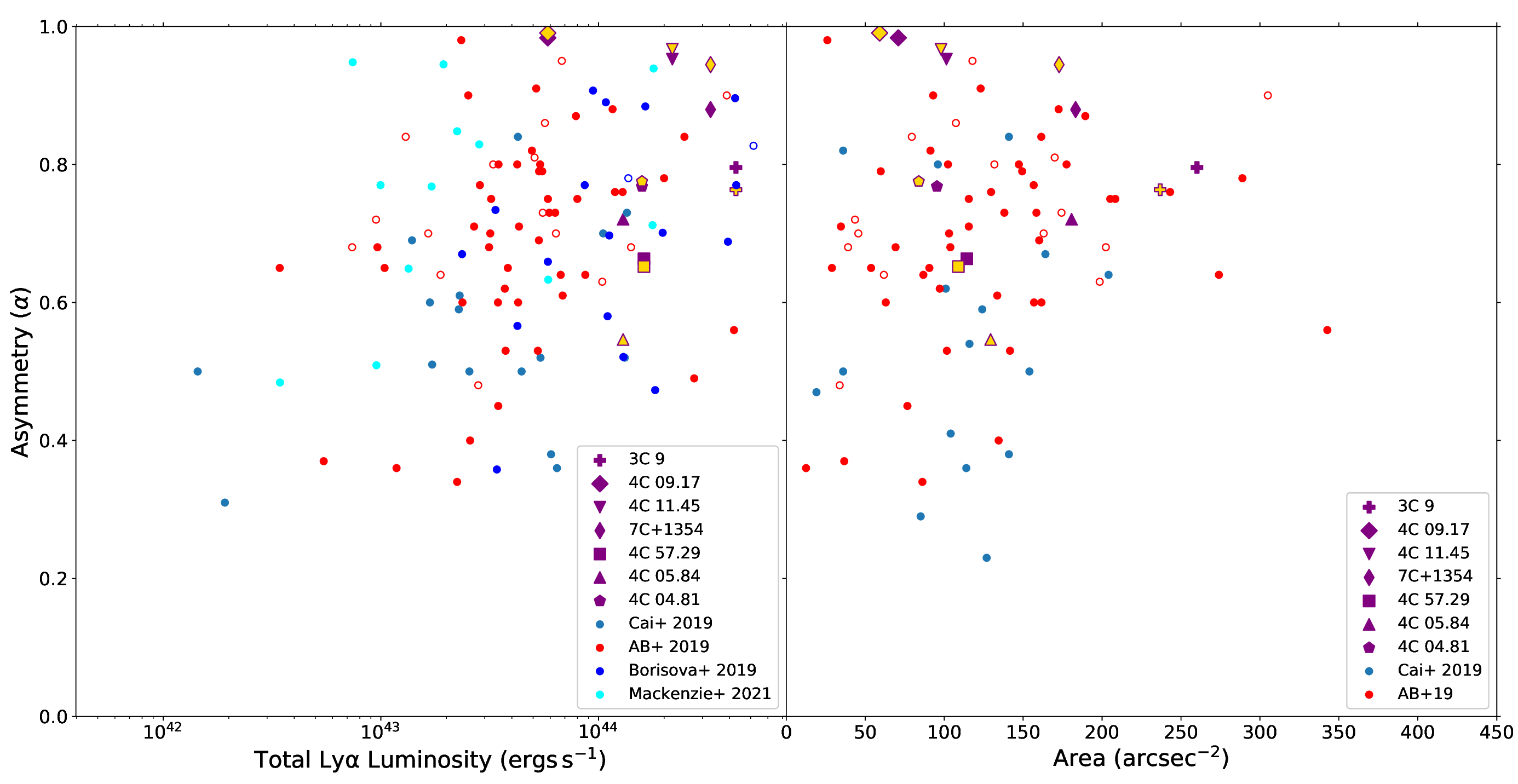}
    \caption{Asymmetry of \lya\, nebulae as a function of total luminosity and area on sky. Open circles denote radio-loud sources in the other samples where data is available. The gold markers for QUART sources denote Asymmetry calculated when considering spaxels with a minimum Surface Brightness threshold of 2 $\times$ 10$^{-18}$ \fergarc.}
    \label{fig:Asymmetry}
\end{figure*}

We calculate the second order moment of the observed flux \citep{Stoughton2002}. First we define $M_{xx}$, $M_{yy}$ and $M_{xy}$ as
\begin{equation}
    M_{xx} = \Bigg \langle \frac{(x-x_{Neb})^2}{r^2} \Bigg\rangle_f ; \,
    M_{yy} = \Bigg\langle \frac{(y-y_{Neb})^2}{r^2} \Bigg\rangle_f 
\end{equation}

\begin{equation}
        M_{xy} = \Bigg \langle \frac{(x-x_{Neb})(y-y_{Neb})}{r^2} \Bigg \rangle_f
\end{equation}

Where $x_{Neb}$ and $y_{Neb}$ are flux weighted centroids for the $2\sigma$ isophotes of the nebulae and $r$ is the distance of a given pixel from the centroid. We then calculate the Stokes parameters $Q$ and $U$ defined as 
\begin{equation}
    Q = M_{xx} - M_{yy} ;
    U = 2 M_{xy}
\end{equation}
The ratio between the minor and major axes of the nebula gives the asymmetry $\alpha$, and the angle $\phi$ is the angle between the major axis and the nearest X or Y axis given by
\begin{equation}
    \alpha = \frac{b}{a} = \frac{1 - \sqrt{Q^2 + U^2}}{1 + \sqrt{Q^2 + U^2}} 
\end{equation}
\begin{equation}\label{eq:phi}
    \phi = arctan \Bigg ( \frac{U}{Q} \Bigg )
\end{equation}

This asymmetry $\alpha$ is compared to the total \lya\, luminosity, and the on-sky projected area of the quasars in Figure \ref{fig:Asymmetry}. Due to the exquisite sensitivity of the KCWI data, we need to compare the asymmetry in other literature sources at a lower sensitivity of 2 $\times 10^{-18}$ \ferg. These data points are indicated for the QUART sample with a gold fill, showcasing the change in asymmetry and on-sky projected area as a result of higher sensitivity. 

We make an exception in the calculation of the angle of major axis for 4C 09.17 and the \heii\, and \civ\, nebulae around 4C 57.29 due to the highly asymmetric nature of the emission. In this case, we compute the centroid of emission for \lya, \heii, and \civ\, emission, and construct a major axis by joining the centroid with the location of the quasar.

The calculated angle of the moment is used to compare to the radio jet axis, so we clarify the uncertainty of the angle here. Injecting the moment map with noise realizations to get a spread of angular values gives a very small spread in the angle, as the moment calculation is flux weighted, and the high signal to noise flux close to quasar dominates the angular calculation. The pixelization of the maps is a larger uncertainty, and we consider the angular extent subtended by a single pixel at a distance of $\sim$ 30 kpc to be the uncertainty in our calculation, which is $\sim 5^{\circ}$. We choose a value of 30 kpc since this close to the radius of the 2$\sigma$ isophotes used to compute the angle for all \heii\, and \civ\, nebulae, and is also a reasonable extent for the high signal to noise \lya\, emission. Larger extents will only make the uncertainty smaller (which could be true for 7C+1354), but we choose a smaller extent to keep our uncertainty estimate uniform and conservative across all sources. 

We find that the asymmetry and on-sky area do not change significantly due to improved sensitivity except for three sources, 3C 9, 4C 05.84 and 7C 1354+2552. We note that the observed asymmetry does not systematically increase or decrease with improvement in sensitivity, as both behaviors are observed in our sample, even when QSOs with evidence for companions are excluded. These sources have the highest exposure times in the QUART sample and also display signatures of companion galaxies in the CGM. The observed \textit{increase} in asymmetry for 3C 9 and 4C 05.84 when reducing sensitivity is due to emission from the CGM subhalos being distinct and detectable at low sensitivity, whereas improved sensitivity detects intermediate lower SNR features, making the observed emission more symmetric. In the case of 7C 1354+2552, no distinct CGM subhalo is seen, even though the system contains companion galaxies, and as a result, displays lower symmetry at higher sensitivity. This could be the result of the extended \lya\, emission filling the KCWI FoV and the flux weighted symmetry calculated for this detection. 

\subsection{Luminosity}
\begin{figure*}[ht!]
    \centering
    \includegraphics[width=\linewidth]{  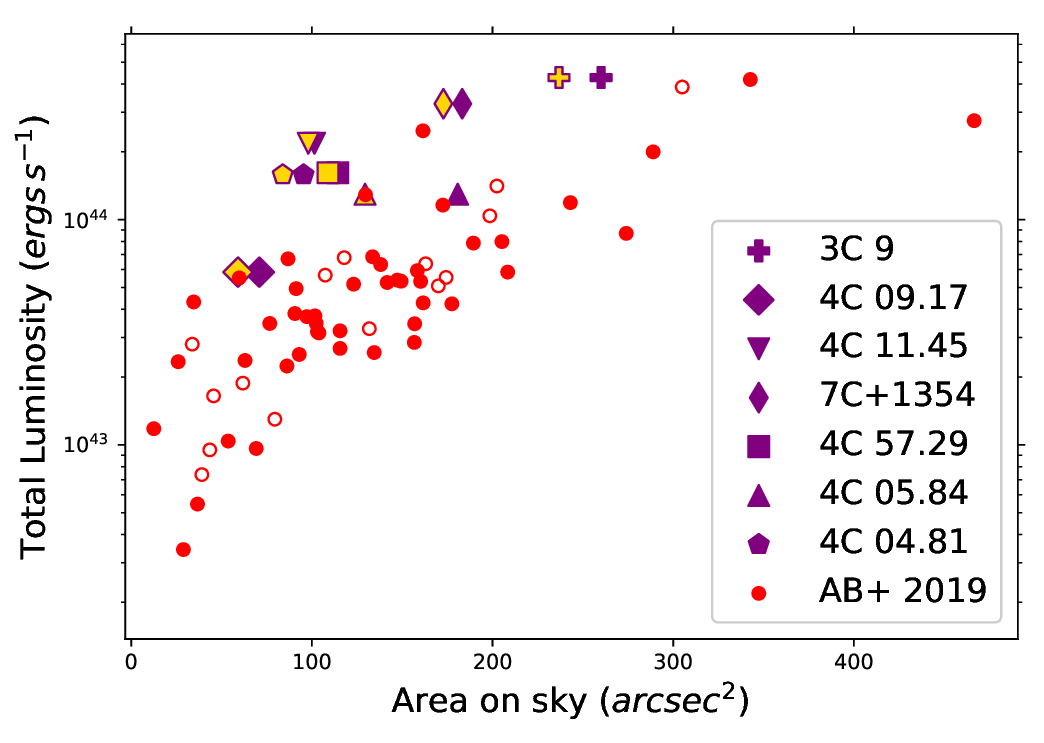}
    \caption{Total Luminosity of \lya\, emission as a function of area on sky. Open circles denote radio-loud sources in the \cite{AB2019} sample. The gold markers for QUART sources denote total luminosity when considering spaxels with a minimum Surface Brightness threshold of 2 $\times$ 10$^{-18}$ \fergarc. \label{fig:LuminosityArea} }
 \end{figure*}

Figure \ref{fig:LuminosityArea} shows the dependence of total \lya\, luminosity on the observed on-sky area in square arcseconds. We reproduce the observed increase in total luminosity as a function of area as seen in literature, showing the sources from \citep{AB2019}, where the radio-loud sources are indicated with open circles. The QUART sources do indicate a higher luminosity compared to the \citet{AB2019} sample for the same area. This comparison gets stronger when we compare for reduced sensitivity of 2 $\times 10^{-18}$ \ferg, as shown with gold-filled symbols for the QUART sample. 

\section{Discussion} \label{sec:Discussion}

\begin{figure*}[ht!]
	\centering
	\includegraphics[width=\linewidth]{ 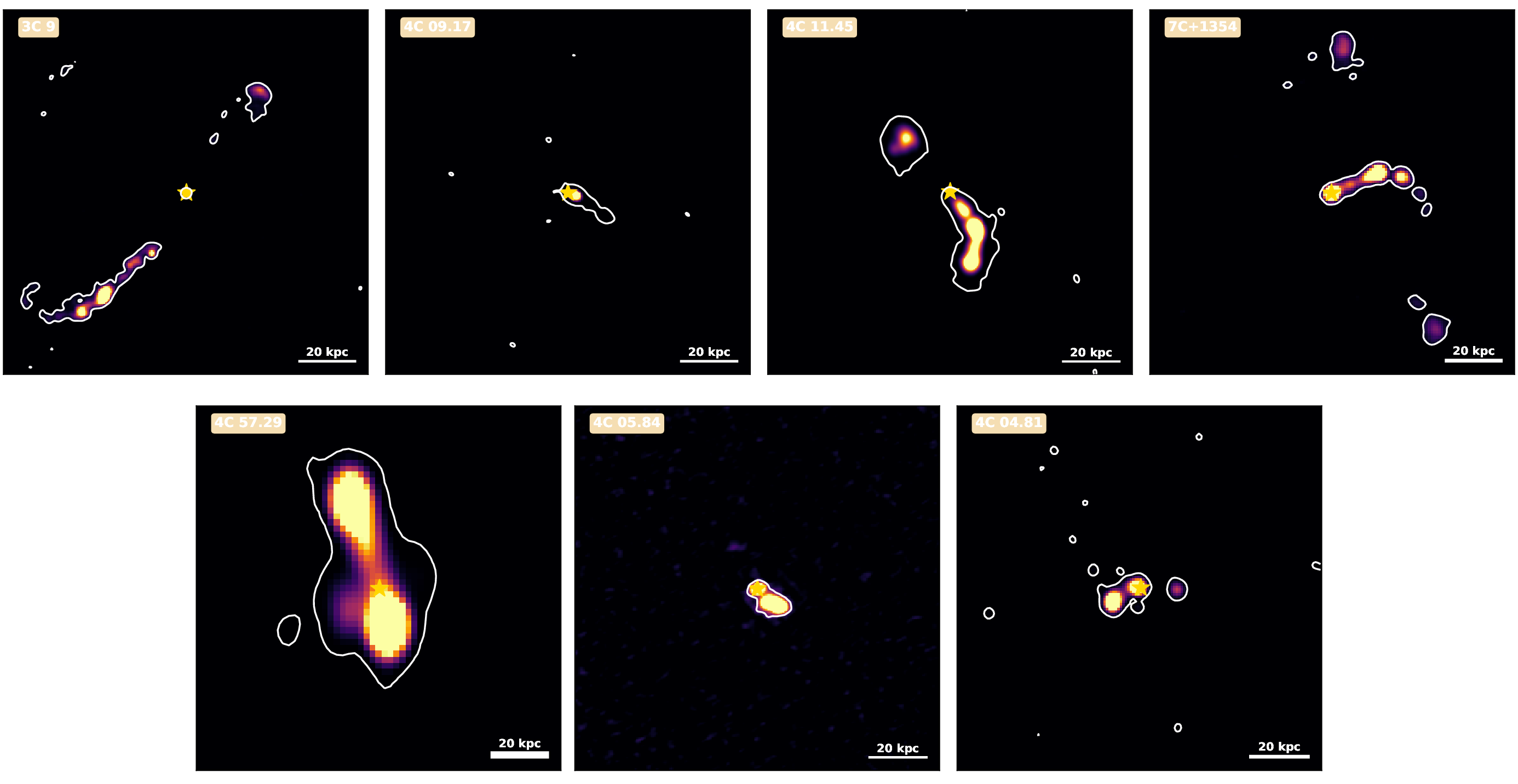}
	\caption{Radio jets around the QUART sample. The location of the quasar is denoted with a gold star. The data is described in Table \ref{tab:QUART_Radio}. The outermost contours represent a 5$\sigma$ SNR extraction limit.  }
	\label{fig:Radio_Mom0}
\end{figure*}

The radio jets around the QUART sample are shown in Figure \ref{fig:Radio_Mom0} and described in Table \ref{tab:QUART_Members}. Originally presented in the 3C \citep{Catalog_3C}, 4C \citep{Catalog_4C} and 7C \citep{Catalog_7C} radio catalogs, these sources all consist of a compact FRII radio jet \citep{Lonsdale1993}. The jets are fully contained within the KCWI field of view, making it possible to correlate the jet emission with observed nebular emission in the CGM. All jets show hotspots brighter than the emission at the core far away from the quasar at on-sky projected distances of 8 -- 30 kpc away \citep{Lonsdale1993,Barthel1988}. Due to Doppler boosting, the brighter hotspot can be identified in all the sources, which informs us that the jet is traveling towards the observer along the line of sight, in the foreground of the host galaxy \citep{Bridle1984,Barthel1988}. This allows us to identify the morphology and geometry of radio jet emission with respect to the quasar host galaxy and will be useful when comparing with nebular emissions. 

\subsection{Major axes of radio and CGM emission} \label{subsec:MajorAxes}

A comparison between the astrometry of radio jet emission and the nebular emission observed using KCWI is used to investigate if the two are related or if there is insufficient data to make a determination. The underlying theoretical rationale for this exercise is as follows. In \cite{Sabhlok2024a} we discussed how the observed nebular emissions for 3C 9 and 4C 05.84 can be described by using a two-component model, where the Inner CGM (with on-sky projection $<$ 30 kpc) is ionized directly by the quasar, whereas the outer CGM is only visible in \lya, likely due to resonant scattering. 
A radio jet by itself is unlikely to create the extended nebular emission seen in these sources, even taking into account the bow shocks that can be driven out of a radio jet. It is therefore useful to consider the radio jets as tracers of the quasar ionization cone, which can have half opening angles of $ \mathrm{<} \, 70 ^{\circ}$ depending on the individual source, but regardless of opening angle are well aligned with the radio jet \citep{Wilson1996:IonizationCones}. In contrast, more recent simulation work by \cite{Obreja2024} finds that the CGM profiles are much better approximated by half angles of $30 ^{\circ}$ or a biconical opening angle of $60 ^{\circ}$. 

We consider the CGM emission to be aligned with the ionization cone if the angular difference between the major axis of the CGM and the radio jet are less than a half-opening angle of $\sim$ 45$^{\circ}$, consistent with ionization cones seen in high-redshift AGN \citep{Wylezalek2016:IonizationCones, Zhuang2018:TorusCones}. This half opening angle is wider than those approximated by \cite{Obreja2024} since radio jets are only proxies for tracing the AGN ionization cone. This larger opening angle allows for a misalignment between the radio jet and ionization cone, consistent with previous work \citep{Wilson1996:IonizationCones}. If the radio jets are indeed aligned with the ionization cone of the quasar, then we would expect to see the nebular emission aligned with the radio jet major axis.

\begin{figure*}[ht!]
	\centering
	\includegraphics[width=\linewidth]{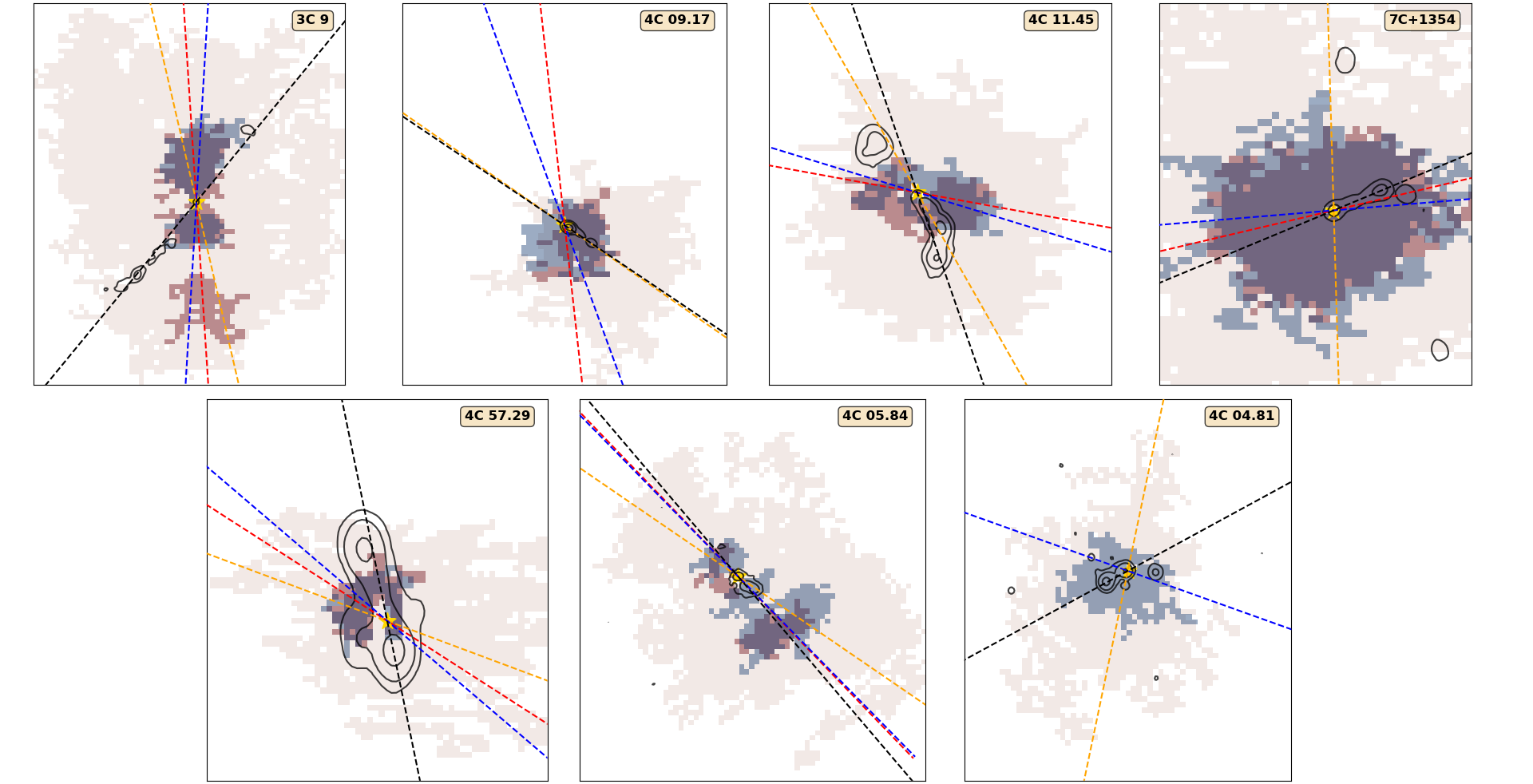}
	\caption{The major axes for the radio and nebular emission are shown around all QUART sources. The radio contours are shown in black, with the flux weighted major axis shown in black dashed lines. Similarly, \lya, \heii, and \civ\, are shown in orange, red and blue shades, with the major axes plotted as dashed lines with the same colors respectively. }
	\label{fig:All_axes}
\end{figure*}

The calculation of the major axes orientation for nebular emission results from the asymmetry estimation as described in Section \ref{subsec:Asymmetry} following the procedure described in \cite{Sabhlok2024a}. Figure \ref{fig:All_axes} shows the major axes derived for \lya\, (orange), \heii\, (red) and \civ\, (blue) emission. We further plot the the major axis of the radio jet emission (black) by plotting the vector from the quasar to the flux weighted centroid of radio emission. Due to the extended morphologies of all radio sources, the calculated centroids are not dominated by the emission from the core. 

We calculate the value of the angle between \lya\, emission and the radio axis from the results of \cite{Heckman1991a}, and their designated errors, shown in Table \ref{tab:QUART_Axis_Angles}. We find a reasonable agreement between the two sets of values.

\begin{deluxetable}{lcccc}
\tablenum{3}
\caption{Angles (in degrees) between major axes of CGM emission and the radio jet emission. All angles have an uncertainty of }$\sim 5 \degree$ \label{tab:QUART_Axis_Angles}
\tablehead{\colhead{Source Name} &
\colhead{\lya} &
\colhead{\heii} &
\colhead{\civ} & 
\colhead{\lya [H91] \tablenotemark{a}} 
}

\startdata
3C 9     & 52.5 & 43.2 & 36.1 & 36 $\pm$ 15  \\
4C 09.17 & 0.8 & 49.7 & 36.  & 15 $\pm$ 30\\
4C 11.45 & 10.5 & 60.6 & 53.9  & 9 $\pm$ 15 \\
7C +1354 & 69.1 & 9.4  & 17.8  & 37 $\pm$ 30\\
4C 57.29 & 57.9 & 45.6 & 37.9  & - \tablenotemark{b} \\
4C 05.84 & 15.2 & 3.5  & 4.0 & 2 $\pm$ 30 \\
4C 04.81 & 49.7 & -    & 48.2 & 33 $\pm$ 30 \\
\enddata
\tablenotetext{a}{Values from \cite{Heckman1991a}}
\tablenotetext{b}{Denoted as too round by \cite{Heckman1991a} to calculate an angle for \lya\, emission}
\end{deluxetable}
We find the \lya\, is aligned with the radio jet emission in 3 out of 7 sources respectively. The \heii\, emission is aligned for 4 out of 6 and \civ\, emission in 5 out of the 7 sources. Given the uncertainty in the major axis calculation which we have assumed to be $5^{\circ}$, we note that the major axes of 4C 04.81, and the \heii\, emission in 4C 09.17 are essentially within the uncertainty when compared to the radio jet axis. As the \heii, and \civ\, nebulae are spatially resolved in all of these sources, so the agreement is not an artifact of PSF subtraction of the quasar emission. This is discussed further in Section \ref{subsec:LyaEmissionAtJets}. Unlike \lya, \heii\, and \civ\, emission must be powered largely by ionizing radiation from the host galaxy AGN. 

The detection of extended \heii\, and \civ\, emission around most of our sample and the alignment of major axes of emission for 5 out of 7 sources in the case of \civ\, emission with the radio jet emission indicate the emission is powered by quasar photoionization. We note that the cases where \civ\, major axis is aligned with the radio jet emission and \heii\, is not, the nebulae surface brightness maps show an overlap between the two emission lines, consistent with a mutual source of ionization.   

\subsection{\heii\, and \civ\, emission around the QUART sample} \label{subsec:heiiemission}
Extended \heii\, emission is detected around 6 out of 7 QUART sources whereas \civ\, emission is detected around all 7 sources. The \heii\, and \civ\, nebulae extend from $\sim$ 20 kpc (4C 57.29) to $\sim$ 80 kpc (7C+1354). The nebulae range in velocity from -800 to +800 \kms as indicated by the moment maps.

Comparing the nebulae to the location of the radio jets in each quasar offers an avenue to explore the role of geometry and photoionization in powering these nebulae. Given $\psi$ as the ratio of intensities between the brighter and fainter jet measured at the same distance from the parent object at low transverse resolution (to minimize the influence on $\psi$ of differences in their expansion rates), the jets are classified as one sided if $\psi > 4$ and two sided for which $\psi < 4$ everywhere. Since all radio jets in the QUART sample are FR-II one sided jets \citep{Bridle1984, Barthel1988, Lonsdale1993}, the direction of the jet with respect to the quasar indicates the near side beaming towards the observer. If the observed \heii\, and \civ\, emission is due to quasar photoionization, the emission should be coincident with the direction of the jet, since the photoionization cone of the quasar should be aligned with the radio jet. This photoionization cone can be clearly seen in \heii\, and \civ\, emission in 3C 9, and in \civ\, emission in 4C 05.84. These photoionization cones are consistent with the direction of the radio jets, notwithstanding projection effects along the line of sight. While the majority of \heii\, and \civ\, emission is in the same direction as the radio jet, there are a few exceptions -
\begin{enumerate}
    \item 3C 9 shows a large 15--30 kpc \heii\, nebula south of the quasar offset from the radio jet emission, with no spatially resolved \civ\, emission. This is the CGM subhalo of the companion galaxy of the quasar as discussed in \cite{Sabhlok2024a}. Moreover, while the jet in 3C 9 is brighter southeast of the quasar, the biconical structure seen in \civ\, and \heii\, is in fact more extended north of the quasar. 
    \item While the radio jet in 4C 09.17 points southwest of the quasar, the \heii\, and \civ\, emission is located to the North of the quasar.
    \item 7C+1354 shows the most extended \heii\, nebula in our sample. However, the radio jet points west of the nebula, with no eastern lobe detected. Additionally, two radio jet relics can be seen North and South of the quasar, $\sim$ 40 kpc away from the quasar. If these are radio relics of past quasar activity, they indicate a significant change in the direction of the radio jet. If the AGN ionization cone is also presumed to have changed direction in the same time, that could explain the significantly larger \heii\, and \civ\, emission around the quasar and relatively symmetric nature of this emission compared to the other sources in the sample.  
\end{enumerate}

\subsection{Sources of Ionization for the nebulae} \label{subsec:sourcesofionization}

We now consider potential sources that can power the emission in the CGM around these nebulae. There are three potential sources we consider, i.e. photoionization by the quasar, shock excitation due to interaction between the radio jet and the CGM, and resonant scattering of photons. The last scenario is only considered for \lya\, radiation.

\begin{figure*}[ht!]
    \centering
    \includegraphics[width=1.0 \linewidth]{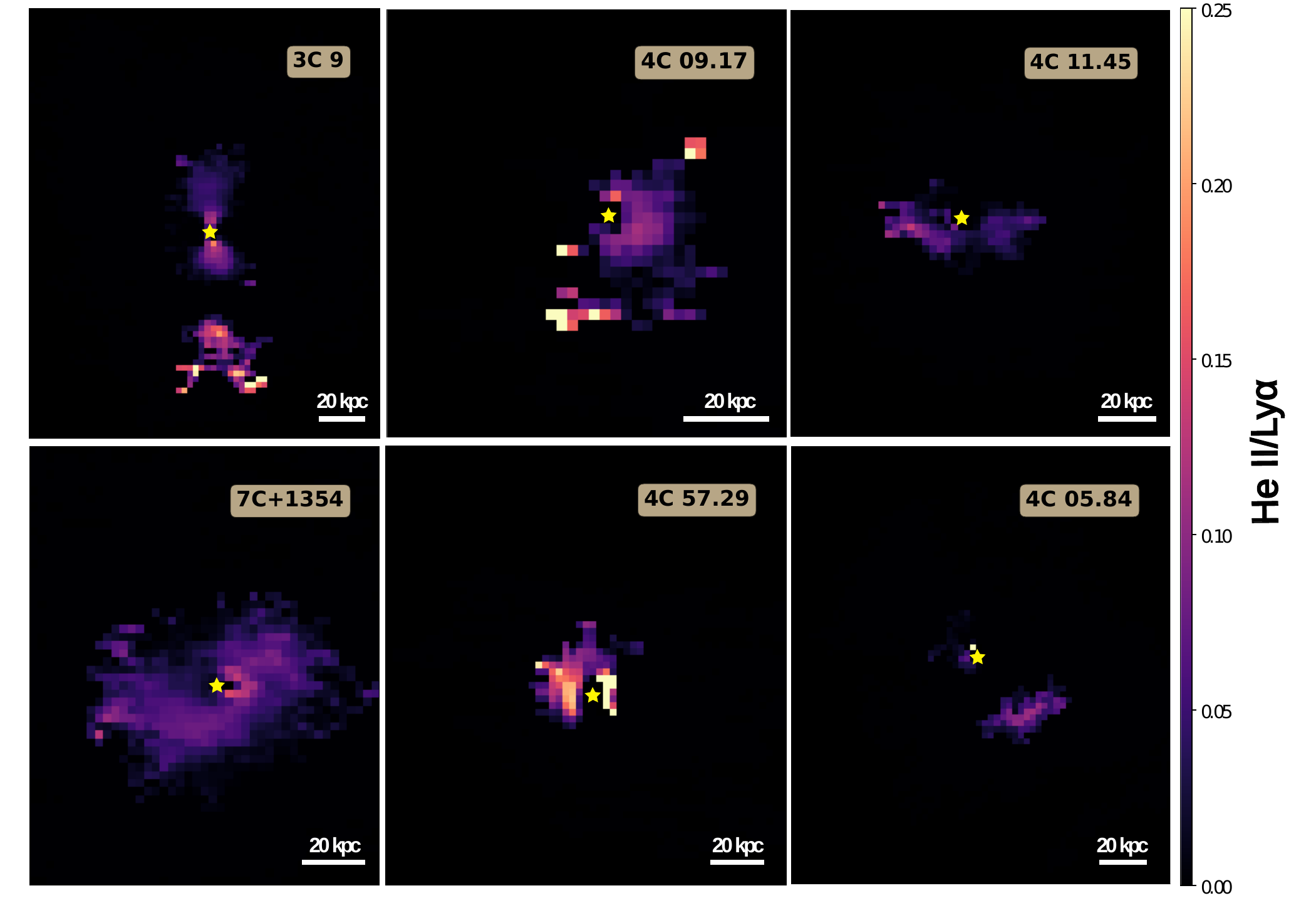}
    \caption{\heii/\lya\, emission ratio around 6 out of 7 quasars. The emission ratios show a radially decreasing gradient around all sources, consistent with photoionzation from the quasar. Section \ref{subsec:sourcesofionization} discusses the details of the calculation. }
    \label{fig:IonizationSource}
\end{figure*}

We investigate quasar photoionization by constructing \heii/\lya\, emission ratio maps as shown in Figure \ref{fig:IonizationSource}. We emphasize that this discussion only investigates the source of ionization where \heii\, emission is observed. The maps shown in Figure \ref{fig:IonizationSource} only show regions where the ratio is observed. We refer the reader to Figure \ref{fig:All_axes} for relative size of the \heii\, nebulae and the larger \lya\, nebulae. We reference the calculation in \citep{Cantalupo2019} to consider the observed ratio of $\langle F_{\mathrm{He\,II}} \rangle/\langle F_{\mathrm{Ly}\alpha} \rangle$, which we briefly summarize here.

\begin{equation} 
\frac{\langle F_{\mathrm{He\,II}} \rangle}{\langle F_{\mathrm{Ly}\alpha} \rangle} = \frac{h\nu_{\mathrm{He\,II}}\,\alpha^{\mathrm{eff}}_{\mathrm{He\,II}}(T)\, \langle n_e n_{\mathrm{He\,III}} \rangle} {h\nu_{\mathrm{Ly}\alpha}\,\alpha^{\mathrm{eff}}_{\mathrm{Ly}\alpha}(T)\, \langle n_e n_p \rangle} 
\end{equation}

Assuming that the gas is at a uniform temperature, this reduces to -

\begin{equation} \frac{\langle F_{\mathrm{He\,II}} \rangle}{\langle F_{\mathrm{Ly}\alpha} \rangle} \simeq R_0(T)\, \frac{\langle n_H^2\, x_{\mathrm{He\,III}}\, x_{\mathrm{H\,II}} \rangle} {\langle n_H^2\, x_{\mathrm{H\,II}}^2 \rangle} \end{equation}

where $R_0(T) \sim 0.23$ for Case A reionization conditions and $R_0(T) \sim 0.3$ for Case B reionization conditions, at a Temperature of $\sim 2 \times 10^4 K$. \cite{Cantalupo2019} find a critical hydrogen density of -

\begin{equation}
n_{\mathrm{H}}^{\mathrm{HI}, \mathrm{crit}} \simeq \frac{\Gamma_{\mathrm{HL}}}{\alpha_{\mathrm{HII}}} \simeq 1500\left(\frac{r}{500 \mathrm{kpc}}\right)^{-2} \mathrm{~cm}^{-3}
\end{equation}

above which Hydrogen becomes mostly neutral. This relies on the strength of the ionizing radiation from the quasar. Since the sources in the QUART sample have a bolometric luminosity within a factor of 10--100, and the critical density is higher at a lower threshold of 100 kpc (more representative of the QUART sample), it is reasonable to assume that the phase of hydrogen probed by \lya\, radiation consists of mostly ionized hydrogen. This means the expression for observed \heii/\lya\, ratio reduces to 

\begin{equation}
\begin{aligned}
\frac{\langle F_{\mathrm{He\,II}} \rangle}{\langle F_{\mathrm{Ly}\alpha} \rangle}
&\simeq
R_0(T)\,
\frac{\langle n_H^2\, x_{\mathrm{He\,III}} \rangle}
{\langle n_H^2 \rangle} \\
&=
R_0(T)\,
\frac{\int_V x_{\mathrm{He\,III}}\, n_H^2 \, dV}
{\int_V n_H^2 \, dV}.
\end{aligned}
\end{equation}

This implies the observed ratio is weighted by the square of the density of Hydrogen in the CGM. Given the observed \heii/\lya\, profiles shown in Figure \ref{fig:IonizationSource}, we see that the observed ratio is below the peak value of 0.23 for these CGM observations. The only exceptions are a few pixels south of 3C 9 and a small clump NE and SW of 4C 09.17, both of which are close to the 2$\sigma$ detection limit and likely result of noisy detections. Thus, the observed Helium is not fully ionized around any of the sources. Now, if the observed ratio traces the ionization of Helium, then the ratio must decrease with distance, which is consistent with most observations, except 7C+1354, where there is evidence for an extended region where a radial decrease is not clear and obvious. Given the alignment of radio axes with respect to the \heii\, nebulae, a photoionization and subsequent recombination scenario is consistent with the observations, for regions with \heii\, emission. In the regions where no \heii\, emission is observed, the \heii/\lya\, ratio is low, which can indicate partial illumination. In this case, \lya\, emission from the extended region can be due to other effects, such as resonant scattering. 

We then consider the extended \lya\, emission from the regions beyond the inner 30 kpc. While \lya\, emission is consistent with photoionization, previous work in literature has discussed resonant scattering and disfavored as a major mechanism for emission, since \lya\, photons scatter in velocity space and escape the nebulae within the inner 10 kpc \citep{Cantalupo2005, Dijkstra2006, AB2015}. However, if the radio jets trace out ionization cones, then a much larger fraction of \lya\ photons may escape selectively in the radio jet direction and scatter from cold clouds much further away from the quasar, as the cold cloud fraction is reduced. We see indications of this in Figure \ref{fig:NebularEmissionAtRadioJets}, where 4C 09.17, 7C+1354 and 4C 04.81 all show double peaked \lya\, profiles extracted from the location of radio jets, which would be expected in the case of resonant scattering of \lya\, photons. We note that this is in agreement with simulations of halos where \lya\, radiative transfer has been applied in post processing \citep{Costa2022}, albeit in halos with $z>6$. Future observations of \ha\, at these locations are required to confirm whether the underlying gas distribution agrees with this interpretation.

\begin{figure}[ht!]
    \centering
    \includegraphics[width=0.4 \textwidth]{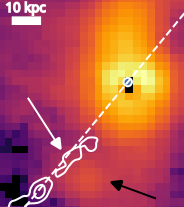}
    \caption{The \lya\, surface brightness map of 3C 9 zoomed in to show the brighter \lya\, emission knot (indicated with a black arrow) near the possible site of a re-collimation shock from the radio jet (indicated with a white arrow). No \heii, \civ\, emission is seen at the location of the knot.} 
    \label{fig:RecollimationShock}
\end{figure}

Lastly, we consider the presence of shocks at CGM scales as a possible mechanism for \lya\, excitation, especially in the presence of radio jets. While radio jet ISM interaction can create a large quantity of ionizing radiation, this likely happens close to the host galaxies, where quasar PSF subtraction residuals affect the observed fluxes. Firstly, we note that the kinematics of all three emission lines do not show high dispersion at the location of the radio jets. Thus if a shock is present, it is not visible as high velocity gas relative to the surroundings. Secondly, if a shock is powering the emission, this should be visible in the \heii/\lya\, ionization tracer, and the maps in Figure \ref{fig:IonizationSource} should show decreasing ratios at increasing transverse distances from the radio jet. This is not what is observed for these sources. The only case where we might see evidence of a shock is a small knot SE of 3C 9, where we see an increased surface brightness of \lya\, transverse to the jet axis as shown in Figure \ref{fig:RecollimationShock}, where no \heii, \civ\, or companion galaxies are seen. While we cannot rule out this being the effect of enhanced emission due to geometric effects, this is the only indication of shock induced \lya\, emission we see in our sample. 

To summarize, these KCWI observations are consistent with a model where the nebular emission is powered by quasar photionization and subsequent recombination, as traced by gas kinematics and the observed \heii/\lya\, ratio ionization tracer in the region where \heii\, emission is observed. In the region with extended \lya\, emission with no \heii\, observation, partial illumination is indicated which could imply other effects such as resonant scattering being responsible for \lya\, emission.

\subsection{\lya\, emission at the location of radio jets} \label{subsec:LyaEmissionAtJets}
\begin{figure*}[ht!]
    \centering
    \includegraphics[width=1.0 \linewidth]{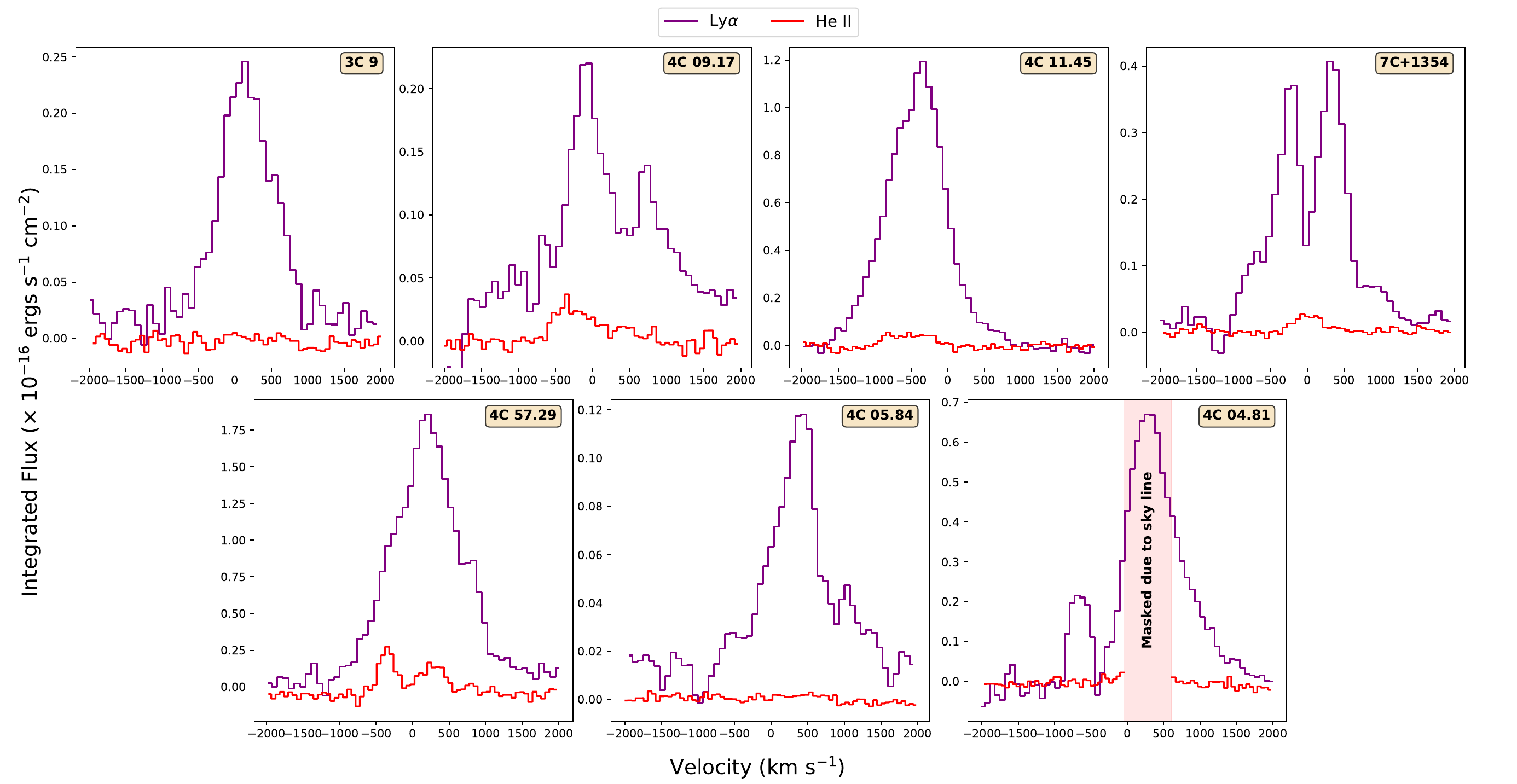}
    \caption{\lya\, and \heii\, emission at the location of the radio jets. The velocity is calculated with respect to the systemic redshift of the quasar. The implications for individual sources are discussed in Section \ref{subsec:LyaEmissionAtJets} }
    \label{fig:NebularEmissionAtRadioJets}
\end{figure*}

We extract \lya\, and \heii\, spectra at the location of the radio jets by taking the 5$\sigma$ contours from radio maps and extracting nebular emission from the KCWI data from those regions. Data from spaxels that are less than 3 pixels away from the quasar are excluded to remove any residual emission from the AGN. The spectra thus represent the integrated emission from the CGM at the location of the radio jets. Figure \ref{fig:NebularEmissionAtRadioJets} shows these spectra. 

Before we discuss individual sources, it is useful to consider what is expected from the nebular emission, based on radio jet models. All the jets considered in this sample are FR-II jets, which implies their emission from the hotspots far away from the AGN dominates the emission from the core \citep{Bridle1984, Lonsdale1993}. As FR-II jets typically have one lobe appear brighter than the other due to Doppler boosting of the synchrotron radiation, it establishes whether that side of the jet is in front (if brighter) or behind (if fainter) the quasar. Thus, if these jets are driving outflows, the emissions from outflows in front of the galaxy should appear to be blueshifted, whereas outflows behind the galaxy should be redshifted. This is particularly helpful since many of the jets in this sample are largely one sided, and thus the expectation is for the blueshifted component of nebular emission to dominate. However, as \lya\, is the primary emission line being observed, it is difficult to directly ascribe gas motions to observed components due to the resonant nature of the line that can lead to scattering. Or to rephrase, the broadening of \lya\, due to the presence of an outflow is degenerate with the broadening due to the random walk in frequency/wavelength space experienced by the photons as a result of repeated scattering throughout the medium. 

The presence of \heii\, emission from the same region helps resolve this degeneracy to some extent, as the presence of \heii\, likely indicates the true gas velocities in the CGM. The absence of \heii\, in these spectra is thus indicative of either lower density gas being primarily responsible for the \lya\, emission, due to the squared gas weighting of the observed \heii/\lya\, ratio as described in Section \ref{subsec:sourcesofionization}, or larger contributions due to scattering of \lya in the absence of shocks.

An additional problem in identifying components occurs due to the degeneracy in distinguishing two separate kinematic components or one component with a foreground absorber. \cite{Vayner2023} and \cite{Wang23} have shown these foreground absorbers cover a large fraction of the \lya\, nebula and can be spatially resolved by mapping the absorption depth and width across the nebula. The presence of \heii\, emission can help distinguish between these degeneracies as well, as multiple \lya\, peaks with corresponding \heii\, peaks are likely to be two independent components, whereas the presence of a single \heii\, component in between two \lya\, components likely indicates the absorption scenario. 

The \civ\, emission provides further probes regarding the ionization state and metallicity of the CGM. As \heii\, and \civ\, are both high ionization lines, given an ionizing radiation field, \civ\, emission traces the metallicity of the CGM. Whereas Helium can have a primordial origin, Carbon must necessarily originate from star formation in the host galaxies. Thus, areas indicating \civ\, and \heii\, emission simultaneously with similar kinematics likely indicate the presence of the quasar ionizing radiation, whereas observations of \heii\, without \civ\, may indicate the presence of pristine primordial CGM gas as seen in 3C 9 \citep{Sabhlok2024a}. Lastly, regions where we see only \civ\, emission and no \heii\, emission are likely rare, as \heii\, is present in the CGM when \civ\, is seen in high redshift radio galaxies \citep{VillarMartin2007a, Travascio2020}, the opposite being more likely for CGM around radio-quiet quasars \citep{Borisova2016}. 

We now consider individual sources and comment on the nebular emission, given the observed radio jet emission.

\textit{3C 9}: 3C 9 shows \lya\, emission centered at the quasar redshift. There is no evidence of significant \heii\, emission corresponding to the \lya\, emission at the location of the radio jet, indicating that the \lya\, emission largely originates from photoionization and subsequent recombination. We do see the presence of a \lya\, knot that is aligned transverse to the jet axis SW of the quasar. This emission could be powered by ionizing photons created from a recollimation shock as discussed in Section \ref{subsec:sourcesofionization}.

\textit{4C 09.17}: 4C 09.17 shows distinct red and blueshifted components with a potential redshifted broad component in the spectrum. Since the on-sky projection of the radio jet is small and relatively close to the quasar, the nebular emission originates in the CGM close to the quasar host galaxy. It is interesting to note that the presence of a red and blueshifted \heii\, component corresponding to the \lya\, emission, indicating these are two independent components and not a single component with foreground absorption. The \lya\, nebula around 4C 09.17 and the radio jet are both largely on the same side of the quasar, and thus the expectation is for \lya\, to be originating in front of the quasar along the line of sight. This suggests the redshifted component corresponds to foreground gas moving towards to quasar. 

\textit{4C 11.45}: 4C 11.45 is a really interesting source due to the presence of a bent radio jet. The \lya\, emission here shows only a blueshifted component with a small contribution from \heii, indicating that the \heii\, only represents a fraction of the gas giving rise to the \lya\, emission which points to the role of scattering in \lya\, emission. Radio jets typically bend in the presence of hot gas from the intergalactic medium. The fact that the bend here is seen at the boundary of both the \heii\, and \civ\, emission detection indicates that the \heii\, and \civ\, emission boundary is tracing a change in the CGM pressure, as a bent jet at this distance from a quasar is likely encountering ram pressure due to the motion of the host galaxy through the Intergalactic medium. This observation complements our assertion of a two component model of the CGM, traced here as an abrupt change in gas pressure.

\textit{7C 1354}: The \heii\, emission around this source is by far the largest individual \heii\, nebula around any of the QUART sources. The \lya\, emission from the radio jet regions shows a bimodal distribution centered at the systemic velocity of the quasar. The \heii\, emission is centered at the quasar redshift. These results are in agreement with the picture presented in \cite{Vayner2023} that the quasar sits behind a system of foreground absorbers. 

\textit{4C 57.29}: 4C 57.29 shows the smallest \heii\, and \civ\, nebulae in the QUART sample with the projected on-sky radio jet emission extending beyond the nebulae. Nevertheless, the \heii\, emission spectra at the location of the jet indicate the presence of two distinct blue and redshifted kinematic components. 

\textit{4C 05.84}: 4C 05.84 is an interesting source that shows only a redshifted component, with no contribution from \heii\, at the location of the jet. The quasar system is home to companion galaxies SW of the quasar \citep{Sabhlok2024a}, aligned with the direction of the radio jet. Since the jet is largely one-sided, the spatial region from which the spectrum has been extracted likely traces the foreground emission, hence the largely redshifted component indicates the gas at the location of the jet is likely moving towards the quasar, potentially due to tidal interactions with the CGM subhalo of the companion galaxy. However, due to the resonant scattering of \lya\, photons, the kinematics only trace the surface of last scattering, so the indication of inflowing material is tentative and needs further investigation.

\textit{4C 04.81}: 4C 04.81 shows a distinct asymmetric profile at the location of the radio jet. The jet itself is interesting since the on-sky projection remains close to the quasar host galaxy, and yet both hotspots have comparable peak intensities. We note that this is exactly the kind of spectrum we would expect to see if \lya\, were to exactly trace the gas motion, and if the extraction was from a brighter foreground (redshifted) component and a fainter background (blueshifted) component. However, due to the lack of \heii\, detection around the source, the determination cannot be made definitively. 

To summarize, we use the assumption that the nebular emission from the locations of the radio jets originates from the foreground or the background of the galaxy, and use the \heii\, emission to confirm whether the \lya\, emission is originating from multiple kinematic components or a single component with foreground absorbers. Under these assumptions, we detect evidence for cases with multiple kinematic components in the quasars 4C 09.17 and 4C 57.29, whereas the emission in 4C 11.45 is due to a single component. The quasars 3C 9 and 4C 05.84 show no \heii\, emission at the location of the radio jets, indicating the origin of the \lya\, radiation to be from lower density cold CGM with significant contributions from scattering. We see in the case of the quasar 7C 1354+2552 that the \lya\, emission is likely due to a single component at the redshift of the quasar and is absorbed by foreground absorbers. For 4C 04.81, we detect two components but are unable to comment on their origins due to inability to extract \heii\, emission from the regions. We see the potential impact of a bent jet on the CGM in the case of 4C 11.45, where the bend occurs precisely at the boundary of \heii\, and \civ\, emission. Thus, 3 quasars show evidence of multiple kinematic components 2 show evidence of a single component, and for 2 sources the origin cannot be determined based on the available data.

\subsection{How does the CGM around blue quasars with resolved radio jets compare with CGM around other types of quasars?}
The results we have obtained are largely in agreement with the work originally done on these quasars in \cite{Heckman1991a}, who also found extended \lya\, nebulae, and the alignment with the radio jet axis. Subsequent spectroscopy also confirmed the largely gravitational motion of the gas around the nebulae, where the authors also assume companion galaxies might be responsible for some of the extended \lya\, features around the galaxies.

We now compare our results with other studies of \lya\, CGM emission in the literature, namely \cite{Cai2019} at z$=$2, \cite{Borisova2016} and \cite{AB2019} at z$=$3, which have established the ubiquity of large \lya\, nebulae around high z quasars. We find a 100\% detection rate for \lya, \heii\, and \civ emission and are able to extract \heii\, emission for 6 out of 7 sources in our sample. The detected \lya\, nebulae range in radius from 70 to 110 kpc, whereas the \heii\, and \civ\, nebulae range from 10 to 40 kpc. 

The QUART sample differs from these studies due to deeper observations, probing lower 2$\sigma$ sky residual noise (2.2 $\times 10^{-19}$ \fergarc), dedicated targeting of radio-loud sources with compact jets and the presence of extensive multi-wavelength data including spatially resolved observations of the quasar host galaxies, which we compare in Section \ref{subsec:OSIRIS_Comparison}. We detect \heii\, and \civ\, emission around all of our sources, although \heii\, detection around 4C 04.81 could not be extracted effectively due to proximity with a sky line. The brightness of the nebulae indicates the detections are not solely due to longer exposure times and improved sensitivity. Lastly, since the gas motions associated with \lya\, are difficult to constrain from Moment 1 maps alone, we associate the radio jet orientation with \lya\, emission to constrain the geometry of the \lya\, emission. 

We suggest that the visibility of radio jets on sky implies that all these sources are being viewed at an angle that allows the ionization cone to be projected on sky beyond the seeing disk, which allows us to view the \heii\, and \civ\, nebulae being ionized around the quasars. Confirming this will require comparing to a larger sample of sources where the radio emission can be categorized as either extended or compact. It is possible that the low rate of \heii\, and \civ\, detection in the other sources is due to the radio source being confined to the seeing disk around the quasar and thus not distinguishable from the quasar PSF. Some Extremely Red Quasars (ERQs) and a few radio-quiet type-1 quasars show \heii\ and \civ\ emission \citep{Lau2022}. If the visibility of \heii\, and \civ\, was solely due to the viewing angle, then we might expect to detect nebulae around radio-quiet sources and ERQs. 

We compare our results with the recent observations of ERQs \citep{Lau2022,Gillette23}. This is particularly useful since ERQs are AGN viewed through the dusty torus. We note that the W3 magnitudes for our sources are comparable to those for ERQs in these samples and the differences between the samples are due to the color selection for ERQs. We point out that the ERQs can be probed much closer to the quasar and the authors definition of the ``inner'' CGM extends closer to the quasar than the 10--30 kpc on-sky projected radial distance considered in this work. If the ERQs are viewed through the dusty torus, then the ionization cones must be projected on the sky and \heii\, and \civ\, emissions must be visible in the inner CGM. While \cite{Lau2022} do find \heii\, and \civ\, emission, across a broader sample, \heii\, and \civ\, is not detected by \citep{Gillette23}. \cite{Gillette23} argue based on the observed \heii/\lya\, ratios that the ERQ sample is dissimilar from the larger Type-II quasar samples based on more than differences due to viewing angles. They further argue that the lack of \heii\, and \civ\, is due to the lack of UV photons escaping to large O(kpc) distances due to AGN obscuration. Given our observations of \civ\, emission being largely aligned with the radio jet, and the lack of observations in the \citep{Gillette23} sample attributed to a lack of UV photons suggest that the radio jet plays an important role in clearing a path for the ionizing UV photons to escape to circumgalactic regions, where the high ionization lines are observed. If so, this implies the ionizing radiation escapes the galaxy along the radio jet axis, creating an important distinction between the CGM observed around jet vs. non jet quasars. 

Recent work by \cite{Shukla2022} compared the \lya\, emission to the radio emission of radio loud quasars from $2.7<z<3.3$ using long slit spectroscopy. They find a radio detection rate that correlates strongly with the extent of the radio source, which we confirm here as all our sources are extended and show \lya\, emission. Their detection rate of \heii\, and \civ\, emission is much lower than ours, but it is possible that due to longslit spectroscopy, emission near the quasar has been missed. 

We compare our results with recent observations of $2.9 < \mathrm{z} < 4.5$ radio galaxies in \cite{Wang23} (W23), to compare the \lya\, emission around quasars with radio jets at higher redshifts. They find a higher asymmetry in comparison to other samples, which is also true for comparison with the QUART sample, we have shown that deeper observations can change the asymmetry of the observed nebula. However there is a caveat when comparing the QUART sample with the results in W23. The maps used for asymmetry and radio jet axis comparison with nebular major axes are corrected for intrinsic absorption by foreground absorbers. We have masked the PSF region to remove any issues due to PSF residuals, so the two samples are not strictly comparable. With that caveat in mind, it is reasonable that while W23 find all radio jets are aligned with the \lya\, emission, we only find alignment for 3 out of 7 sources. Moreover, W23 propose a model for \lya\, emission in the CGM being elongated along the ionization cone axis, arguing that the asymmetry for \lya\, nebulae for cases where the jets are pointed towards the observer should be lower (more symmetric nebulae, centered on the jet) or projected on-sky for cases where the jets are visible (less symmetric due to a rugby ball like ionization along the jet axis). The problem here is that \lya\, scatters resonantly and thus the observed \lya\, surface brightness does not necessarily describe gas motion. We consider a similar argument for \heii\, and \civ\, instead of \lya, and find only \civ\, to be aligned with the radio jet axes in our sample compared to the \lya\, and \heii\, emission. This is supported by the two component CGM model. If we assume that the inner CGM is largely influenced by galactic dynamics and the outer CGM is influenced by the larger environment, then it stands to reason that \textit{only the inner CGM asymmetry and orientation should be aligned with the radio jet}. The recent work by \cite{Gonzales2025} which investigated $z \sim 3$ quasars using VLT MUSE find an increased \lya\, dispersion width in the inner 40 kpc is in agreement with such a two component model. Future work on larger samples at other redshifts comparing the emission to the radio jet emission will be required to confirm this hypothesis. 

\subsection{Comparison with OSIRIS observations of quasar host galaxies} \label{subsec:OSIRIS_Comparison}

We compare the integrated luminosities of nebular emissions in the CGM with measurements of the quasars bolometric luminosity $Q_{bol}$, and the total host galaxy \ha\, and \oiii\, emissions (after subtracting emissions from the PSF) as calculated in \cite{QUART1} and \cite{QUART2}. We find that for our sample, the \lya\, luminosity shows no trends with the quasar bolometric luminosity or the host galaxy's \ha\, and \oiii\, emissions. As the quasar bolometric luminosity is representative of the ionizing radiation due to the AGN and the \ha\, and \oiii\, emission from the host galaxy represent ionizing radiation from potential star formation in the host galaxies, the independence of the total \lya\, emission from the CGM indicates that scattering likely plays an important role in the total luminosity, as discussed previously in \cite{Sabhlok2024a}.

\section{Summary} \label{sec:Summary}
We have carried out an optical IFS survey of 7 radio-loud quasars with radio jet emissions at z$=$2 using Keck KCWI, with total exposure times on each source ranging from 100 minutes (4C 11.45) to 560 minutes (4C 05.84) and present our conclusions here -

\begin{enumerate}
    \item We detect spatially extended \lya, \heii, and \civ\, nebulae around 100\% of our sources. We detect \heii\, around all of our sources but the extraction of moment maps is only possible around 6 out of the 7 sources due to the presence of the sky line.
    \item The \lya\, nebulae have diameters from 80 -- 120 kpc, whereas the \heii\, and \civ\, nebulae have diameters ranging from 10 -- 40 kpc. 
    \item We find radial profiles for \lya\, have higher surface brightness than previous studies in the inner 30 kpc. We also find the median of the integrated total \lya\, luminosities is brighter than other sample medians including those at higher redshifts.
    \item We compute the asymmetry for the \lya\, nebulae and compare them to previous works, finding our sample to be more symmetric than previous studies. We find that improved sensitivity can increase or decrease the observed asymmetry of the nebulae, and therefore future studies may need deeper observations to investigate any underlying asymmetries.
    \item We infer that the \heii\, and \civ\, emission is likely caused due to photoionization by the quasar and represents the extent of the inner CGM around the quasar host galaxies, where gas dynamics are influenced by host galaxy processes, as compared to the outer CGM where gas dynamics are likely more affected by the larger environment and gas turbulence.
    \item We compare the major axes of CGM nebular emission with the axes of the radio jets and find that the \heii\, emission is aligned within the ionization cone of the quasar for 4 out of 6 (\heii) and 5 out of 7 (\civ) sources. The alignment of the \civ\, nebular emission in the CGM with the radio jets indicates that the inner CGM is aligned with the radio jets. We also note that a choice of a wider opening angle of $\sim 60^{\circ}$ would align all \heii\, and \civ\, axes with the radio jet, and while this half opening angle has been noted for ionization cones \citep{Wilson1996:IonizationCones}, we have kept our definition narrower to agree with expected profiles from simulations \citep{Obreja2024}.
    \item The \lya\, emission is found to be aligned only with the radio jet for 3 out of 7 sources and the lack of alignment can be attributed to either the \lya\, nebula filling the KCWI field of view or due to the scattering of radiation in the outer CGM decoupling the \lya\, radiation emission from the quasar photoionization cone. 

    \item For 4C 11.45, we find that the bent radio jet location coincides with the edge of \heii\, and \civ\, nebulae, indicating a transition from the inner CGM influenced by galactic processes to an outer CGM influenced by the larger environment. Since radio jets are bent due to environmental effects, the coincidence with the \heii\, boundary indicates detected \heii nebulae could similarly indicate a change of CGM pressure and temperatures.  
\end{enumerate}

By combining observations of \lya, \heii, and \civ\, emission with radio jet emissions, we confirm that the CGM emission is likely powered by the quasars with the ionizing radiation escaping along the direction of the radio jets. We uncover a two component model where the inner CGM ($< 30$ kpc) is directly influenced by the host galaxy through ionization and gas kinematics. Whereas the larger outer CGM ($> 30$ kpc) is relatively undisturbed, with gas kinematics only being affected by interactions with companion galaxies. 

\section*{Acknowledgments}
The authors wish to thank Jim Lyke, Randy Campbell, Sherry Yeh and other SAs at the Keck Observatory for their assistance in obtaining the data presented in this work. This paper makes use of the following ALMA data: ADS/JAO.ALMA 2017.1.01527.S. ALMA is a partnership of ESO (representing its member states), NSF (USA), and NINS (Japan), together with NRC (Canada), MOST and ASIAA (Taiwan), and KASI (Republic of Korea), in cooperation with the Republic of Chile. The Joint ALMA Observatory is operated by ESO, AUI/NRAO, and NAOJ. The National Radio Astronomy Observatory is a facility of the National Science Foundation operated under cooperative agreement by Associated Universities, Inc. The data presented herein were obtained at the W.M. Keck Observatory, which is operated as a scientific partnership among the California Institute of Technology, the University of California, and the National Aeronautics and Space Administration. The Observatory was made possible by the generous financial support of the W.M. Keck Foundation. The authors wish to recognize and acknowledge the very significant cultural role and reverence that the summit of Maunakea has always had within the indigenous Hawaiian community. We are most fortunate to have the opportunity to conduct observations from this mountain.

\bibliography{Lyman-alpha-KCWI}{}

@ARTICLE{Chen25,
       author = {{Chen}, Mandy C. and {Chen}, Hsiao-Wen and {Rauch}, Michael and {Vayner}, Andrey and {Liu}, Weizhe and {Rupke}, David S.~N. and {Greene}, Jenny E. and {Zakamska}, Nadia L. and {Wylezalek}, Dominika and {Liu}, Guilin and {Veilleux}, Sylvain and {Nesvadba}, Nicole P.~H. and {Bertemes}, Caroline},
        title = "{Resolving Turbulence Drivers in Two Luminous Obscured Quasars with JWST/NIRSpec Integral Field Unit}",
      journal = {\apjl},
         year = 2025,
        month = jan,
       volume = {978},
       number = {2},
          eid = {L18},
        pages = {L18},
          doi = {10.3847/2041-8213/ad9bac},
archivePrefix = {arXiv},
       eprint = {2410.14785},
 primaryClass = {astro-ph.GA},
       adsurl = {https://ui.adsabs.harvard.edu/abs/2025ApJ...978L..18C}
}

@ARTICLE{Ginolfi22,
       author = {{Ginolfi}, M. and {Piconcelli}, E. and {Zappacosta}, L. and {Jones}, G.~C. and {Pentericci}, L. and {Maiolino}, R. and {Travascio}, A. and {Menci}, N. and {Carniani}, S. and {Rizzo}, F. and {Arrigoni Battaia}, F. and {Cantalupo}, S. and {De Breuck}, C. and {Graziani}, L. and {Knudsen}, K. and {Laursen}, P. and {Mainieri}, V. and {Schneider}, R. and {Stanley}, F. and {Valiante}, R. and {Verhamme}, A.},
        title = "{Detection of companion galaxies around hot dust-obscured hyper-luminous galaxy W0410-0913}",
      journal = {Nature Communications},
         year = 2022,
        month = aug,
       volume = {13},
          eid = {4574},
        pages = {4574},
          doi = {10.1038/s41467-022-32297-x},
archivePrefix = {arXiv},
       eprint = {2208.03248},
 primaryClass = {astro-ph.GA},
       adsurl = {https://ui.adsabs.harvard.edu/abs/2022NatCo..13.4574G}
}

@ARTICLE{Vayner24,
       author = {{Vayner}, Andrey and {D{\'\i}az-Santos}, Tanio and {Eisenhardt}, Peter R.~M. and {Stern}, Daniel and {Armus}, Lee and {Angl{\'e}s-Alc{\'a}zar}, Daniel and {Assef}, Roberto J. and {Fern{\'a}ndez Aranda}, Rom{\'a}n and {Blain}, Andrew W. and {Jun}, Hyunsung D. and {Tsai}, Chao-Wei and {Roy}, Niranjan Chandra and {Brisbin}, Drew and {Ferkinhoff}, Carl D. and {Aravena}, Manuel and {Gonz{\'a}lez-L{\'o}pez}, Jorge and {Li}, Guodong and {Liao}, Mai and {Shobhana}, Devika and {Wu}, Jingwen and {Zewdie}, Dejene},
        title = "{Powerful nuclear outflows and circumgalactic medium shocks driven by the most luminous known obscured quasar in the Universe}",
      journal = {arXiv e-prints},
         year = 2024,
        month = dec,
          eid = {arXiv:2412.02862},
        pages = {arXiv:2412.02862},
          doi = {10.48550/arXiv.2412.02862},
archivePrefix = {arXiv},
       eprint = {2412.02862},
 primaryClass = {astro-ph.GA},
       adsurl = {https://ui.adsabs.harvard.edu/abs/2024arXiv241202862V}
}

@ARTICLE{Wang23,
       author = {{Wang}, Wuji and {Wylezalek}, Dominika and {Vernet}, Jo{\"e}l and {De Breuck}, Carlos and {Gullberg}, Bitten and {Swinbank}, Mark and {Villar Mart{\'\i}n}, Montserrat and {Lehnert}, Matthew D. and {Drouart}, Guillaume and {Arrigoni Battaia}, Fabrizio and {Humphrey}, Andrew and {Noirot}, Ga{\"e}l and {Kolwa}, Sthabile and {Seymour}, Nick and {Lagos}, Patricio},
        title = "{3D tomography of the giant Ly{\ensuremath{\alpha}} nebulae of z {\ensuremath{\approx}} 3-5 radio-loud AGN}",
      journal = {\aap},
         year = 2023,
        month = dec,
       volume = {680},
          eid = {A70},
        pages = {A70},
          doi = {10.1051/0004-6361/202346415},
archivePrefix = {arXiv},
       eprint = {2309.15144},
 primaryClass = {astro-ph.GA},
       adsurl = {https://ui.adsabs.harvard.edu/abs/2023A&A...680A..70W}
}

@ARTICLE{Gillette23,
       author = {{Gillette}, Jarred and {Lau}, Marie Wingyee and {Hamann}, Fred and {Perrotta}, Serena and {Rupke}, David S.~N. and {Wylezalek}, Dominika and {Zakamska}, Nadia L. and {Vayner}, Andrey},
        title = "{Compact and quiescent circumgalactic medium and Ly {\ensuremath{\alpha}} haloes around extremely red quasars}",
      journal = {\mnras},
         year = 2023,
        month = dec,
       volume = {526},
       number = {2},
        pages = {2578-2595},
          doi = {10.1093/mnras/stad2923},
archivePrefix = {arXiv},
       eprint = {2303.12835},
 primaryClass = {astro-ph.GA},
       adsurl = {https://ui.adsabs.harvard.edu/abs/2023MNRAS.526.2578G}
}

@ARTICLE{QUART_ALMA,
       author = {{Vayner}, Andrey and {Zakamska}, Nadia and {Wright}, Shelley A. and {Armus}, Lee and {Murray}, Norman and {Walth}, Gregory},
        title = "{Multiphase Outflows in High-redshift Quasar Host Galaxies}",
      journal = {\apj},
         year = 2021,
        month = dec,
       volume = {923},
       number = {1},
          eid = {59},
        pages = {59},
          doi = {10.3847/1538-4357/ac2b9e},
archivePrefix = {arXiv},
       eprint = {2110.00019},
 primaryClass = {astro-ph.GA},
       adsurl = {https://ui.adsabs.harvard.edu/abs/2021ApJ...923...59V}
}

@ARTICLE{Gould1996,
       author = {{Gould}, Andrew and {Weinberg}, David H.},
        title = "{Imaging the Forest of Lyman Limit Systems}",
      journal = {\apj},
         year = 1996,
        month = sep,
       volume = {468},
        pages = {462},
          doi = {10.1086/177707},
archivePrefix = {arXiv},
       eprint = {astro-ph/9512138},
 primaryClass = {astro-ph},
       adsurl = {https://ui.adsabs.harvard.edu/abs/1996ApJ...468..462G}
}

@ARTICLE{Hogan1987,
       author = {{Hogan}, Craig J. and {Weymann}, Ray J.},
        title = "{Lyman-alpha emission from the Lyman-alpha forest}",
      journal = {\mnras},
         year = 1987,
        month = mar,
       volume = {225},
        pages = {1P-5P},
          doi = {10.1093/mnras/225.1.1P},
       adsurl = {https://ui.adsabs.harvard.edu/abs/1987MNRAS.225P...1H}
}

@ARTICLE{Cantalupo2005,
       author = {{Cantalupo}, Sebastiano and {Porciani}, Cristiano and {Lilly}, Simon J. and
         {Miniati}, Francesco},
        title = "{Fluorescent Ly{\ensuremath{\alpha}} Emission from the High-Redshift Intergalactic Medium}",
      journal = {\apj},
         year = 2005,
        month = jul,
       volume = {628},
       number = {1},
        pages = {61-75},
          doi = {10.1086/430758},
archivePrefix = {arXiv},
       eprint = {astro-ph/0504015},
 primaryClass = {astro-ph},
       adsurl = {https://ui.adsabs.harvard.edu/abs/2005ApJ...628...61C}
}

@ARTICLE{Kollmeier2010,
       author = {{Kollmeier}, Juna A. and {Zheng}, Zheng and {Dav{\'e}}, Romeel and
         {Gould}, Andrew and {Katz}, Neal and {Miralda-Escud{\'e}}, Jordi and
         {Weinberg}, David H.},
        title = "{Ly{\ensuremath{\alpha}} Emission from Cosmic Structure. I. Fluorescence}",
      journal = {\apj},
         year = 2010,
        month = jan,
       volume = {708},
       number = {2},
        pages = {1048-1075},
          doi = {10.1088/0004-637X/708/2/1048},
archivePrefix = {arXiv},
       eprint = {0907.0704},
 primaryClass = {astro-ph.CO},
       adsurl = {https://ui.adsabs.harvard.edu/abs/2010ApJ...708.1048K}
}

@ARTICLE{Cai2019,
       author = {{Cai}, Zheng and {Cantalupo}, Sebastiano and {Prochaska}, J. Xavier and
         {Arrigoni Battaia}, Fabrizio and {Burchett}, Joe and {Li}, Qiong and
         {Chisholm}, John and {Bundy}, Kevin and {Hennawi}, Joseph F.},
        title = "{Evolution of the Cool Gas in the Circumgalactic Medium of Massive Halos: A Keck Cosmic Web Imager Survey of Ly{\ensuremath{\alpha}} Emission around QSOs at z {\ensuremath{\approx}} 2}",
      journal = {\apjs},
         year = 2019,
        month = dec,
       volume = {245},
       number = {2},
          eid = {23},
        pages = {23},
          doi = {10.3847/1538-4365/ab4796},
archivePrefix = {arXiv},
       eprint = {1909.11098},
 primaryClass = {astro-ph.GA},
       adsurl = {https://ui.adsabs.harvard.edu/abs/2019ApJS..245...23C}
}

@ARTICLE{Cantalupo2019,
       author = {{Cantalupo}, Sebastiano and {Pezzulli}, Gabriele and {Lilly}, Simon J. and
         {Marino}, Raffaella Anna and {Gallego}, Sofia G. and {Schaye}, Joop and
         {Bacon}, Roland and {Feltre}, Anna and {Kollatschny}, Wolfram and
         {Nanayakkara}, Themiya and {Richard}, Johan and {Wendt}, Martin and
         {Wisotzki}, Lutz and {Prochaska}, J. Xavier},
        title = "{The large- and small-scale properties of the intergalactic gas in the Slug Ly {\ensuremath{\alpha}} nebula revealed by MUSE He II emission observations}",
      journal = {\mnras},
         year = 2019,
        month = mar,
       volume = {483},
       number = {4},
        pages = {5188-5204},
          doi = {10.1093/mnras/sty3481},
archivePrefix = {arXiv},
       eprint = {1811.11783},
 primaryClass = {astro-ph.GA},
       adsurl = {https://ui.adsabs.harvard.edu/abs/2019MNRAS.483.5188C}
}

@ARTICLE{QUART_OSIRIS,
       author = {{Vayner}, Andrey and {Wright}, Shelley A. and {Do}, Tuan and {Larkin}, James E. and {Armus}, Lee and {Gallagher}, S.~C.},
        title = "{Providing Stringent Star Formation Rate Limits of z {\ensuremath{\sim}} 2 QSO Host Galaxies at High Angular Resolution}",
      journal = {\apj},
         year = 2016,
        month = apr,
       volume = {821},
       number = {1},
          eid = {64},
        pages = {64},
          doi = {10.3847/0004-637X/821/1/64},
archivePrefix = {arXiv},
       eprint = {1410.4229},
 primaryClass = {astro-ph.GA},
       adsurl = {https://ui.adsabs.harvard.edu/abs/2016ApJ...821...64V}
}

@ARTICLE{CWIToolsRef,
       author = {{O'Sullivan}, Donal and {Chen}, Yuguang},
        title = "{CWITools: A Python3 Data Analysis Pipeline for the Cosmic Web Imager Instruments}",
      journal = {arXiv e-prints},
         year = 2020,
        month = nov,
          eid = {arXiv:2011.05444},
        pages = {arXiv:2011.05444},
archivePrefix = {arXiv},
       eprint = {2011.05444},
 primaryClass = {astro-ph.IM},
       adsurl = {https://ui.adsabs.harvard.edu/abs/2020arXiv201105444O}
}

@ARTICLE{GAIARef,
       author = {{Gaia Collaboration} and {Brown}, A.~G.~A. and {Vallenari}, A. and {Prusti}, T. and {de Bruijne}, J.~H.~J. and {Babusiaux}, C. and {Bailer-Jones}, C.~A.~L. and {Biermann}, M. and {Evans}, D.~W. and {Eyer}, L. and {Jansen}, F. and {Jordi}, C. and {Klioner}, S.~A. and {Lammers}, U. and {Lindegren}, L. and {Luri}, X. and {Mignard}, F. and {Panem}, C. and {Pourbaix}, D. and {Randich}, S. and {Sartoretti}, P. and {Siddiqui}, H.~I. and {Soubiran}, C. and {van Leeuwen}, F. and {Walton}, N.~A. and {Arenou}, F. and {Bastian}, U. and {Cropper}, M. and {Drimmel}, R. and {Katz}, D. and {Lattanzi}, M.~G. and {Bakker}, J. and {Cacciari}, C. and {Casta{\~n}eda}, J. and {Chaoul}, L. and {Cheek}, N. and {De Angeli}, F. and {Fabricius}, C. and {Guerra}, R. and {Holl}, B. and {Masana}, E. and {Messineo}, R. and {Mowlavi}, N. and {Nienartowicz}, K. and {Panuzzo}, P. and {Portell}, J. and {Riello}, M. and {Seabroke}, G.~M. and {Tanga}, P. and {Th{\'e}venin}, F. and {Gracia-Abril}, G. and {Comoretto}, G. and {Garcia-Reinaldos}, M. and {Teyssier}, D. and {Altmann}, M. and {Andrae}, R. and {Audard}, M. and {Bellas-Velidis}, I. and {Benson}, K. and {Berthier}, J. and {Blomme}, R. and {Burgess}, P. and {Busso}, G. and {Carry}, B. and {Cellino}, A. and {Clementini}, G. and {Clotet}, M. and {Creevey}, O. and {Davidson}, M. and {De Ridder}, J. and {Delchambre}, L. and {Dell'Oro}, A. and {Ducourant}, C. and {Fern{\'a}ndez-Hern{\'a}ndez}, J. and {Fouesneau}, M. and {Fr{\'e}mat}, Y. and {Galluccio}, L. and {Garc{\'\i}a-Torres}, M. and {Gonz{\'a}lez-N{\'u}{\~n}ez}, J. and {Gonz{\'a}lez-Vidal}, J.~J. and {Gosset}, E. and {Guy}, L.~P. and {Halbwachs}, J. -L. and {Hambly}, N.~C. and {Harrison}, D.~L. and {Hern{\'a}ndez}, J. and {Hestroffer}, D. and {Hodgkin}, S.~T. and {Hutton}, A. and {Jasniewicz}, G. and {Jean-Antoine-Piccolo}, A. and {Jordan}, S. and {Korn}, A.~J. and {Krone-Martins}, A. and {Lanzafame}, A.~C. and {Lebzelter}, T. and {L{\"o}ffler}, W. and {Manteiga}, M. and {Marrese}, P.~M. and {Mart{\'\i}n-Fleitas}, J.~M. and {Moitinho}, A. and {Mora}, A. and {Muinonen}, K. and {Osinde}, J. and {Pancino}, E. and {Pauwels}, T. and {Petit}, J. -M. and {Recio-Blanco}, A. and {Richards}, P.~J. and {Rimoldini}, L. and {Robin}, A.~C. and {Sarro}, L.~M. and {Siopis}, C. and {Smith}, M. and {Sozzetti}, A. and {S{\"u}veges}, M. and {Torra}, J. and {van Reeven}, W. and {Abbas}, U. and {Abreu Aramburu}, A. and {Accart}, S. and {Aerts}, C. and {Altavilla}, G. and {{\'A}lvarez}, M.~A. and {Alvarez}, R. and {Alves}, J. and {Anderson}, R.~I. and {Andrei}, A.~H. and {Anglada Varela}, E. and {Antiche}, E. and {Antoja}, T. and {Arcay}, B. and {Astraatmadja}, T.~L. and {Bach}, N. and {Baker}, S.~G. and {Balaguer-N{\'u}{\~n}ez}, L. and {Balm}, P. and {Barache}, C. and {Barata}, C. and {Barbato}, D. and {Barblan}, F. and {Barklem}, P.~S. and {Barrado}, D. and {Barros}, M. and {Barstow}, M.~A. and {Bartholom{\'e} Mu{\~n}oz}, S. and {Bassilana}, J. -L. and {Becciani}, U. and {Bellazzini}, M. and {Berihuete}, A. and {Bertone}, S. and {Bianchi}, L. and {Bienaym{\'e}}, O. and {Blanco-Cuaresma}, S. and {Boch}, T. and {Boeche}, C. and {Bombrun}, A. and {Borrachero}, R. and {Bossini}, D. and {Bouquillon}, S. and {Bourda}, G. and {Bragaglia}, A. and {Bramante}, L. and {Breddels}, M.~A. and {Bressan}, A. and {Brouillet}, N. and {Br{\"u}semeister}, T. and {Brugaletta}, E. and {Bucciarelli}, B. and {Burlacu}, A. and {Busonero}, D. and {Butkevich}, A.~G. and {Buzzi}, R. and {Caffau}, E. and {Cancelliere}, R. and {Cannizzaro}, G. and {Cantat-Gaudin}, T. and {Carballo}, R. and {Carlucci}, T. and {Carrasco}, J.~M. and {Casamiquela}, L. and {Castellani}, M. and {Castro-Ginard}, A. and {Charlot}, P. and {Chemin}, L. and {Chiavassa}, A. and {Cocozza}, G. and {Costigan}, G. and {Cowell}, S. and {Crifo}, F. and {Crosta}, M. and {Crowley}, C. and {Cuypers}, J. and {Dafonte}, C. and {Damerdji}, Y. and {Dapergolas}, A. and {David}, P. and {David}, M. and {de Laverny}, P. and {De Luise}, F. and {De March}, R. and {de Martino}, D. and {de Souza}, R. and {de Torres}, A. and {Debosscher}, J. and {del Pozo}, E. and {Delbo}, M. and {Delgado}, A. and {Delgado}, H.~E. and {Di Matteo}, P. and {Diakite}, S. and {Diener}, C. and {Distefano}, E. and {Dolding}, C. and {Drazinos}, P. and {Dur{\'a}n}, J. and {Edvardsson}, B. and {Enke}, H. and {Eriksson}, K. and {Esquej}, P. and {Eynard Bontemps}, G. and {Fabre}, C. and {Fabrizio}, M. and {Faigler}, S. and {Falc{\~a}o}, A.~J. and {Farr{\`a}s Casas}, M. and {Federici}, L. and {Fedorets}, G. and {Fernique}, P. and {Figueras}, F. and {Filippi}, F. and {Findeisen}, K. and {Fonti}, A. and {Fraile}, E. and {Fraser}, M. and {Fr{\'e}zouls}, B. and {Gai}, M. and {Galleti}, S. and {Garabato}, D. and {Garc{\'\i}a-Sedano}, F. and {Garofalo}, A. and {Garralda}, N. and {Gavel}, A. and {Gavras}, P. and {Gerssen}, J. and {Geyer}, R. and {Giacobbe}, P. and {Gilmore}, G. and {Girona}, S. and {Giuffrida}, G. and {Glass}, F. and {Gomes}, M. and {Granvik}, M. and {Gueguen}, A. and {Guerrier}, A. and {Guiraud}, J. and {Guti{\'e}rrez-S{\'a}nchez}, R. and {Haigron}, R. and {Hatzidimitriou}, D. and {Hauser}, M. and {Haywood}, M. and {Heiter}, U. and {Helmi}, A. and {Heu}, J. and {Hilger}, T. and {Hobbs}, D. and {Hofmann}, W. and {Holland}, G. and {Huckle}, H.~E. and {Hypki}, A. and {Icardi}, V. and {Jan{\ss}en}, K. and {Jevardat de Fombelle}, G. and {Jonker}, P.~G. and {Juh{\'a}sz}, {\'A}. L. and {Julbe}, F. and {Karampelas}, A. and {Kewley}, A. and {Klar}, J. and {Kochoska}, A. and {Kohley}, R. and {Kolenberg}, K. and {Kontizas}, M. and {Kontizas}, E. and {Koposov}, S.~E. and {Kordopatis}, G. and {Kostrzewa-Rutkowska}, Z. and {Koubsky}, P. and {Lambert}, S. and {Lanza}, A.~F. and {Lasne}, Y. and {Lavigne}, J. -B. and {Le Fustec}, Y. and {Le Poncin-Lafitte}, C. and {Lebreton}, Y. and {Leccia}, S. and {Leclerc}, N. and {Lecoeur-Taibi}, I. and {Lenhardt}, H. and {Leroux}, F. and {Liao}, S. and {Licata}, E. and {Lindstr{\o}m}, H.~E.~P. and {Lister}, T.~A. and {Livanou}, E. and {Lobel}, A. and {L{\'o}pez}, M. and {Managau}, S. and {Mann}, R.~G. and {Mantelet}, G. and {Marchal}, O. and {Marchant}, J.~M. and {Marconi}, M. and {Marinoni}, S. and {Marschalk{\'o}}, G. and {Marshall}, D.~J. and {Martino}, M. and {Marton}, G. and {Mary}, N. and {Massari}, D. and {Matijevi{\v{c}}}, G. and {Mazeh}, T. and {McMillan}, P.~J. and {Messina}, S. and {Michalik}, D. and {Millar}, N.~R. and {Molina}, D. and {Molinaro}, R. and {Moln{\'a}r}, L. and {Montegriffo}, P. and {Mor}, R. and {Morbidelli}, R. and {Morel}, T. and {Morris}, D. and {Mulone}, A.~F. and {Muraveva}, T. and {Musella}, I. and {Nelemans}, G. and {Nicastro}, L. and {Noval}, L. and {O'Mullane}, W. and {Ord{\'e}novic}, C. and {Ord{\'o}{\~n}ez-Blanco}, D. and {Osborne}, P. and {Pagani}, C. and {Pagano}, I. and {Pailler}, F. and {Palacin}, H. and {Palaversa}, L. and {Panahi}, A. and {Pawlak}, M. and {Piersimoni}, A.~M. and {Pineau}, F. -X. and {Plachy}, E. and {Plum}, G. and {Poggio}, E. and {Poujoulet}, E. and {Pr{\v{s}}a}, A. and {Pulone}, L. and {Racero}, E. and {Ragaini}, S. and {Rambaux}, N. and {Ramos-Lerate}, M. and {Regibo}, S. and {Reyl{\'e}}, C. and {Riclet}, F. and {Ripepi}, V. and {Riva}, A. and {Rivard}, A. and {Rixon}, G. and {Roegiers}, T. and {Roelens}, M. and {Romero-G{\'o}mez}, M. and {Rowell}, N. and {Royer}, F. and {Ruiz-Dern}, L. and {Sadowski}, G. and {Sagrist{\`a} Sell{\'e}s}, T. and {Sahlmann}, J. and {Salgado}, J. and {Salguero}, E. and {Sanna}, N. and {Santana-Ros}, T. and {Sarasso}, M. and {Savietto}, H. and {Schultheis}, M. and {Sciacca}, E. and {Segol}, M. and {Segovia}, J.~C. and {S{\'e}gransan}, D. and {Shih}, I. -C. and {Siltala}, L. and {Silva}, A.~F. and {Smart}, R.~L. and {Smith}, K.~W. and {Solano}, E. and {Solitro}, F. and {Sordo}, R. and {Soria Nieto}, S. and {Souchay}, J. and {Spagna}, A. and {Spoto}, F. and {Stampa}, U. and {Steele}, I.~A. and {Steidelm{\"u}ller}, H. and {Stephenson}, C.~A. and {Stoev}, H. and {Suess}, F.~F. and {Surdej}, J. and {Szabados}, L. and {Szegedi-Elek}, E. and {Tapiador}, D. and {Taris}, F. and {Tauran}, G. and {Taylor}, M.~B. and {Teixeira}, R. and {Terrett}, D. and {Teyssandier}, P. and {Thuillot}, W. and {Titarenko}, A. and {Torra Clotet}, F. and {Turon}, C. and {Ulla}, A. and {Utrilla}, E. and {Uzzi}, S. and {Vaillant}, M. and {Valentini}, G. and {Valette}, V. and {van Elteren}, A. and {Van Hemelryck}, E. and {van Leeuwen}, M. and {Vaschetto}, M. and {Vecchiato}, A. and {Veljanoski}, J. and {Viala}, Y. and {Vicente}, D. and {Vogt}, S. and {von Essen}, C. and {Voss}, H. and {Votruba}, V. and {Voutsinas}, S. and {Walmsley}, G. and {Weiler}, M. and {Wertz}, O. and {Wevers}, T. and {Wyrzykowski}, {\L}. and {Yoldas}, A. and {{\v{Z}}erjal}, M. and {Ziaeepour}, H. and {Zorec}, J. and {Zschocke}, S. and {Zucker}, S. and {Zurbach}, C. and {Zwitter}, T.},
        title = "{Gaia Data Release 2. Summary of the contents and survey properties}",
      journal = {\aap},
         year = 2018,
        month = aug,
       volume = {616},
          eid = {A1},
        pages = {A1},
          doi = {10.1051/0004-6361/201833051},
archivePrefix = {arXiv},
       eprint = {1804.09365},
 primaryClass = {astro-ph.GA},
       adsurl = {https://ui.adsabs.harvard.edu/abs/2018A&A...616A...1G}
}

@ARTICLE{QUART1,
       author = {{Vayner}, Andrey and {Wright}, Shelley A. and {Murray}, Norman and {Armus}, Lee and {Boehle}, Anna and {Cosens}, Maren and {Larkin}, James E. and {Mieda}, Etsuko and {Walth}, Gregory},
        title = "{A Spatially Resolved Survey of Distant Quasar Host Galaxies. I. Dynamics of Galactic Outflows}",
      journal = {\apj},
         year = 2021,
        month = oct,
       volume = {919},
       number = {2},
          eid = {122},
        pages = {122},
          doi = {10.3847/1538-4357/ac0f56},
archivePrefix = {arXiv},
       eprint = {2106.08337},
 primaryClass = {astro-ph.GA},
       adsurl = {https://ui.adsabs.harvard.edu/abs/2021ApJ...919..122V}
}

@ARTICLE{QUART2,
       author = {{Vayner}, Andrey and {Wright}, Shelley A. and {Murray}, Norman and {Armus}, Lee and {Boehle}, Anna and {Cosens}, Maren and {Larkin}, James E. and {Mieda}, Etsuko and {Walth}, Gregory},
        title = "{A Spatially Resolved Survey of Distant Quasar Host Galaxies. II. Photoionization and Kinematics of the ISM}",
      journal = {\apj},
         year = 2021,
        month = mar,
       volume = {910},
       number = {1},
          eid = {44},
        pages = {44},
          doi = {10.3847/1538-4357/abddc1},
archivePrefix = {arXiv},
       eprint = {2101.08291},
 primaryClass = {astro-ph.GA},
       adsurl = {https://ui.adsabs.harvard.edu/abs/2021ApJ...910...44V}
}

@ARTICLE{AB2019,
       author = {{Arrigoni Battaia}, Fabrizio and {Hennawi}, Joseph F. and {Prochaska}, J. Xavier and {O{\~n}orbe}, Jose and {Farina}, Emanuele P. and {Cantalupo}, Sebastiano and {Lusso}, Elisabeta},
        title = "{QSO MUSEUM I: a sample of 61 extended Ly {\ensuremath{\alpha}}-emission nebulae surrounding z {\ensuremath{\sim}} 3 quasars}",
      journal = {\mnras},
         year = 2019,
        month = jan,
       volume = {482},
       number = {3},
        pages = {3162-3205},
          doi = {10.1093/mnras/sty2827},
archivePrefix = {arXiv},
       eprint = {1808.10857},
 primaryClass = {astro-ph.GA},
       adsurl = {https://ui.adsabs.harvard.edu/abs/2019MNRAS.482.3162A}
}

@ARTICLE{Stoughton2002,
       author = {{Stoughton}, Chris and {Lupton}, Robert H. and {Bernardi}, Mariangela and {Blanton}, Michael R. and {Burles}, Scott and {Castander}, Francisco J. and {Connolly}, A.~J. and {Eisenstein}, Daniel J. and {Frieman}, Joshua A. and {Hennessy}, G.~S. and {Hindsley}, Robert B. and {Ivezi{\'c}}, {\v{Z}}eljko and {Kent}, Stephen and {Kunszt}, Peter Z. and {Lee}, Brian C. and {Meiksin}, Avery and {Munn}, Jeffrey A. and {Newberg}, Heidi Jo and {Nichol}, R.~C. and {Nicinski}, Tom and {Pier}, Jeffrey R. and {Richards}, Gordon T. and {Richmond}, Michael W. and {Schlegel}, David J. and {Smith}, J. Allyn and {Strauss}, Michael A. and {SubbaRao}, Mark and {Szalay}, Alexander S. and {Thakar}, Aniruddha R. and {Tucker}, Douglas L. and {Vanden Berk}, Daniel E. and {Yanny}, Brian and {Adelman}, Jennifer K. and {Anderson}, John E., Jr. and {Anderson}, Scott F. and {Annis}, James and {Bahcall}, Neta A. and {Bakken}, J.~A. and {Bartelmann}, Matthias and {Bastian}, Steven and {Bauer}, Amanda and {Berman}, Eileen and {B{\"o}hringer}, Hans and {Boroski}, William N. and {Bracker}, Steve and {Briegel}, Charlie and {Briggs}, John W. and {Brinkmann}, J. and {Brunner}, Robert and {Carey}, Larry and {Carr}, Michael A. and {Chen}, Bing and {Christian}, Damian and {Colestock}, Patrick L. and {Crocker}, J.~H. and {Csabai}, Istv{\'a}n and {Czarapata}, Paul C. and {Dalcanton}, Julianne and {Davidsen}, Arthur F. and {Davis}, John Eric and {Dehnen}, Walter and {Dodelson}, Scott and {Doi}, Mamoru and {Dombeck}, Tom and {Donahue}, Megan and {Ellman}, Nancy and {Elms}, Brian R. and {Evans}, Michael L. and {Eyer}, Laurent and {Fan}, Xiaohui and {Federwitz}, Glenn R. and {Friedman}, Scott and {Fukugita}, Masataka and {Gal}, Roy and {Gillespie}, Bruce and {Glazebrook}, Karl and {Gray}, Jim and {Grebel}, Eva K. and {Greenawalt}, Bruce and {Greene}, Gretchen and {Gunn}, James E. and {de Haas}, Ernst and {Haiman}, Zolt{\'a}n and {Haldeman}, Merle and {Hall}, Patrick B. and {Hamabe}, Masaru and {Hansen}, Brad and {Harris}, Frederick H. and {Harris}, Hugh and {Harvanek}, Michael and {Hawley}, Suzanne L. and {Hayes}, J.~J.~E. and {Heckman}, Timothy M. and {Helmi}, Amina and {Henden}, Arne and {Hogan}, Craig J. and {Hogg}, David W. and {Holmgren}, Donald J. and {Holtzman}, Jon and {Huang}, Chih-Hao and {Hull}, Charles and {Ichikawa}, Shin-Ichi and {Ichikawa}, Takashi and {Johnston}, David E. and {Kauffmann}, Guinevere and {Kim}, Rita S.~J. and {Kimball}, Tim and {Kinney}, E. and {Klaene}, Mark and {Kleinman}, S.~J. and {Klypin}, Anatoly and {Knapp}, G.~R. and {Korienek}, John and {Krolik}, Julian and {Kron}, Richard G. and {Krzesi{\'n}ski}, Jurek and {Lamb}, D.~Q. and {Leger}, R. French and {Limmongkol}, Siriluk and {Lindenmeyer}, Carl and {Long}, Daniel C. and {Loomis}, Craig and {Loveday}, Jon and {MacKinnon}, Bryan and {Mannery}, Edward J. and {Mantsch}, P.~M. and {Margon}, Bruce and {McGehee}, Peregrine and {McKay}, Timothy A. and {McLean}, Brian and {Menou}, Kristen and {Merelli}, Aronne and {Mo}, H.~J. and {Monet}, David G. and {Nakamura}, Osamu and {Narayanan}, Vijay K. and {Nash}, Thomas and {Neilsen}, Eric H., Jr. and {Newman}, Peter R. and {Nitta}, Atsuko and {Odenkirchen}, Michael and {Okada}, Norio and {Okamura}, Sadanori and {Ostriker}, Jeremiah P. and {Owen}, Russell and {Pauls}, A. George and {Peoples}, John and {Peterson}, R.~S. and {Petravick}, Donald and {Pope}, Adrian and {Pordes}, Ruth and {Postman}, Marc and {Prosapio}, Angela and {Quinn}, Thomas R. and {Rechenmacher}, Ron and {Rivetta}, Claudio H. and {Rix}, Hans-Walter and {Rockosi}, Constance M. and {Rosner}, Robert and {Ruthmansdorfer}, Kurt and {Sandford}, Dale and {Schneider}, Donald P. and {Scranton}, Ryan and {Sekiguchi}, Maki and {Sergey}, Gary and {Sheth}, Ravi and {Shimasaku}, Kazuhiro and {Smee}, Stephen and {Snedden}, Stephanie A. and {Stebbins}, Albert and {Stubbs}, Christopher and {Szapudi}, Istv{\'a}n and {Szkody}, Paula and {Szokoly}, Gyula P. and {Tabachnik}, Serge and {Tsvetanov}, Zlatan and {Uomoto}, Alan and {Vogeley}, Michael S. and {Voges}, Wolfgang and {Waddell}, Patrick and {Walterbos}, Ren{\'e} and {Wang}, Shu-i. and {Watanabe}, Masaru and {Weinberg}, David H. and {White}, Richard L. and {White}, Simon D.~M. and {Wilhite}, Brian and {Wolfe}, David and {Yasuda}, Naoki and {York}, Donald G. and {Zehavi}, Idit and {Zheng}, Wei},
        title = "{Sloan Digital Sky Survey: Early Data Release}",
      journal = {\aj},
         year = 2002,
        month = jan,
       volume = {123},
       number = {1},
        pages = {485-548},
          doi = {10.1086/324741},
       adsurl = {https://ui.adsabs.harvard.edu/abs/2002AJ....123..485S}
}

@ARTICLE{Chen2020,
       author = {{Chen}, Yuguang and {Steidel}, Charles C. and {Hummels}, Cameron B. and {Rudie}, Gwen C. and {Dong}, Bili and {Trainor}, Ryan F. and {Bogosavljevi{\'c}}, Milan and {Erb}, Dawn K. and {Pettini}, Max and {Reddy}, Naveen A. and {Shapley}, Alice E. and {Strom}, Allison L. and {Theios}, Rachel L. and {Faucher-Gigu{\`e}re}, Claude-Andr{\'e} and {Hopkins}, Philip F. and {Kere{\v{s}}}, Du{\v{s}}an},
        title = "{The Keck Baryonic Structure Survey: using foreground/background galaxy pairs to trace the structure and kinematics of circumgalactic neutral hydrogen at z   2}",
      journal = {\mnras},
         year = 2020,
        month = dec,
       volume = {499},
       number = {2},
        pages = {1721-1746},
          doi = {10.1093/mnras/staa2808},
archivePrefix = {arXiv},
       eprint = {2006.13236},
 primaryClass = {astro-ph.GA},
       adsurl = {https://ui.adsabs.harvard.edu/abs/2020MNRAS.499.1721C}
}

@ARTICLE{Morrissey2018KCWI,
       author = {{Morrissey}, Patrick and {Matuszewski}, Matuesz and {Martin}, D. Christopher and {Neill}, James D. and {Epps}, Harland and {Fucik}, Jason and {Weber}, Bob and {Darvish}, Behnam and {Adkins}, Sean and {Allen}, Steve and {Bartos}, Randy and {Belicki}, Justin and {Cabak}, Jerry and {Callahan}, Shawn and {Cowley}, Dave and {Crabill}, Marty and {Deich}, Willian and {Delecroix}, Alex and {Doppman}, Greg and {Hilyard}, David and {James}, Ean and {Kaye}, Steve and {Kokorowski}, Michael and {Kwok}, Shui and {Lanclos}, Kyle and {Milner}, Steve and {Moore}, Anna and {O'Sullivan}, Donal and {Parihar}, Prachi and {Park}, Sam and {Phillips}, Andrew and {Rizzi}, Luca and {Rockosi}, Constance and {Rodriguez}, Hector and {Salaun}, Yves and {Seaman}, Kirk and {Sheikh}, David and {Weiss}, Jason and {Zarzaca}, Ray},
        title = "{The Keck Cosmic Web Imager Integral Field Spectrograph}",
      journal = {\apj},
         year = 2018,
        month = sep,
       volume = {864},
       number = {1},
          eid = {93},
        pages = {93},
          doi = {10.3847/1538-4357/aad597},
archivePrefix = {arXiv},
       eprint = {1807.10356},
 primaryClass = {astro-ph.IM},
       adsurl = {https://ui.adsabs.harvard.edu/abs/2018ApJ...864...93M}
}

@INPROCEEDINGS{Bacon2010MUSE,
       author = {{Bacon}, R. and {Accardo}, M. and {Adjali}, L. and {Anwand}, H. and {Bauer}, S. and {Biswas}, I. and {Blaizot}, J. and {Boudon}, D. and {Brau-Nogue}, S. and {Brinchmann}, J. and {Caillier}, P. and {Capoani}, L. and {Carollo}, C.~M. and {Contini}, T. and {Couderc}, P. and {Daguis{\'e}}, E. and {Deiries}, S. and {Delabre}, B. and {Dreizler}, S. and {Dubois}, J. and {Dupieux}, M. and {Dupuy}, C. and {Emsellem}, E. and {Fechner}, T. and {Fleischmann}, A. and {Fran{\c{c}}ois}, M. and {Gallou}, G. and {Gharsa}, T. and {Glindemann}, A. and {Gojak}, D. and {Guiderdoni}, B. and {Hansali}, G. and {Hahn}, T. and {Jarno}, A. and {Kelz}, A. and {Koehler}, C. and {Kosmalski}, J. and {Laurent}, F. and {Le Floch}, M. and {Lilly}, S.~J. and {Lizon}, J. -L. and {Loupias}, M. and {Manescau}, A. and {Monstein}, C. and {Nicklas}, H. and {Olaya}, J. -C. and {Pares}, L. and {Pasquini}, L. and {P{\'e}contal-Rousset}, A. and {Pell{\'o}}, R. and {Petit}, C. and {Popow}, E. and {Reiss}, R. and {Remillieux}, A. and {Renault}, E. and {Roth}, M. and {Rupprecht}, G. and {Serre}, D. and {Schaye}, J. and {Soucail}, G. and {Steinmetz}, M. and {Streicher}, O. and {Stuik}, R. and {Valentin}, H. and {Vernet}, J. and {Weilbacher}, P. and {Wisotzki}, L. and {Yerle}, N.},
        title = "{The MUSE second-generation VLT instrument}",
    booktitle = {Ground-based and Airborne Instrumentation for Astronomy III},
         year = 2010,
       editor = {{McLean}, Ian S. and {Ramsay}, Suzanne K. and {Takami}, Hideki},
       series = {Society of Photo-Optical Instrumentation Engineers (SPIE) Conference Series},
       volume = {7735},
        month = jul,
          eid = {773508},
        pages = {773508},
          doi = {10.1117/12.856027},
archivePrefix = {arXiv},
       eprint = {2211.16795},
 primaryClass = {astro-ph.IM},
       adsurl = {https://ui.adsabs.harvard.edu/abs/2010SPIE.7735E..08B}
}

@ARTICLE{Langen2023,
       author = {{Langen}, Vivienne and {Cantalupo}, Sebastiano and {Steidel}, Charles C. and {Chen}, Yuguang and {Pezzulli}, Gabriele and {Gallego}, Sofia G.},
        title = "{Characterizing the circumgalactic medium of quasars at z   2.2 through H {\ensuremath{\alpha}} and Ly {\ensuremath{\alpha}} emission}",
      journal = {\mnras},
         year = 2023,
        month = mar,
       volume = {519},
       number = {4},
        pages = {5099-5113},
          doi = {10.1093/mnras/stac3205},
archivePrefix = {arXiv},
       eprint = {2303.05531},
 primaryClass = {astro-ph.GA},
       adsurl = {https://ui.adsabs.harvard.edu/abs/2023MNRAS.519.5099L}
}

@ARTICLE{Planck2016,
       author = {{Planck Collaboration} and {Ade}, P.~A.~R. and {Aghanim}, N. and {Arnaud}, M. and {Ashdown}, M. and {Aumont}, J. and {Baccigalupi}, C. and {Banday}, A.~J. and {Barreiro}, R.~B. and {Bartlett}, J.~G. and {Bartolo}, N. and {Battaner}, E. and {Battye}, R. and {Benabed}, K. and {Beno{\^\i}t}, A. and {Benoit-L{\'e}vy}, A. and {Bernard}, J. -P. and {Bersanelli}, M. and {Bielewicz}, P. and {Bock}, J.~J. and {Bonaldi}, A. and {Bonavera}, L. and {Bond}, J.~R. and {Borrill}, J. and {Bouchet}, F.~R. and {Boulanger}, F. and {Bucher}, M. and {Burigana}, C. and {Butler}, R.~C. and {Calabrese}, E. and {Cardoso}, J. -F. and {Catalano}, A. and {Challinor}, A. and {Chamballu}, A. and {Chary}, R. -R. and {Chiang}, H.~C. and {Chluba}, J. and {Christensen}, P.~R. and {Church}, S. and {Clements}, D.~L. and {Colombi}, S. and {Colombo}, L.~P.~L. and {Combet}, C. and {Coulais}, A. and {Crill}, B.~P. and {Curto}, A. and {Cuttaia}, F. and {Danese}, L. and {Davies}, R.~D. and {Davis}, R.~J. and {de Bernardis}, P. and {de Rosa}, A. and {de Zotti}, G. and {Delabrouille}, J. and {D{\'e}sert}, F. -X. and {Di Valentino}, E. and {Dickinson}, C. and {Diego}, J.~M. and {Dolag}, K. and {Dole}, H. and {Donzelli}, S. and {Dor{\'e}}, O. and {Douspis}, M. and {Ducout}, A. and {Dunkley}, J. and {Dupac}, X. and {Efstathiou}, G. and {Elsner}, F. and {En{\ss}lin}, T.~A. and {Eriksen}, H.~K. and {Farhang}, M. and {Fergusson}, J. and {Finelli}, F. and {Forni}, O. and {Frailis}, M. and {Fraisse}, A.~A. and {Franceschi}, E. and {Frejsel}, A. and {Galeotta}, S. and {Galli}, S. and {Ganga}, K. and {Gauthier}, C. and {Gerbino}, M. and {Ghosh}, T. and {Giard}, M. and {Giraud-H{\'e}raud}, Y. and {Giusarma}, E. and {Gjerl{\o}w}, E. and {Gonz{\'a}lez-Nuevo}, J. and {G{\'o}rski}, K.~M. and {Gratton}, S. and {Gregorio}, A. and {Gruppuso}, A. and {Gudmundsson}, J.~E. and {Hamann}, J. and {Hansen}, F.~K. and {Hanson}, D. and {Harrison}, D.~L. and {Helou}, G. and {Henrot-Versill{\'e}}, S. and {Hern{\'a}ndez-Monteagudo}, C. and {Herranz}, D. and {Hildebrandt}, S.~R. and {Hivon}, E. and {Hobson}, M. and {Holmes}, W.~A. and {Hornstrup}, A. and {Hovest}, W. and {Huang}, Z. and {Huffenberger}, K.~M. and {Hurier}, G. and {Jaffe}, A.~H. and {Jaffe}, T.~R. and {Jones}, W.~C. and {Juvela}, M. and {Keih{\"a}nen}, E. and {Keskitalo}, R. and {Kisner}, T.~S. and {Kneissl}, R. and {Knoche}, J. and {Knox}, L. and {Kunz}, M. and {Kurki-Suonio}, H. and {Lagache}, G. and {L{\"a}hteenm{\"a}ki}, A. and {Lamarre}, J. -M. and {Lasenby}, A. and {Lattanzi}, M. and {Lawrence}, C.~R. and {Leahy}, J.~P. and {Leonardi}, R. and {Lesgourgues}, J. and {Levrier}, F. and {Lewis}, A. and {Liguori}, M. and {Lilje}, P.~B. and {Linden-V{\o}rnle}, M. and {L{\'o}pez-Caniego}, M. and {Lubin}, P.~M. and {Mac{\'\i}as-P{\'e}rez}, J.~F. and {Maggio}, G. and {Maino}, D. and {Mandolesi}, N. and {Mangilli}, A. and {Marchini}, A. and {Maris}, M. and {Martin}, P.~G. and {Martinelli}, M. and {Mart{\'\i}nez-Gonz{\'a}lez}, E. and {Masi}, S. and {Matarrese}, S. and {McGehee}, P. and {Meinhold}, P.~R. and {Melchiorri}, A. and {Melin}, J. -B. and {Mendes}, L. and {Mennella}, A. and {Migliaccio}, M. and {Millea}, M. and {Mitra}, S. and {Miville-Desch{\^e}nes}, M. -A. and {Moneti}, A. and {Montier}, L. and {Morgante}, G. and {Mortlock}, D. and {Moss}, A. and {Munshi}, D. and {Murphy}, J.~A. and {Naselsky}, P. and {Nati}, F. and {Natoli}, P. and {Netterfield}, C.~B. and {N{\o}rgaard-Nielsen}, H.~U. and {Noviello}, F. and {Novikov}, D. and {Novikov}, I. and {Oxborrow}, C.~A. and {Paci}, F. and {Pagano}, L. and {Pajot}, F. and {Paladini}, R. and {Paoletti}, D. and {Partridge}, B. and {Pasian}, F. and {Patanchon}, G. and {Pearson}, T.~J. and {Perdereau}, O. and {Perotto}, L. and {Perrotta}, F. and {Pettorino}, V. and {Piacentini}, F. and {Piat}, M. and {Pierpaoli}, E. and {Pietrobon}, D. and {Plaszczynski}, S. and {Pointecouteau}, E. and {Polenta}, G. and {Popa}, L. and {Pratt}, G.~W. and {Pr{\'e}zeau}, G. and {Prunet}, S. and {Puget}, J. -L. and {Rachen}, J.~P. and {Reach}, W.~T. and {Rebolo}, R. and {Reinecke}, M. and {Remazeilles}, M. and {Renault}, C. and {Renzi}, A. and {Ristorcelli}, I. and {Rocha}, G. and {Rosset}, C. and {Rossetti}, M. and {Roudier}, G. and {Rouill{\'e} d'Orfeuil}, B. and {Rowan-Robinson}, M. and {Rubi{\~n}o-Mart{\'\i}n}, J.~A. and {Rusholme}, B. and {Said}, N. and {Salvatelli}, V. and {Salvati}, L. and {Sandri}, M. and {Santos}, D. and {Savelainen}, M. and {Savini}, G. and {Scott}, D. and {Seiffert}, M.~D. and {Serra}, P. and {Shellard}, E.~P.~S. and {Spencer}, L.~D. and {Spinelli}, M. and {Stolyarov}, V. and {Stompor}, R. and {Sudiwala}, R. and {Sunyaev}, R. and {Sutton}, D. and {Suur-Uski}, A. -S. and {Sygnet}, J. -F. and {Tauber}, J.~A. and {Terenzi}, L. and {Toffolatti}, L. and {Tomasi}, M. and {Tristram}, M. and {Trombetti}, T. and {Tucci}, M. and {Tuovinen}, J. and {T{\"u}rler}, M. and {Umana}, G. and {Valenziano}, L. and {Valiviita}, J. and {Van Tent}, F. and {Vielva}, P. and {Villa}, F. and {Wade}, L.~A. and {Wandelt}, B.~D. and {Wehus}, I.~K. and {White}, M. and {White}, S.~D.~M. and {Wilkinson}, A. and {Yvon}, D. and {Zacchei}, A. and {Zonca}, A.},
        title = "{Planck 2015 results. XIII. Cosmological parameters}",
      journal = {\aap},
         year = 2016,
        month = sep,
       volume = {594},
          eid = {A13},
        pages = {A13},
          doi = {10.1051/0004-6361/201525830},
archivePrefix = {arXiv},
       eprint = {1502.01589},
 primaryClass = {astro-ph.CO},
       adsurl = {https://ui.adsabs.harvard.edu/abs/2016A&A...594A..13P}
}

@ARTICLE{Barthel1988,
       author = {{Barthel}, P.~D. and {Miley}, G.~K.},
        title = "{Evolution of radio structure in quasars: a new probe of protogalaxies?}",
      journal = {\nat},
         year = 1988,
        month = may,
       volume = {333},
       number = {6171},
        pages = {319-325},
          doi = {10.1038/333319a0},
       adsurl = {https://ui.adsabs.harvard.edu/abs/1988Natur.333..319B}
}

@ARTICLE{Shukla2022,
       author = {{Shukla}, Gitika and {Srianand}, Raghunathan and {Gupta}, Neeraj and {Petitjean}, Patrick and {Baker}, Andrew J. and {Krogager}, Jens-Kristian and {Noterdaeme}, Pasquier},
        title = "{Spatially resolved Lyman-{\ensuremath{\alpha}} emission around radio bright quasars}",
      journal = {\mnras},
         year = 2022,
        month = feb,
       volume = {510},
       number = {1},
        pages = {786-806},
          doi = {10.1093/mnras/stab3467},
archivePrefix = {arXiv},
       eprint = {2109.00576},
 primaryClass = {astro-ph.GA},
       adsurl = {https://ui.adsabs.harvard.edu/abs/2022MNRAS.510..786S}
}

@ARTICLE{Borisova2016,
       author = {{Borisova}, Elena and {Cantalupo}, Sebastiano and {Lilly}, Simon J. and {Marino}, Raffaella A. and {Gallego}, Sofia G. and {Bacon}, Roland and {Blaizot}, Jeremy and {Bouch{\'e}}, Nicolas and {Brinchmann}, Jarle and {Carollo}, C. Marcella and {Caruana}, Joseph and {Finley}, Hayley and {Herenz}, Edmund C. and {Richard}, Johan and {Schaye}, Joop and {Straka}, Lorrie A. and {Turner}, Monica L. and {Urrutia}, Tanya and {Verhamme}, Anne and {Wisotzki}, Lutz},
        title = "{Ubiquitous Giant Ly{\ensuremath{\alpha}} Nebulae around the Brightest Quasars at z {\ensuremath{\sim}} 3.5 Revealed with MUSE}",
      journal = {\apj},
         year = 2016,
        month = nov,
       volume = {831},
       number = {1},
          eid = {39},
        pages = {39},
          doi = {10.3847/0004-637X/831/1/39},
archivePrefix = {arXiv},
       eprint = {1605.01422},
 primaryClass = {astro-ph.GA},
       adsurl = {https://ui.adsabs.harvard.edu/abs/2016ApJ...831...39B}
}

@ARTICLE{Heckman1991a,
       author = {{Heckman}, Timothy M. and {Lehnert}, Matthew D. and {van Breugel}, Wil and {Miley}, George K.},
        title = "{Spatially Resolved Optical Images of High-Redshift Quasi-stellar Objects}",
      journal = {\apj},
         year = 1991,
        month = mar,
       volume = {370},
        pages = {78},
          doi = {10.1086/169794},
       adsurl = {https://ui.adsabs.harvard.edu/abs/1991ApJ...370...78H}
}

@ARTICLE{Heckman1991b,
       author = {{Heckman}, Timothy M. and {Lehnert}, Matthew D. and {Miley}, George K. and {van Breugel}, Wil},
        title = "{Spectroscopy of Spatially Extended Material around High-Redshift Radio-loud Quasars}",
      journal = {\apj},
         year = 1991,
        month = nov,
       volume = {381},
        pages = {373},
          doi = {10.1086/170660},
       adsurl = {https://ui.adsabs.harvard.edu/abs/1991ApJ...381..373H}
}

@ARTICLE{Rees1988,
       author = {{Rees}, Martin J.},
        title = "{Quasars as probes of gas in extended protogalaxies.}",
      journal = {\mnras},
         year = 1988,
        month = apr,
       volume = {231},
        pages = {91P-95},
          doi = {10.1093/mnras/231.1.91P},
       adsurl = {https://ui.adsabs.harvard.edu/abs/1988MNRAS.231P..91R}
}

@ARTICLE{Haiman2001,
       author = {{Haiman}, Zolt{\'a}n and {Rees}, Martin J.},
        title = "{Extended Ly{\ensuremath{\alpha}} Emission around Young Quasars: A Constraint on Galaxy Formation}",
      journal = {\apj},
         year = 2001,
        month = jul,
       volume = {556},
       number = {1},
        pages = {87-92},
          doi = {10.1086/321567},
archivePrefix = {arXiv},
       eprint = {astro-ph/0101174},
 primaryClass = {astro-ph},
       adsurl = {https://ui.adsabs.harvard.edu/abs/2001ApJ...556...87H}
}

@ARTICLE{Vayner2023,
       author = {{Vayner}, Andrey and {Zakamska}, Nadia L. and {Sabhlok}, Sanchit and {Wright}, Shelley A. and {Armus}, Lee and {Murray}, Norman and {Walth}, Gregory and {Ishikawa}, Yuzo},
        title = "{Cold mode gas accretion on two galaxy groups at z 2}",
      journal = {\mnras},
         year = 2023,
        month = feb,
       volume = {519},
       number = {1},
        pages = {961-979},
          doi = {10.1093/mnras/stac3537},
archivePrefix = {arXiv},
       eprint = {2212.00152},
 primaryClass = {astro-ph.GA},
       adsurl = {https://ui.adsabs.harvard.edu/abs/2023MNRAS.519..961V}
}

@ARTICLE{VillarMartin2007a,
       author = {{Villar-Mart{\'\i}n}, M. and {Humphrey}, A. and {De Breuck}, C. and {Fosbury}, R. and {Binette}, L. and {Vernet}, J.},
        title = "{Ly{\ensuremath{\alpha}} excess in high-redshift radio galaxies: a signature of star formation}",
      journal = {\mnras},
         year = 2007,
        month = mar,
       volume = {375},
       number = {4},
        pages = {1299-1310},
          doi = {10.1111/j.1365-2966.2006.11371.x},
archivePrefix = {arXiv},
       eprint = {astro-ph/0612116},
 primaryClass = {astro-ph},
       adsurl = {https://ui.adsabs.harvard.edu/abs/2007MNRAS.375.1299V}
}

@ARTICLE{Lau2022,
       author = {{Lau}, Marie Wingyee and {Hamann}, Fred and {Gillette}, Jarred and {Perrotta}, Serena and {Rupke}, David S.~N. and {Wylezalek}, Dominika and {Zakamska}, Nadia L.},
        title = "{Probing the inner circumgalactic medium and quasar illumination around the reddest 'extremely red quasar'}",
      journal = {\mnras},
         year = 2022,
        month = sep,
       volume = {515},
       number = {2},
        pages = {1624-1643},
          doi = {10.1093/mnras/stac1823},
archivePrefix = {arXiv},
       eprint = {2203.06203},
 primaryClass = {astro-ph.GA},
       adsurl = {https://ui.adsabs.harvard.edu/abs/2022MNRAS.515.1624L}
}

@ARTICLE{AB2015,
       author = {{Arrigoni Battaia}, Fabrizio and {Hennawi}, Joseph F. and {Prochaska}, J. Xavier and {Cantalupo}, Sebastiano},
        title = "{Deep He II and C IV Spectroscopy of a Giant Ly{\ensuremath{\alpha}} Nebula: Dense Compact Gas Clumps in the Circumgalactic Medium of a z \raisebox{-0.5ex}\textasciitilde 2 Quasar}",
      journal = {\apj},
         year = 2015,
        month = aug,
       volume = {809},
       number = {2},
          eid = {163},
        pages = {163},
          doi = {10.1088/0004-637X/809/2/163},
archivePrefix = {arXiv},
       eprint = {1504.03688},
 primaryClass = {astro-ph.GA},
       adsurl = {https://ui.adsabs.harvard.edu/abs/2015ApJ...809..163A}
}

@ARTICLE{Sabhlok2024a,
       author = {{Sabhlok}, Sanchit and {Wright}, Shelley A. and {Vayner}, Andrey and {Simonaitis-Boyd}, Sonata and {Murray}, Norman and {Armus}, Lee and {Cosens}, Maren and {Wiley}, James and {Kriek}, Mariska},
        title = "{Circumgalactic Environments around Distant Quasars 3C 9 and 4C 05.84}",
      journal = {arXiv e-prints},
         year = 2024,
        month = jan,
          eid = {arXiv:2401.04284},
        pages = {arXiv:2401.04284},
          doi = {10.48550/arXiv.2401.04284},
archivePrefix = {arXiv},
       eprint = {2401.04284},
 primaryClass = {astro-ph.GA},
       adsurl = {https://ui.adsabs.harvard.edu/abs/2024arXiv240104284S}
}

@ARTICLE{Catalog_7C,
	author = {{McGilchrist}, M.~M. and {Baldwin}, J.~E. and {Riley}, J.~M. and {Titterington}, D.~J. and {Waldram}, E.~M. and {Warner}, P.~J.},
	title = "{The 7C survey of radio sources at 151 MHz - two regions centered at RA 10h 28m, dec. 41 and RA 06h 28m, DEC 45.}",
	journal = {\mnras},
	year = 1990,
	month = sep,
	volume = {246},
	pages = {110-122},
	adsurl = {https://ui.adsabs.harvard.edu/abs/1990MNRAS.246..110M}
}

@ARTICLE{Catalog_3C,
	author = {{Smith}, H.~E. and {Spinrad}, H. and {Smith}, E.~O.},
	title = "{The revised 3C catalog of radio sources: a review of optical identifications and spectroscopy.}",
	journal = {\pasp},
	year = 1976,
	month = oct,
	volume = {88},
	pages = {621-646},
	doi = {10.1086/130001},
	adsurl = {https://ui.adsabs.harvard.edu/abs/1976PASP...88..621S}
}

@dataset{Catalog_4C,
	author = {{Pilkington}, J.~D.~H. and {Scott}, P.~F.},
	title = "{VizieR Online Data Catalog: Fourth Cambridge Survey (4C) (Pilkington+ 1965; Gower+ 1967)}",
	howpublished = {VizieR On-line Data Catalog: VIII/4.  Originally published in: 1965MmRAS..69..183P; 1967MmRAS..71...49G},
	year = 1996,
	month = apr,
	eid = {VIII/4},
	adsurl = {https://ui.adsabs.harvard.edu/abs/1996yCat.8004....0P}
}

@ARTICLE{Bridle1984,
       author = {{Bridle}, Alan H. and {Perley}, Richard A.},
        title = "{Extragalactic Radio Jets}",
      journal = {\araa},
         year = 1984,
        month = jan,
       volume = {22},
        pages = {319-358},
          doi = {10.1146/annurev.aa.22.090184.001535},
       adsurl = {https://ui.adsabs.harvard.edu/abs/1984ARA&A..22..319B}
}

@ARTICLE{Lonsdale1993,
       author = {{Lonsdale}, C.~J. and {Barthel}, P.~D. and {Miley}, G.~K.},
        title = "{The Radio Properties of High-Redshift Quasars. I. Dual-Frequency Observations of 79 Steep-Spectrum Quasars at Z > 1.5}",
      journal = {\apjs},
         year = 1993,
        month = jul,
       volume = {87},
        pages = {63},
          doi = {10.1086/191799},
       adsurl = {https://ui.adsabs.harvard.edu/abs/1993ApJS...87...63L}
}

@ARTICLE{Mackenzie2021,
       author = {{Mackenzie}, Ruari and {Pezzulli}, Gabriele and {Cantalupo}, Sebastiano and {Marino}, Raffaella A. and {Lilly}, Simon and {Muzahid}, Sowgat and {Matthee}, Jorryt and {Schaye}, Joop and {Wisotzki}, Lutz},
        title = "{Revealing the impact of quasar luminosity on giant Ly {\ensuremath{\alpha}} nebulae}",
      journal = {\mnras},
         year = 2021,
        month = mar,
       volume = {502},
       number = {1},
        pages = {494-509},
          doi = {10.1093/mnras/staa3277},
archivePrefix = {arXiv},
       eprint = {2010.12589},
 primaryClass = {astro-ph.GA},
       adsurl = {https://ui.adsabs.harvard.edu/abs/2021MNRAS.502..494M}
}

@ARTICLE{denBrok2020,
       author = {{den Brok}, J.~S. and {Cantalupo}, S. and {Mackenzie}, R. and {Marino}, R.~A. and {Pezzulli}, G. and {Matthee}, J. and {Johnson}, S.~D. and {Krumpe}, M. and {Urrutia}, T. and {Kollatschny}, W.},
        title = "{Probing the AGN unification model at redshift z {\ensuremath{\sim}} 3 with MUSE observations of giant Ly {\ensuremath{\alpha}} nebulae}",
      journal = {\mnras},
         year = 2020,
        month = jun,
       volume = {495},
       number = {2},
        pages = {1874-1887},
          doi = {10.1093/mnras/staa1269},
archivePrefix = {arXiv},
       eprint = {2005.01732},
 primaryClass = {astro-ph.GA},
       adsurl = {https://ui.adsabs.harvard.edu/abs/2020MNRAS.495.1874D}
}

@ARTICLE{Wilson1996:IonizationCones,
       author = {{Wilson}, Andrew S.},
        title = "{Ionization cones}",
      journal = {Vistas in Astronomy},
         year = 1996,
        month = jan,
       volume = {40},
       number = {1},
        pages = {63-70},
          doi = {10.1016/0083-6656(95)00102-6},
       adsurl = {https://ui.adsabs.harvard.edu/abs/1996VA.....40...63W}
}

@ARTICLE{Wylezalek2016:IonizationCones,
       author = {{Wylezalek}, Dominika and {Zakamska}, Nadia L. and {Liu}, Guilin and {Obied}, Georges},
        title = "{Towards a comprehensive picture of powerful quasars, their host galaxies and quasar winds at z {\ensuremath{\sim}} 0.5}",
      journal = {\mnras},
         year = 2016,
        month = mar,
       volume = {457},
       number = {1},
        pages = {745-763},
          doi = {10.1093/mnras/stv3022},
archivePrefix = {arXiv},
       eprint = {1601.02620},
 primaryClass = {astro-ph.GA},
       adsurl = {https://ui.adsabs.harvard.edu/abs/2016MNRAS.457..745W}
}

@ARTICLE{Zhuang2018:TorusCones,
       author = {{Zhuang}, Ming-Yang and {Ho}, Luis C. and {Shangguan}, Jinyi},
        title = "{The Infrared Emission and Opening Angle of the Torus in Quasars}",
      journal = {\apj},
         year = 2018,
        month = aug,
       volume = {862},
       number = {2},
          eid = {118},
        pages = {118},
          doi = {10.3847/1538-4357/aacc2d},
archivePrefix = {arXiv},
       eprint = {1806.03783},
 primaryClass = {astro-ph.GA},
       adsurl = {https://ui.adsabs.harvard.edu/abs/2018ApJ...862..118Z}
}

@ARTICLE{Drake2019,
       author = {{Drake}, Alyssa B. and {Farina}, Emanuele Paolo and {Neeleman}, Marcel and {Walter}, Fabian and {Venemans}, Bram and {Banados}, Eduardo and {Mazzucchelli}, Chiara and {Decarli}, Roberto},
        title = "{Ly{\ensuremath{\alpha}} Halos around z {\ensuremath{\sim}} 6 Quasars}",
      journal = {\apj},
         year = 2019,
        month = aug,
       volume = {881},
       number = {2},
          eid = {131},
        pages = {131},
          doi = {10.3847/1538-4357/ab2984},
archivePrefix = {arXiv},
       eprint = {1906.07197},
 primaryClass = {astro-ph.GA},
       adsurl = {https://ui.adsabs.harvard.edu/abs/2019ApJ...881..131D}
}

@ARTICLE{Farina2019,
       author = {{Farina}, Emanuele Paolo and {Arrigoni-Battaia}, Fabrizio and {Costa}, Tiago and {Walter}, Fabian and {Hennawi}, Joseph F. and {Drake}, Alyssa B. and {Decarli}, Roberto and {Gutcke}, Thales A. and {Mazzucchelli}, Chiara and {Neeleman}, Marcel and {Georgiev}, Iskren and {Eilers}, Anna-Christina and {Davies}, Frederick B. and {Ba{\~n}ados}, Eduardo and {Fan}, Xiaohui and {Onoue}, Masafusa and {Schindler}, Jan-Torge and {Venemans}, Bram P. and {Wang}, Feige and {Yang}, Jinyi and {Rabien}, Sebastian and {Busoni}, Lorenzo},
        title = "{The REQUIEM Survey. I. A Search for Extended Ly{\ensuremath{\alpha}} Nebular Emission Around 31 z > 5.7 Quasars}",
      journal = {\apj},
         year = 2019,
        month = dec,
       volume = {887},
       number = {2},
          eid = {196},
        pages = {196},
          doi = {10.3847/1538-4357/ab5847},
archivePrefix = {arXiv},
       eprint = {1911.08498},
 primaryClass = {astro-ph.GA},
       adsurl = {https://ui.adsabs.harvard.edu/abs/2019ApJ...887..196F}
}

@ARTICLE{Travascio2020,
       author = {{Travascio}, A. and {Zappacosta}, L. and {Cantalupo}, S. and {Piconcelli}, E. and {Arrigoni Battaia}, F. and {Ginolfi}, M. and {Bischetti}, M. and {Vietri}, G. and {Bongiorno}, A. and {D'Odorico}, V. and {Duras}, F. and {Feruglio}, C. and {Vignali}, C. and {Fiore}, F.},
        title = "{The WISSH quasars project. VIII. Outflows and metals in the circum-galactic medium around the hyper-luminous z {\ensuremath{\sim}} 3.6 quasar J1538+08}",
      journal = {\aap},
         year = 2020,
        month = mar,
       volume = {635},
          eid = {A157},
        pages = {A157},
          doi = {10.1051/0004-6361/201936197},
archivePrefix = {arXiv},
       eprint = {2001.07218},
 primaryClass = {astro-ph.GA},
       adsurl = {https://ui.adsabs.harvard.edu/abs/2020A&A...635A.157T}
}

@ARTICLE{Obreja2024,
       author = {{Obreja}, Aura and {Arrigoni Battaia}, Fabrizio and {Macci{\`o}}, Andrea V. and {Buck}, Tobias},
        title = "{AGN radiation imprints on the circumgalactic medium of massive galaxies}",
      journal = {\mnras},
         year = 2024,
        month = jan,
       volume = {527},
       number = {3},
        pages = {8078-8102},
          doi = {10.1093/mnras/stad3410},
archivePrefix = {arXiv},
       eprint = {2311.01503},
 primaryClass = {astro-ph.GA},
       adsurl = {https://ui.adsabs.harvard.edu/abs/2024MNRAS.527.8078O}
}

@ARTICLE{Dijkstra2006,
       author = {{Dijkstra}, Mark and {Haiman}, Zolt{\'a}n and {Spaans}, Marco},
        title = "{Ly{\ensuremath{\alpha}} Radiation from Collapsing Protogalaxies. I. Characteristics of the Emergent Spectrum}",
      journal = {\apj},
         year = 2006,
        month = sep,
       volume = {649},
       number = {1},
        pages = {14-36},
          doi = {10.1086/506243},
archivePrefix = {arXiv},
       eprint = {astro-ph/0510407},
 primaryClass = {astro-ph},
       adsurl = {https://ui.adsabs.harvard.edu/abs/2006ApJ...649...14D}
}

@ARTICLE{Christensen2006,
       author = {{Christensen}, L. and {Jahnke}, K. and {Wisotzki}, L. and {S{\'a}nchez}, S.~F.},
        title = "{Extended Lyman-{\ensuremath{\alpha}} emission around bright quasars}",
      journal = {\aap},
         year = 2006,
        month = dec,
       volume = {459},
       number = {3},
        pages = {717-729},
          doi = {10.1051/0004-6361:20065318},
archivePrefix = {arXiv},
       eprint = {astro-ph/0603835},
 primaryClass = {astro-ph},
       adsurl = {https://ui.adsabs.harvard.edu/abs/2006A&A...459..717C}
}

@ARTICLE{Hennawi2013,
       author = {{Hennawi}, Joseph F. and {Prochaska}, J. Xavier},
        title = "{Quasars Probing Quasars. IV. Joint Constraints on the Circumgalactic Medium from Absorption and Emission}",
      journal = {\apj},
         year = 2013,
        month = mar,
       volume = {766},
       number = {1},
          eid = {58},
        pages = {58},
          doi = {10.1088/0004-637X/766/1/58},
archivePrefix = {arXiv},
       eprint = {1303.2708},
 primaryClass = {astro-ph.CO},
       adsurl = {https://ui.adsabs.harvard.edu/abs/2013ApJ...766...58H}
}

@ARTICLE{Gonzales2025,
       author = {{Gonz{\'a}lez Lobos}, Jay and {Arrigoni Battaia}, Fabrizio and {Obreja}, Aura and {Kauffmann}, Guinevere and {Farina}, Emanuele Paolo and {Costa}, Tiago},
        title = "{QSO MUSEUM III: the circumgalactic medium in Ly$α$ emission around 120 $z\sim3$ quasars covering the SDSS parameter space. Witnessing the instantaneous AGN feedback on halo scales}",
      journal = {arXiv e-prints},
         year = 2025,
        month = jul,
          eid = {arXiv:2507.16898},
        pages = {arXiv:2507.16898},
          doi = {10.48550/arXiv.2507.16898},
archivePrefix = {arXiv},
       eprint = {2507.16898},
 primaryClass = {astro-ph.GA},
       adsurl = {https://ui.adsabs.harvard.edu/abs/2025arXiv250716898G}
}

@ARTICLE{Costa2022,
       author = {{Costa}, Tiago and {Arrigoni Battaia}, Fabrizio and {Farina}, Emanuele P. and {Keating}, Laura C. and {Rosdahl}, Joakim and {Kimm}, Taysun},
        title = "{AGN-driven outflows and the formation of Ly{\ensuremath{\alpha}} nebulae around high-z quasars}",
      journal = {\mnras},
         year = 2022,
        month = dec,
       volume = {517},
       number = {2},
        pages = {1767-1790},
          doi = {10.1093/mnras/stac2432},
archivePrefix = {arXiv},
       eprint = {2203.11232},
 primaryClass = {astro-ph.GA},
       adsurl = {https://ui.adsabs.harvard.edu/abs/2022MNRAS.517.1767C}
}
\bibliographystyle{aasjournal}

\section*{Appendix: All KCWI Maps}
The journal version of the paper includes figure sets which render Figures \ref{fig:KCWI_Lya}, \ref{fig:KCWI_HeII} and \ref{fig:KCWI_CIV} using Figure Sets for the online version of the journal. These figures are reproduced in the appendix here for the Arxiv version. 

\subsection*{\rm{\lya\,} maps}
PSF-subtracted images of the \lya\, surface brightness, Moment 1 and Moment 2 maps for all sources from the QUART sample. The exposure times for the full sample are reported in Table \ref{tab:QUART_Members}. All figures have an on-sky position angle of 0, North is up and East is to the left. The systemic redshift used to calculate the moment maps is used from OSIRIS measurements in \citet{QUART_OSIRIS} and is shown in Table \ref{tab:QUART_Members}. The location of the quasar is denoted with a gold star marker. Companion galaxies shown on the surface brightness profile with a black star.

\begin{figure*}[!ht]
    \centering
    \includegraphics[width=\linewidth]{ 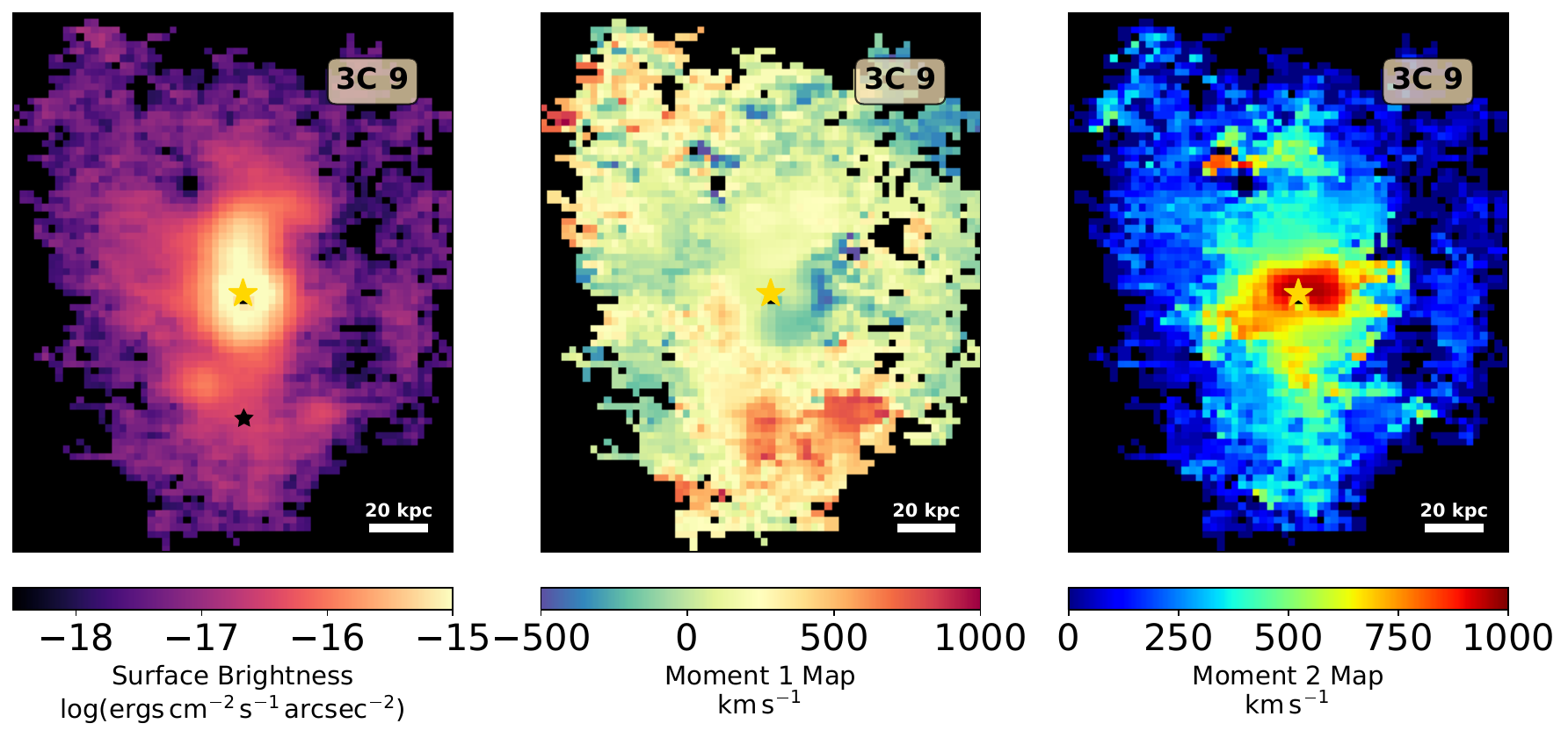}
    
    \caption{3C 9 \lya}
    \label{fig:3c9_Lyalpha}
\end{figure*}

\begin{figure*}[!ht]
    \centering
    \includegraphics[width=\linewidth]{ 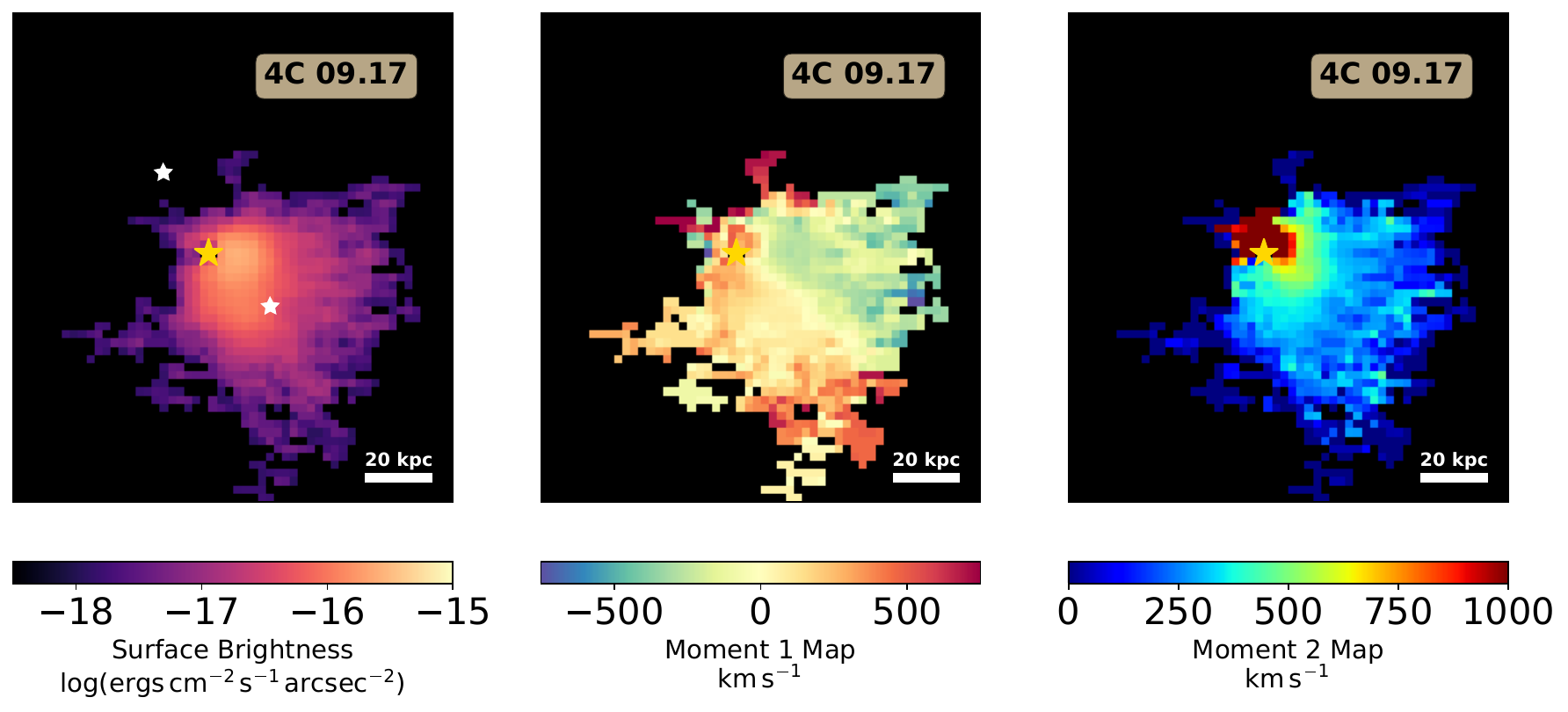}
    
    \caption{4C 09.17 \lya}
    \label{fig:4c0917_Lyalpha}
\end{figure*}

\begin{figure*}[!ht]
    \centering
    \includegraphics[width=\linewidth]{ 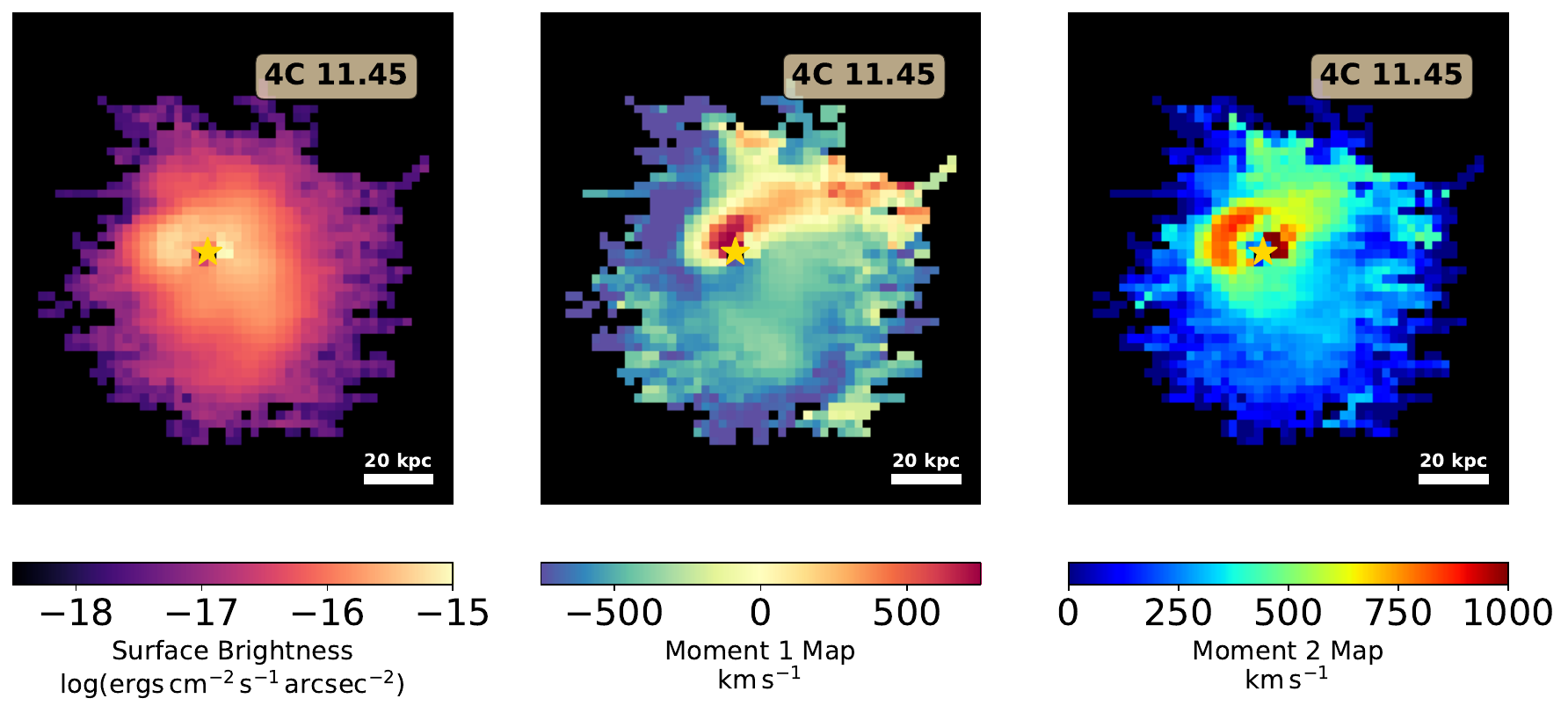}
    
    \caption{4C 11.45 \lya}
    \label{fig:4c1145_Lyalpha}
\end{figure*}

\begin{figure*}[!ht]
    \centering
    \includegraphics[width=\linewidth]{ 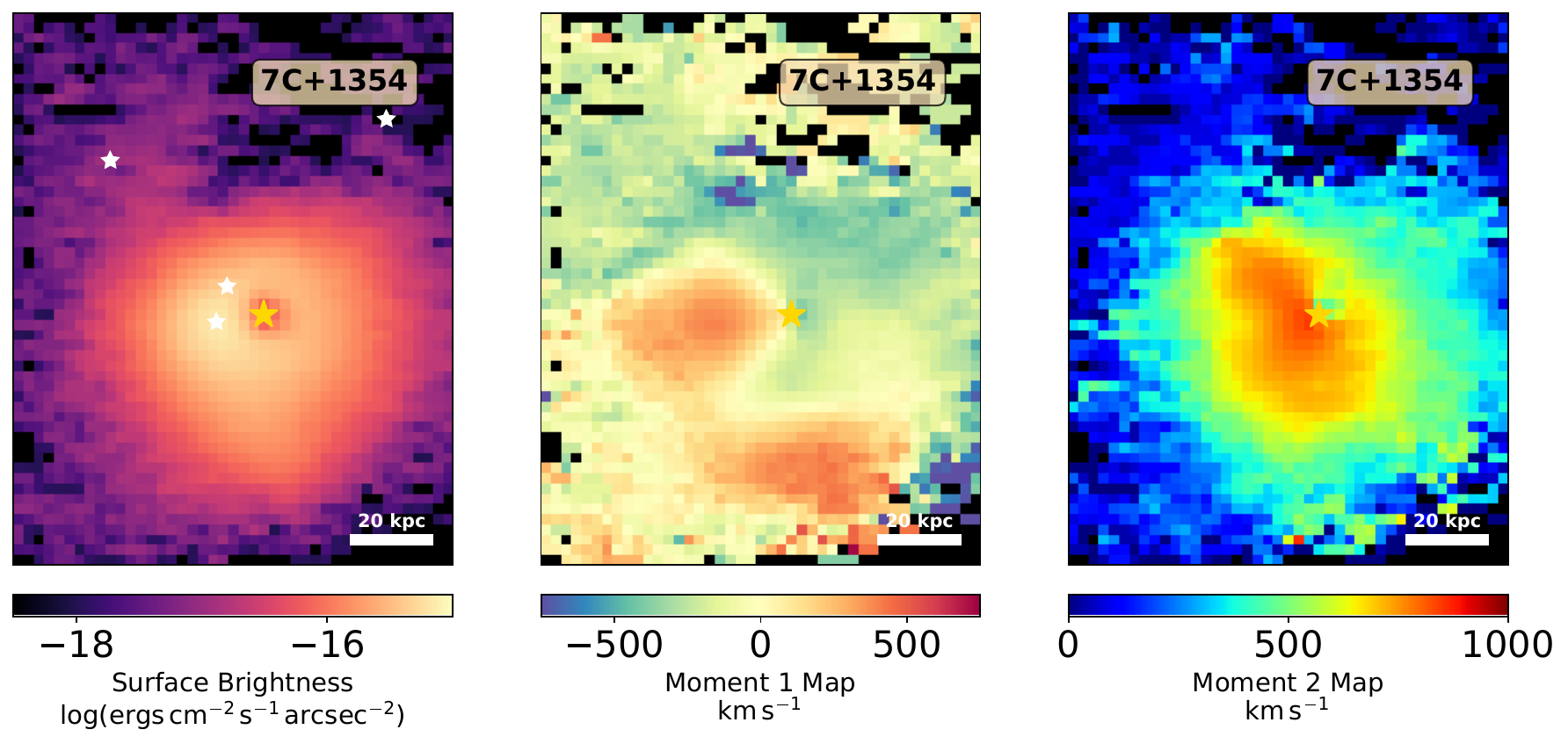}
    
    \caption{7C 1354+2552 \lya}
    \label{fig:7c1354_Lyalpha}
\end{figure*}

\begin{figure*}[!ht]
    \centering
    \includegraphics[width=\linewidth]{ 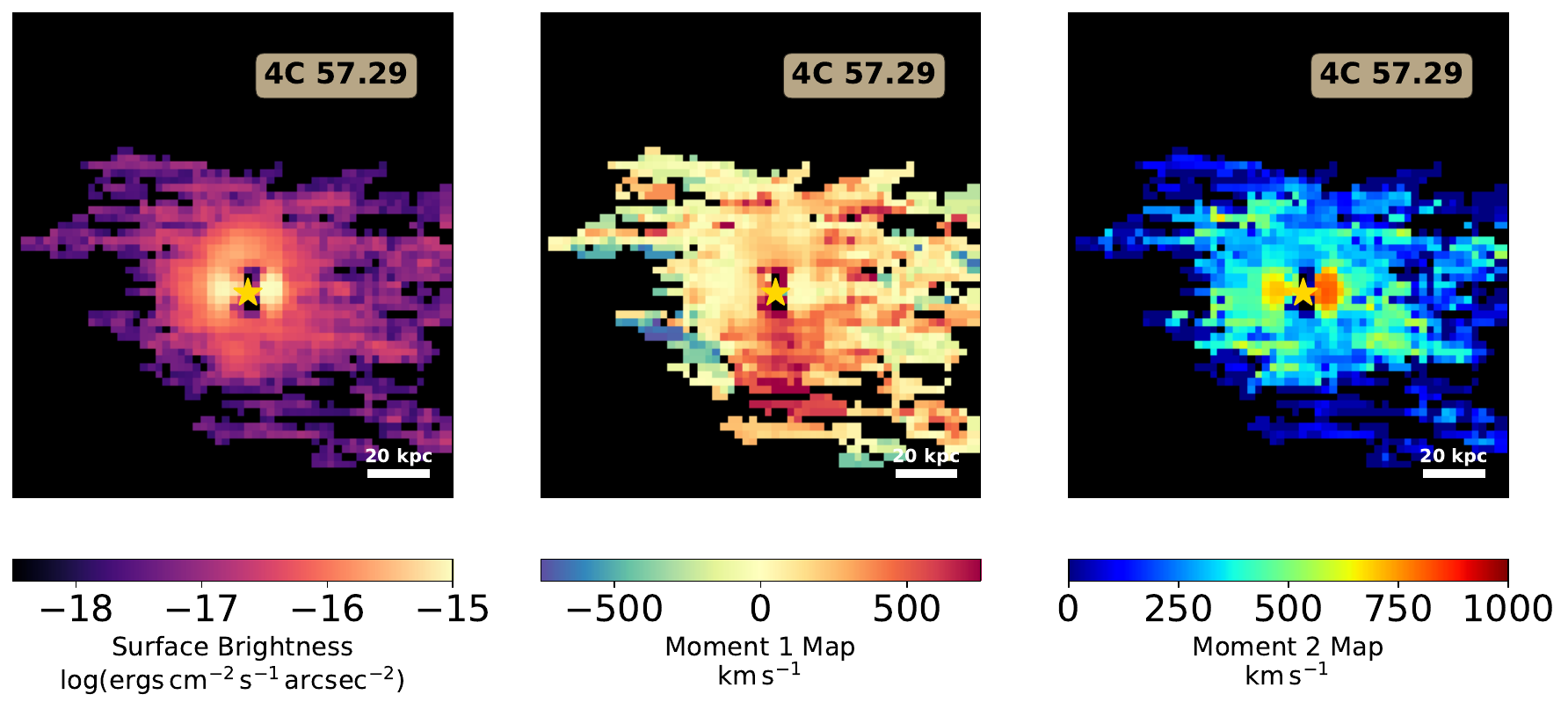}
    
    \caption{4C 57.29 \lya}
    \label{fig:4c5729_Lyalpha}
\end{figure*}

\begin{figure*}[!ht]
    \centering
    \includegraphics[width=\linewidth]{ 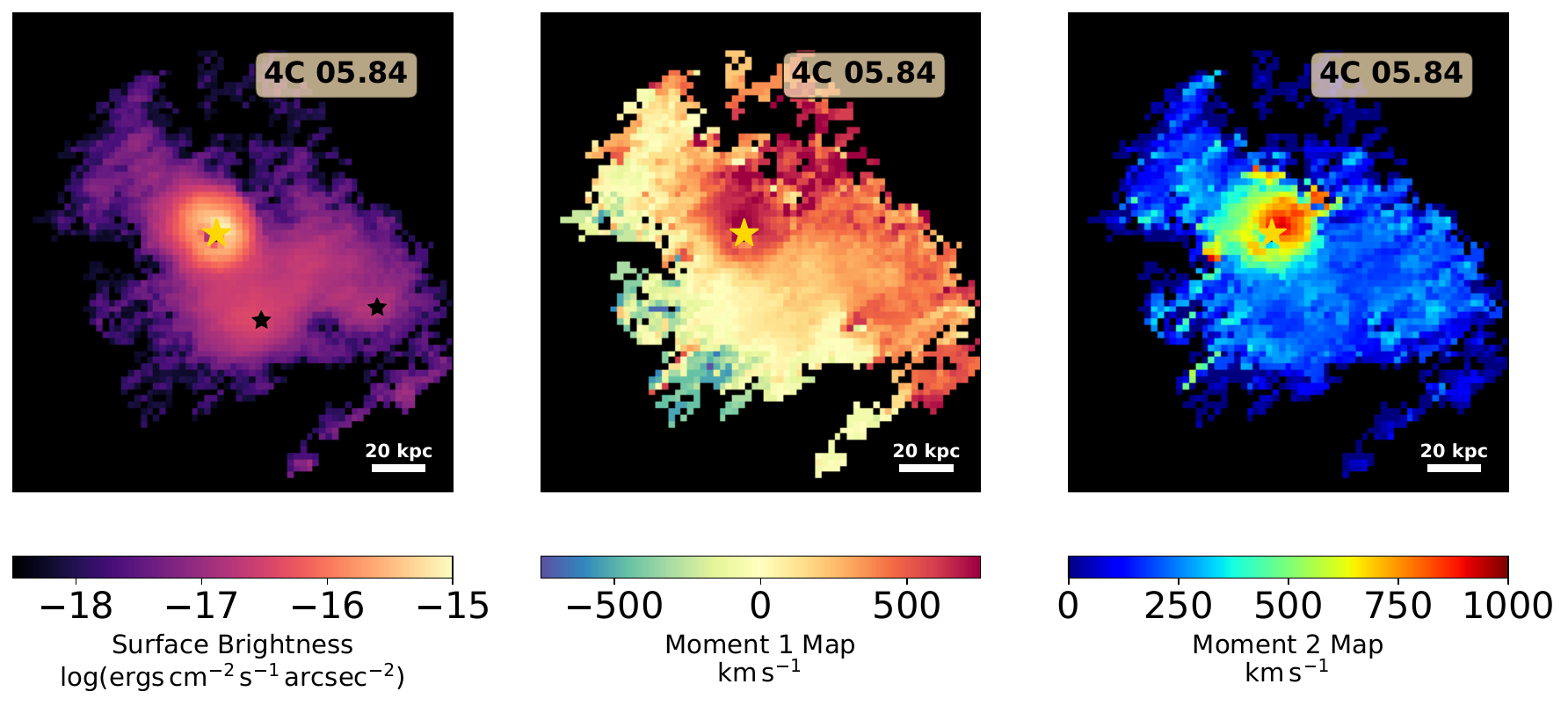}
    
    \caption{4C 05.84 \lya}
    \label{fig:4c0584_Lyalpha}
\end{figure*}

\begin{figure*}[!ht]
    \centering
    \includegraphics[width=\linewidth]{ 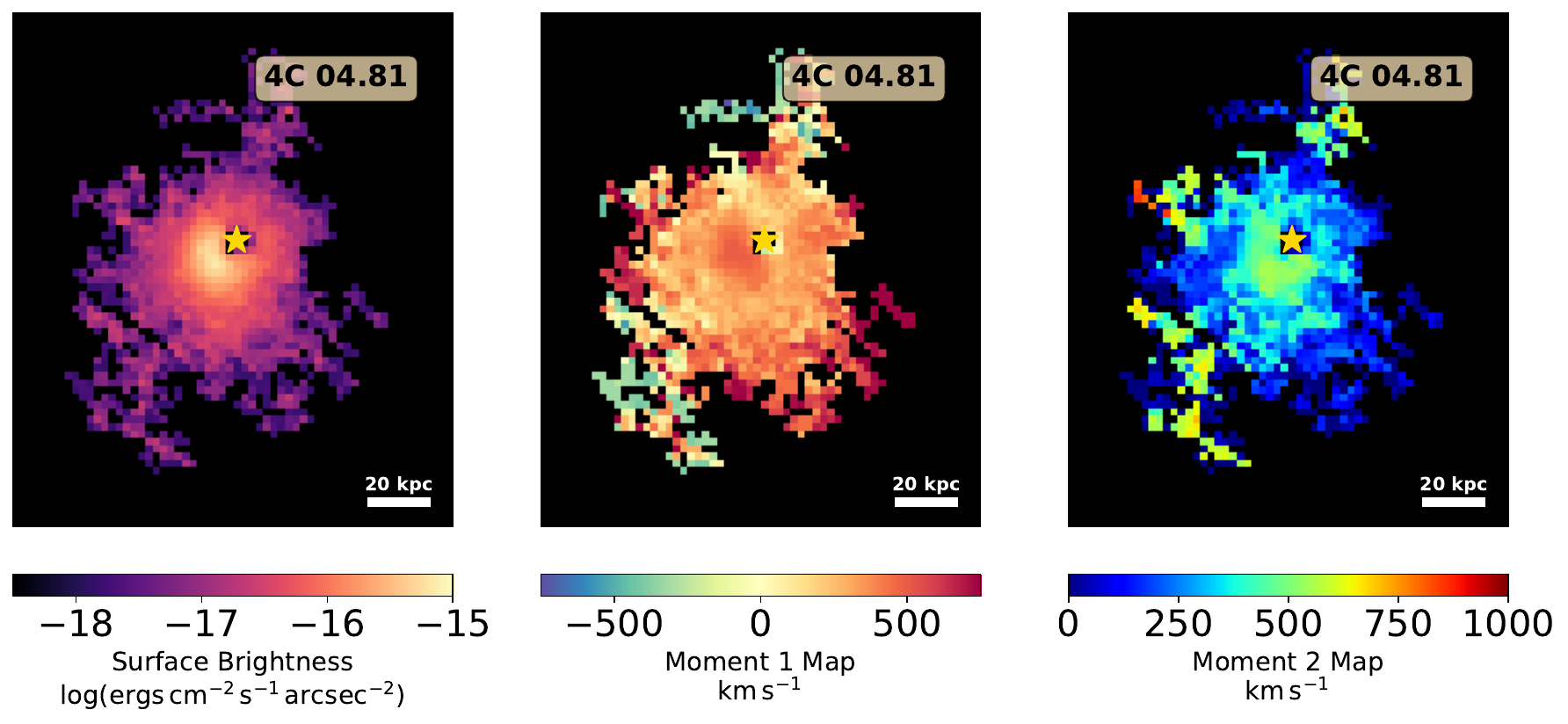}
    
    \caption{4C 04.81 \lya}
    \label{fig:4c0481_Lyalpha}
\end{figure*}

\FloatBarrier
\subsection*{\rm{\heii\,} maps}
PSF-subtracted images of the \heii\, surface brightness, Moment 1 and Moment 2 maps for the all sources. All figures have an on-sky position angle of 0. The systemic redshift used to calculate the moment maps is used from OSIRIS measurements in \citet{QUART_OSIRIS} and as shown in Table \ref{tab:QUART_Members}. The location of the quasar is denoted with a gold star marker.

\begin{figure*}[!ht]
    \centering
    \includegraphics[width=\linewidth]{ 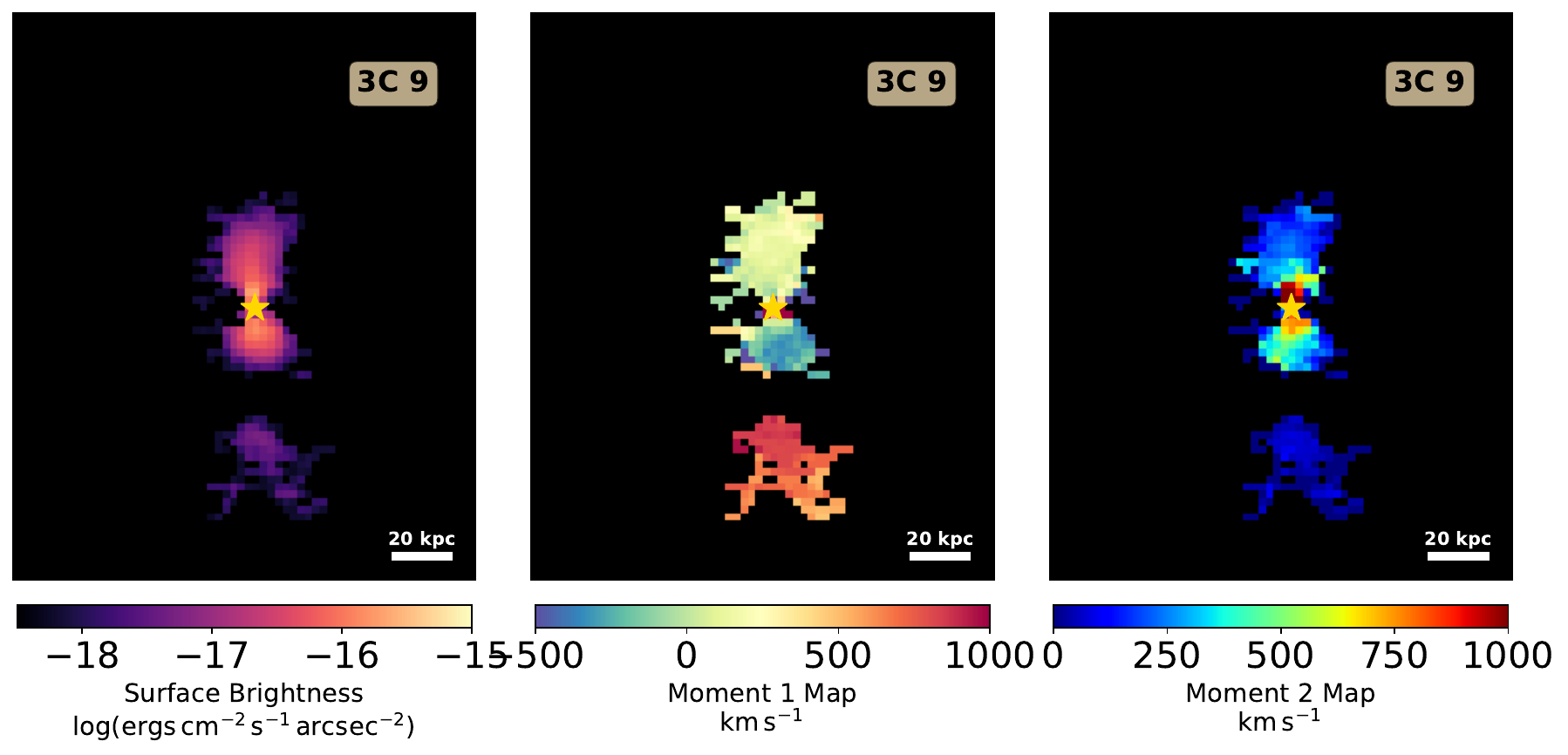}
    
    \caption{3C 9 \heii}
    \label{fig:3c9_HeII}
\end{figure*}

\begin{figure*}[!ht]
    \centering
    \includegraphics[width=\linewidth]{ 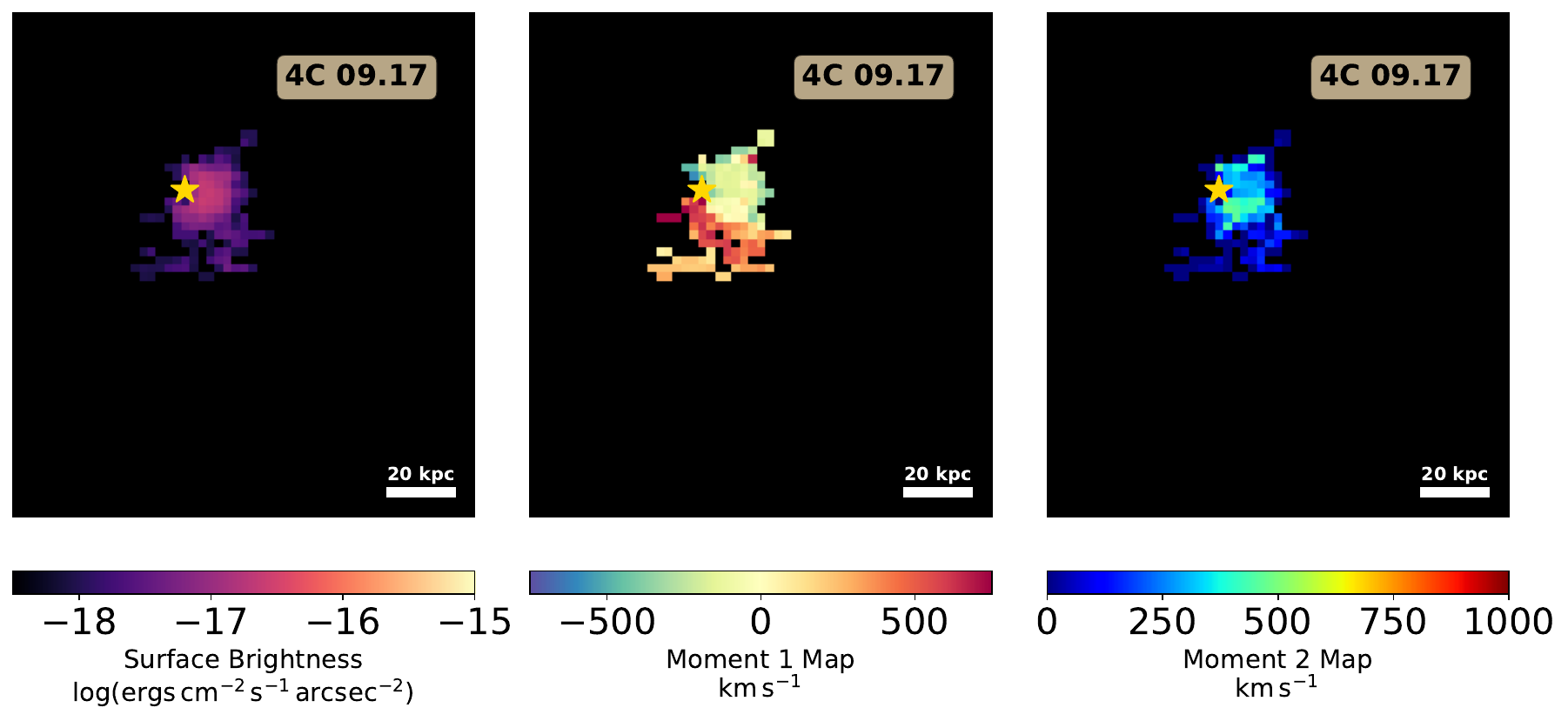}
    
    \caption{4C 09.17 \heii}
    \label{fig:4c0917_HeII}
\end{figure*}

\begin{figure*}[!ht]
    \centering
    \includegraphics[width=\linewidth]{ 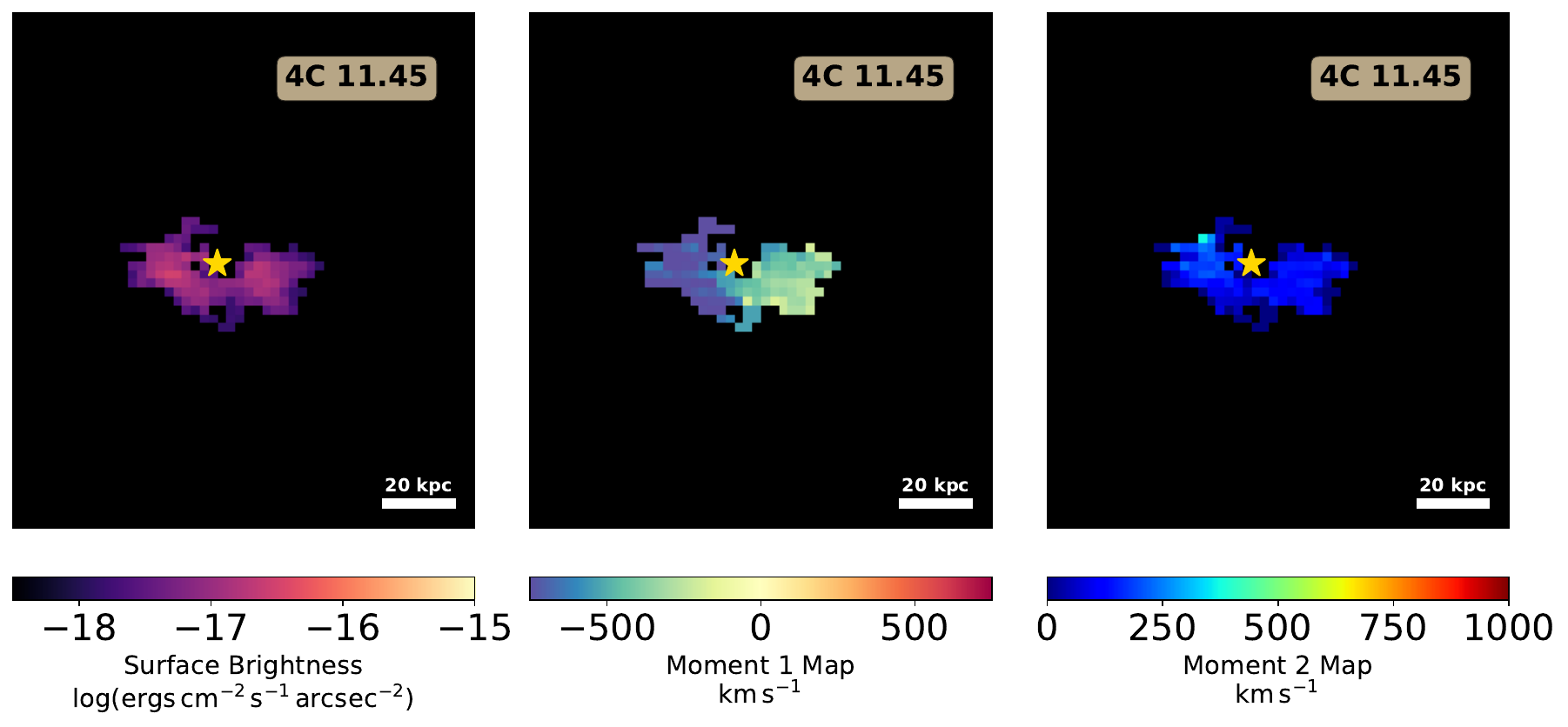}
    
    \caption{4C 11.45 \heii}
    \label{fig:4c1145_HeII}
\end{figure*}

\begin{figure*}[!ht]
    \centering
    \includegraphics[width=\linewidth]{ 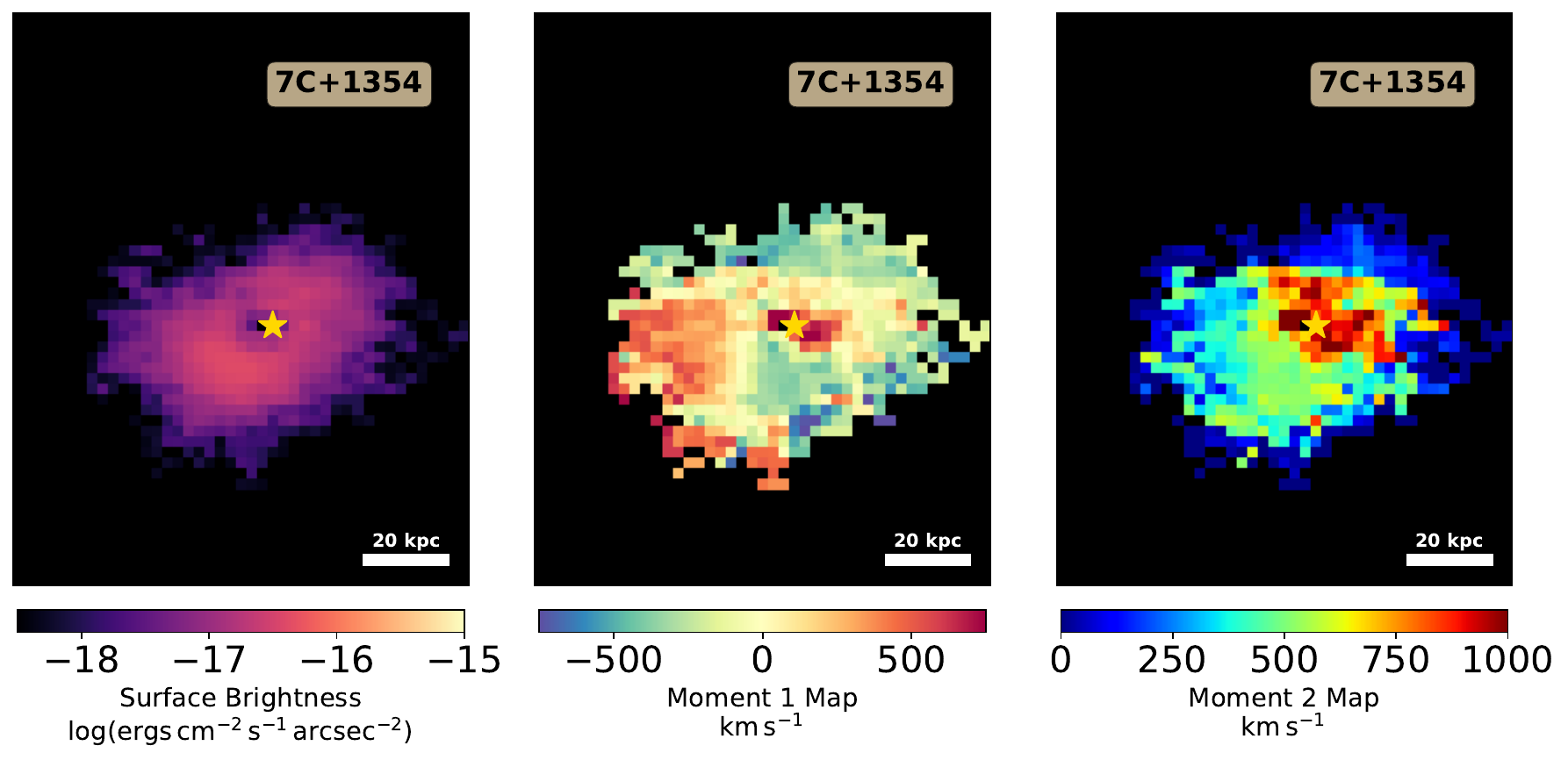}
    
    \caption{7C 1354+2552 \heii}
    \label{fig:7c1354_HeII}
\end{figure*}

\begin{figure*}[!ht]
    \centering
    \includegraphics[width=\linewidth]{ 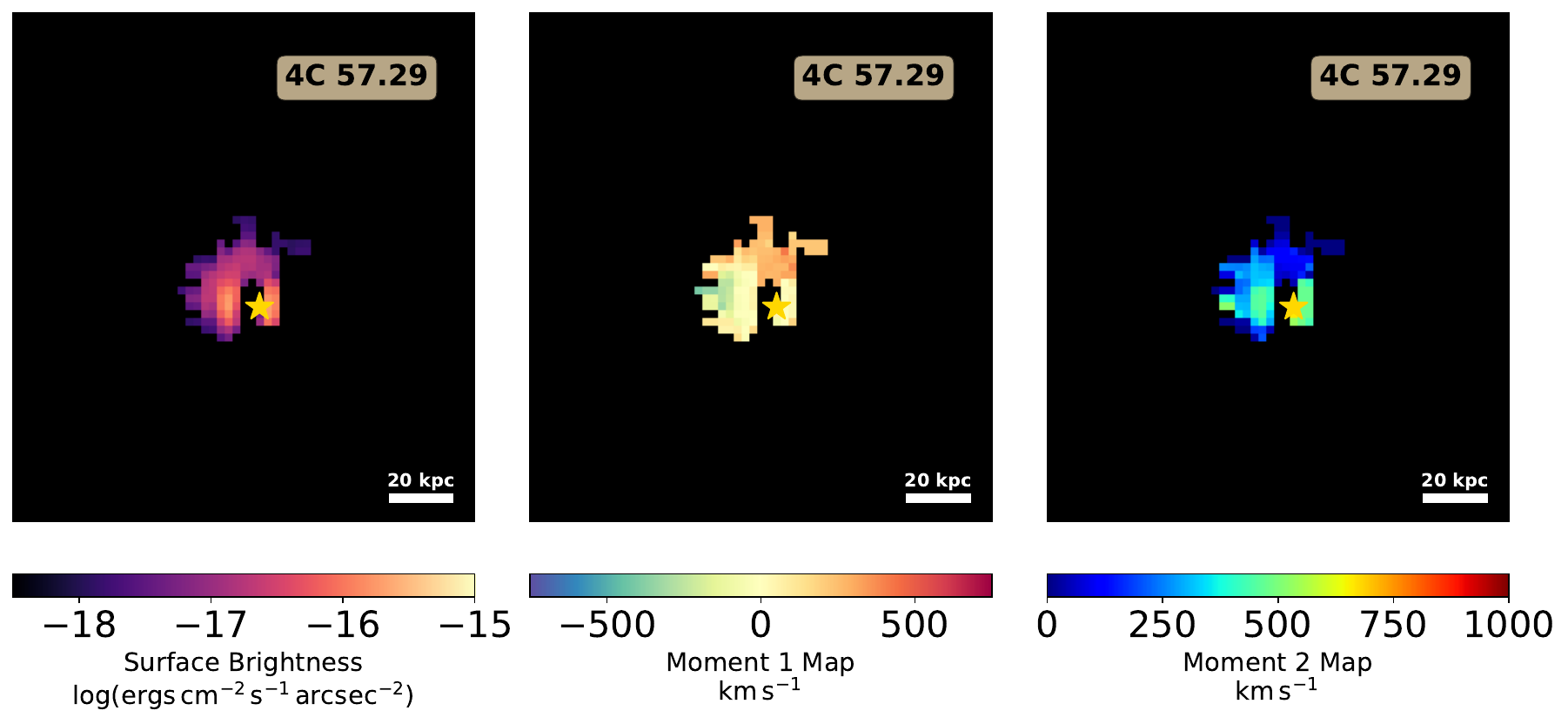}
    
    \caption{4C 57.29 \heii}
    \label{fig:4c5729_HeII}
\end{figure*}

\begin{figure*}[!ht]
    \centering
    \includegraphics[width=\linewidth]{ 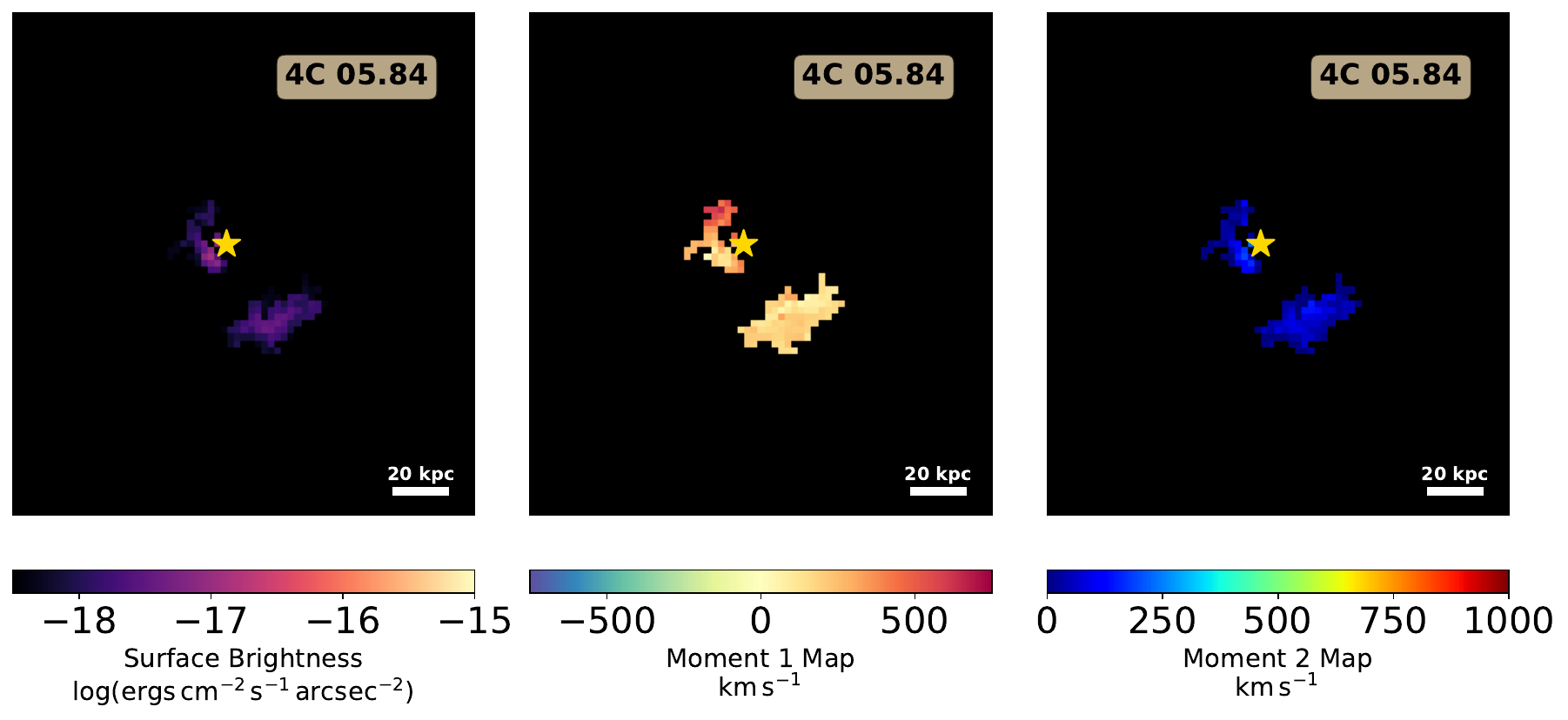}
    
    \caption{4C 05.84 \heii}
    \label{fig:4c0584_HeII}
\end{figure*}

\FloatBarrier
\subsection*{\rm{\civ\,} maps}
PSF-subtracted images of the \civ\, surface brightness, Moment 1 and Moment 2 maps for all sources from the QUART sample. The systemic redshift used to calculate the moment maps is used from OSIRIS measurements in \citet{QUART_OSIRIS} and as shown in Table \ref{tab:QUART_Members}. The location of the quasar is denoted with a gold star marker.

\begin{figure*}[!ht]
    \centering
    \includegraphics[width=\linewidth]{ 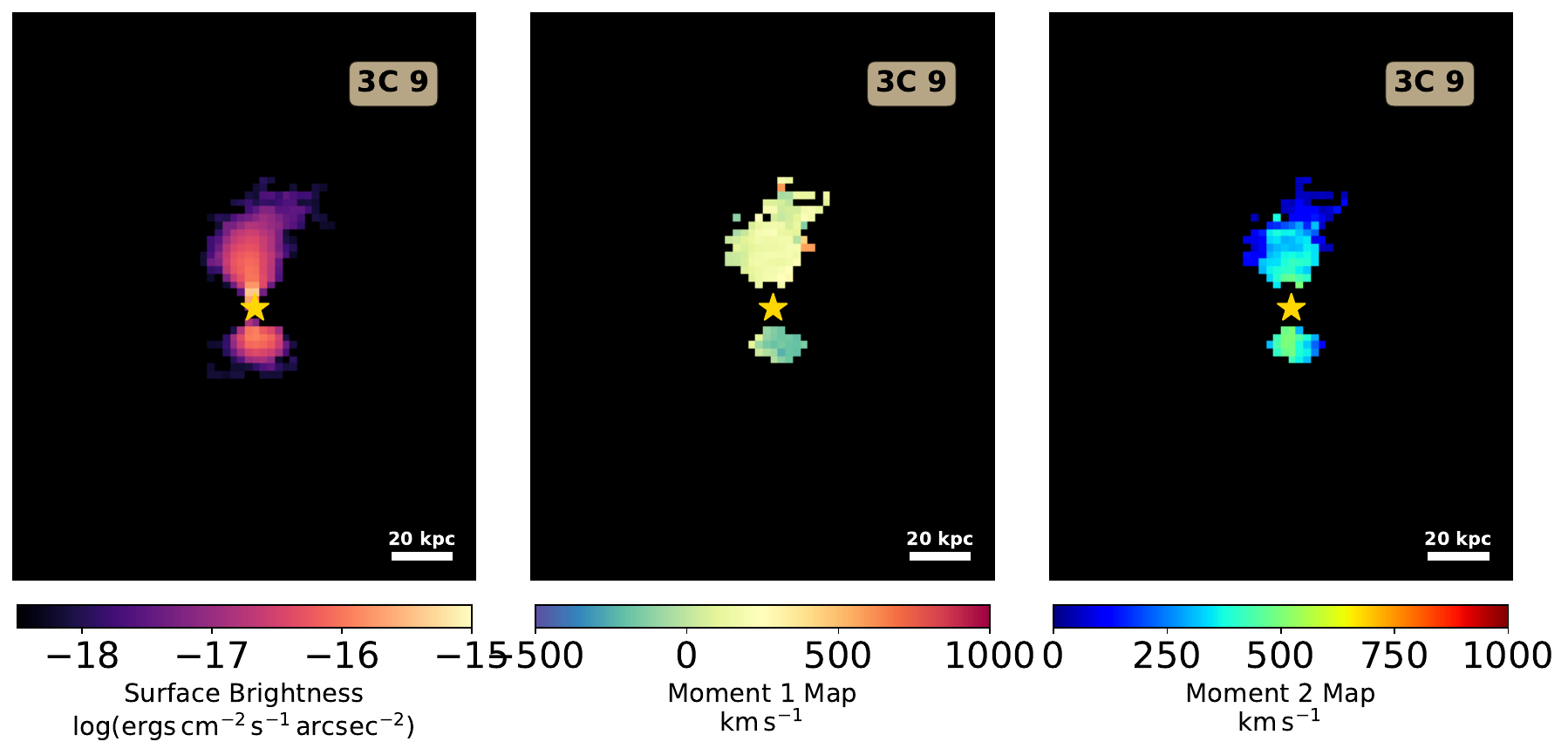}
    
    \caption{3C 9 \civ}
    \label{fig:3c9_CIV}
\end{figure*}

\begin{figure*}[!ht]
    \centering
    \includegraphics[width=\linewidth]{ 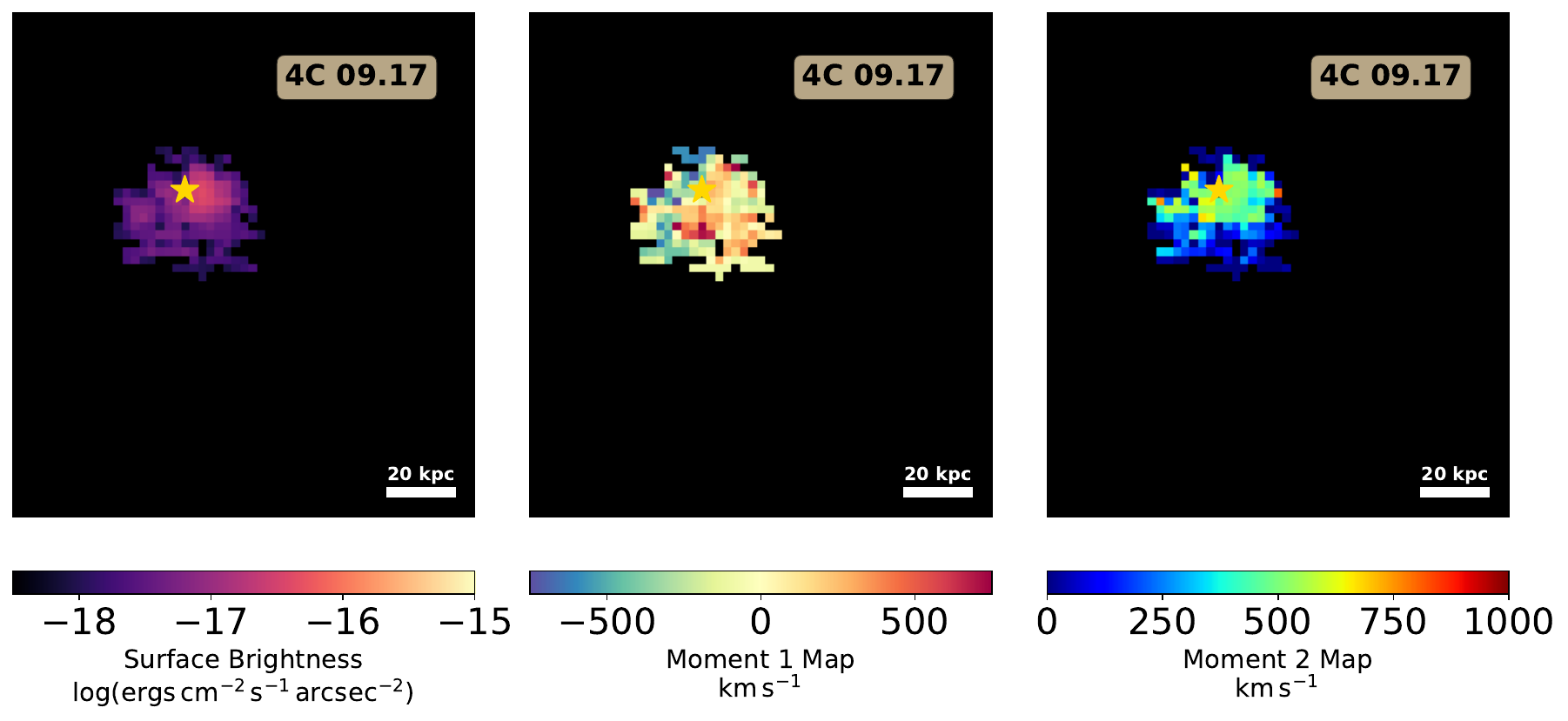}
    
    \caption{4C 09.17 \civ}
    \label{fig:4c0917_CIV}
\end{figure*}

\begin{figure*}[!ht]
    \centering
    \includegraphics[width=\linewidth]{ 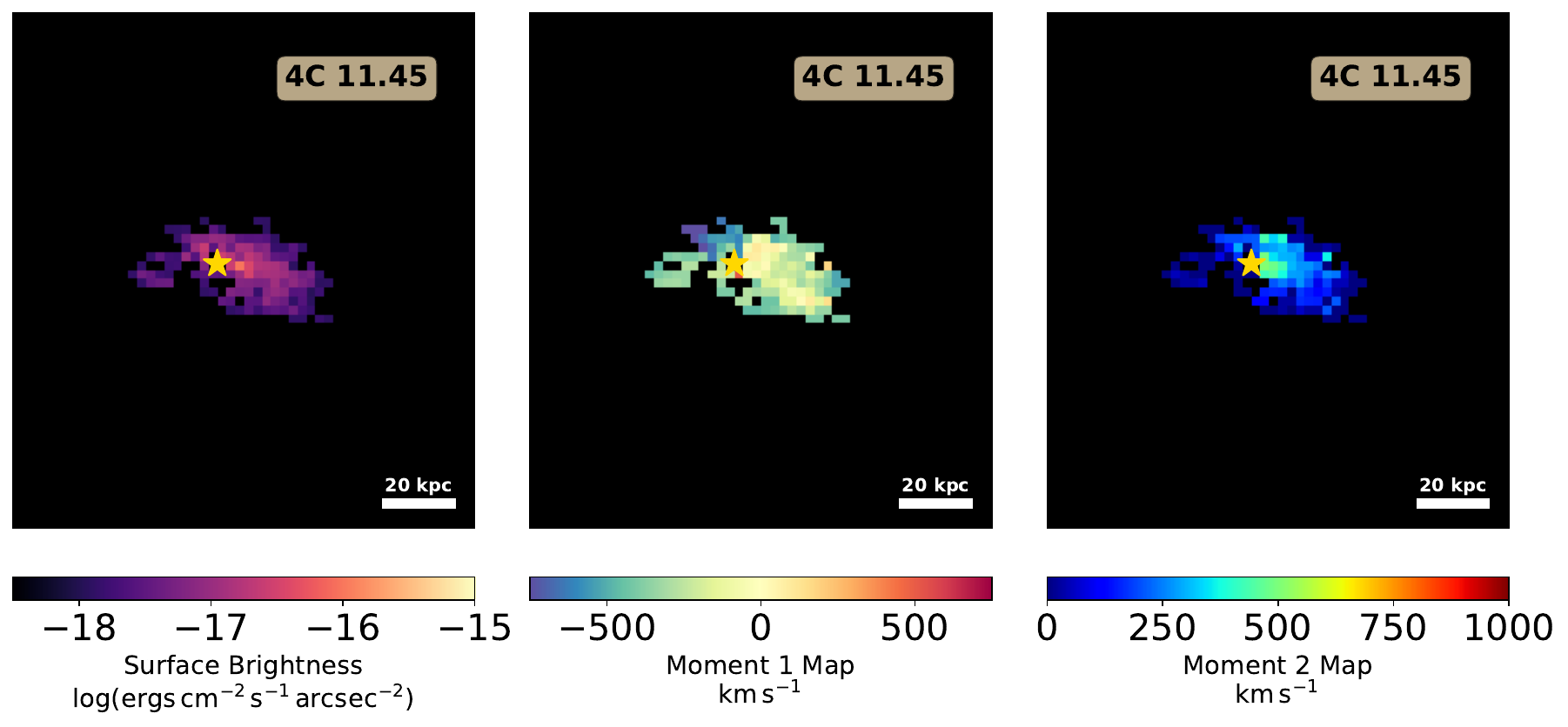}
    
    \caption{4C 11.45 \civ}
    \label{fig:4c1145_CIV}
\end{figure*}

\begin{figure*}[!ht]
    \centering
    \includegraphics[width=\linewidth]{ 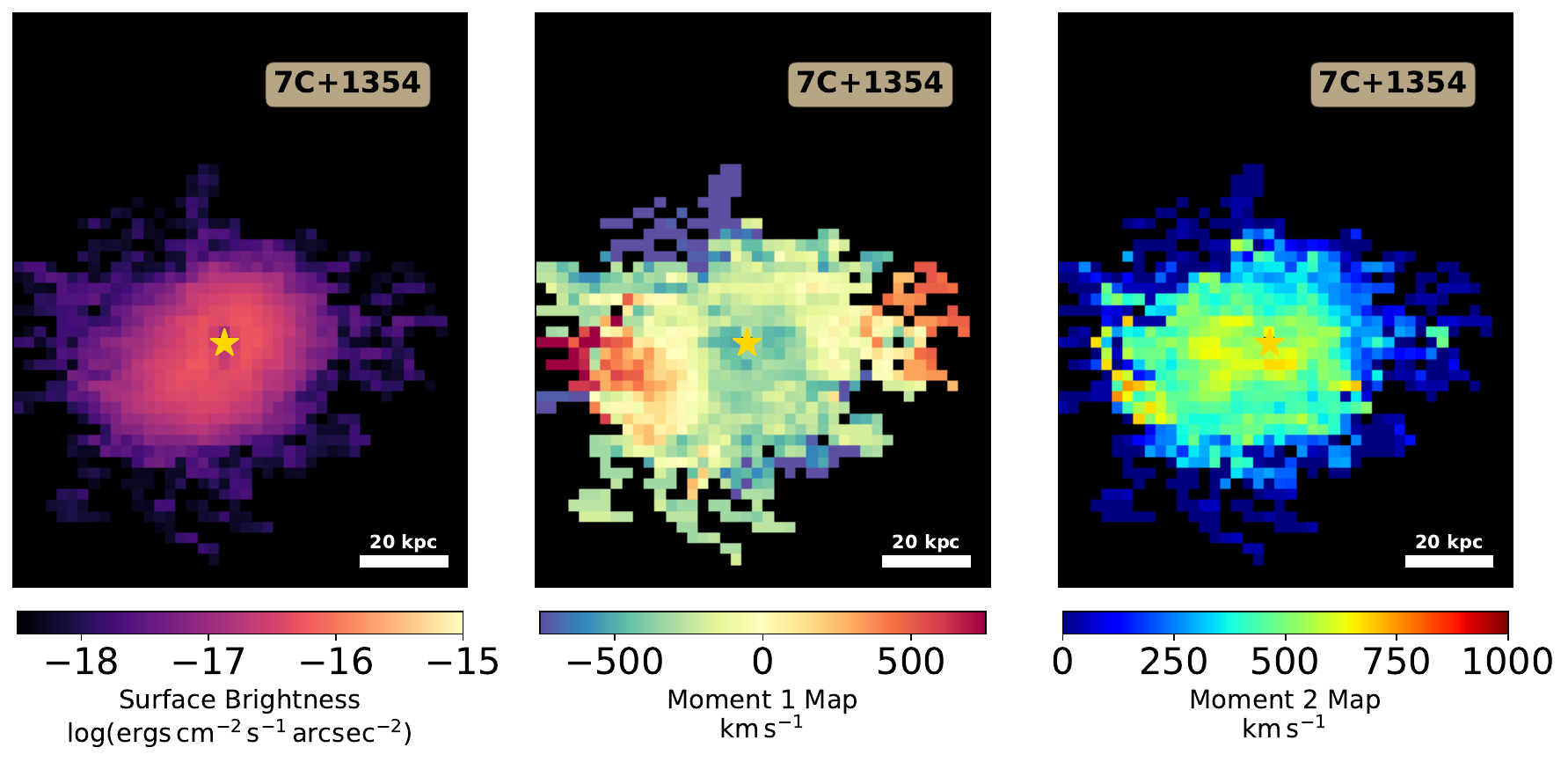}
    
    \caption{7C 1354+2552 \civ }
    \label{fig:7c1354_CIV}
\end{figure*}

\begin{figure*}[!ht]
    \centering
    \includegraphics[width=\linewidth]{ 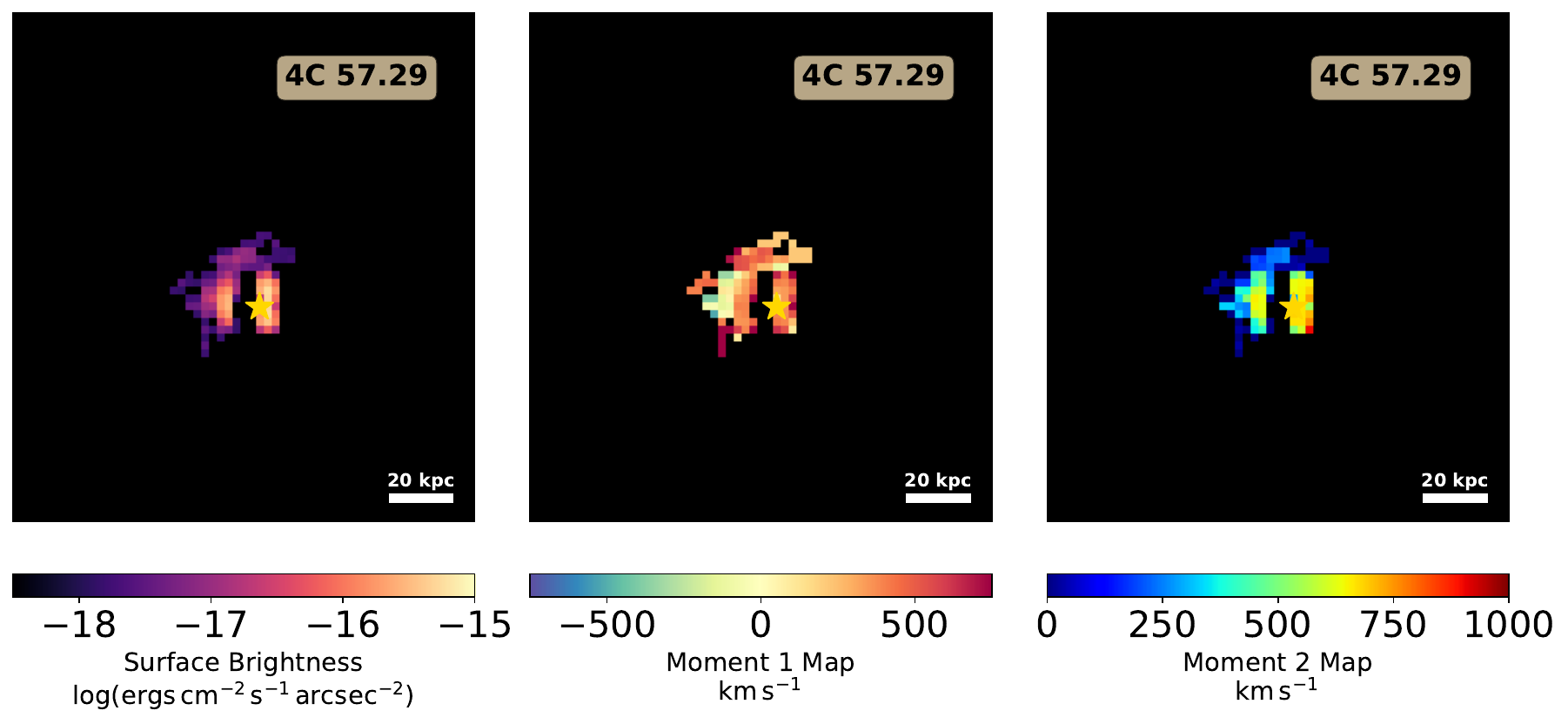}
    
    \caption{4C 57.29 \civ}
    \label{fig:4c5729_CIV}
\end{figure*}

\begin{figure*}[!ht]
    \centering
    \includegraphics[width=\linewidth]{ 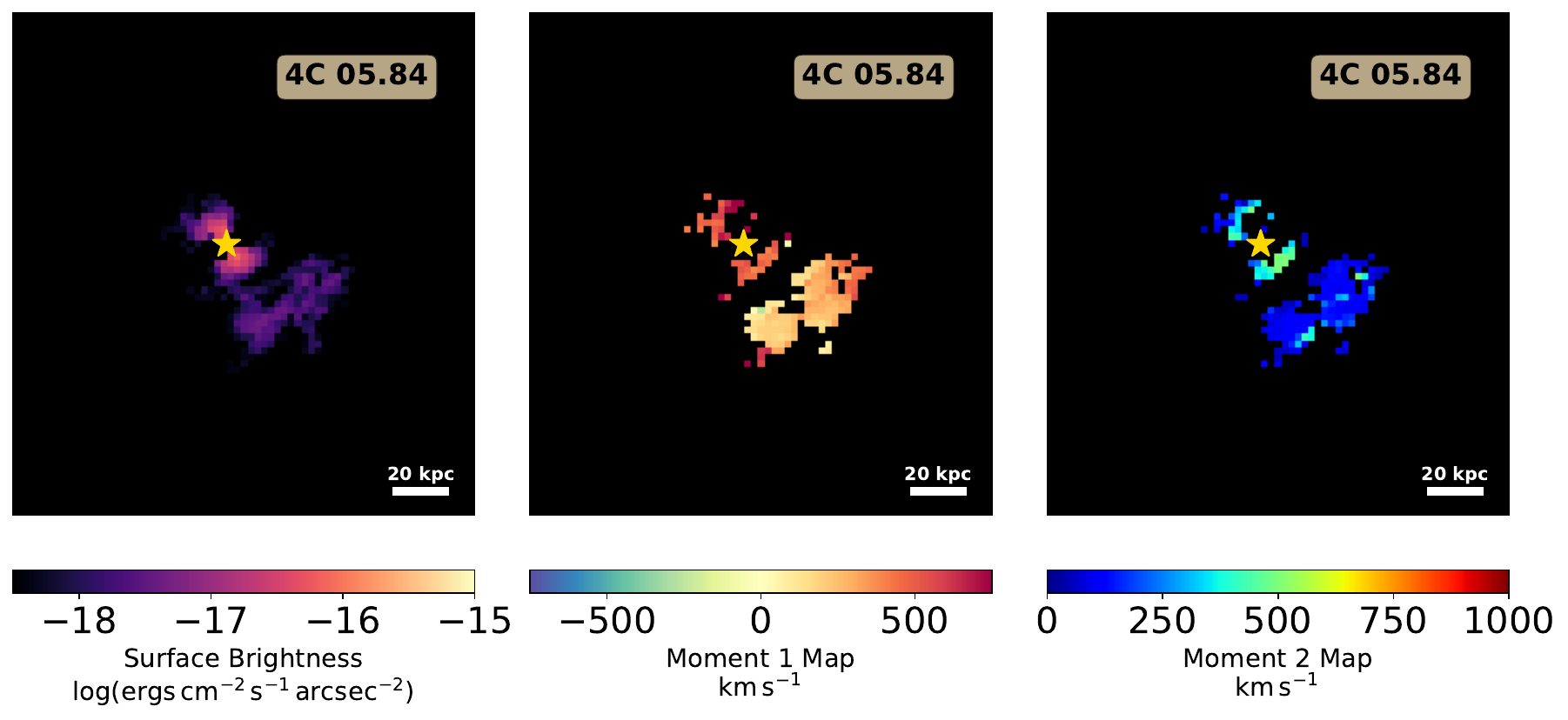}
    
    \caption{4C 05.84 \civ }
    \label{fig:4c0584_CIV}
\end{figure*}

\begin{figure*}[!ht]
    \centering
    \includegraphics[width=\linewidth]{ 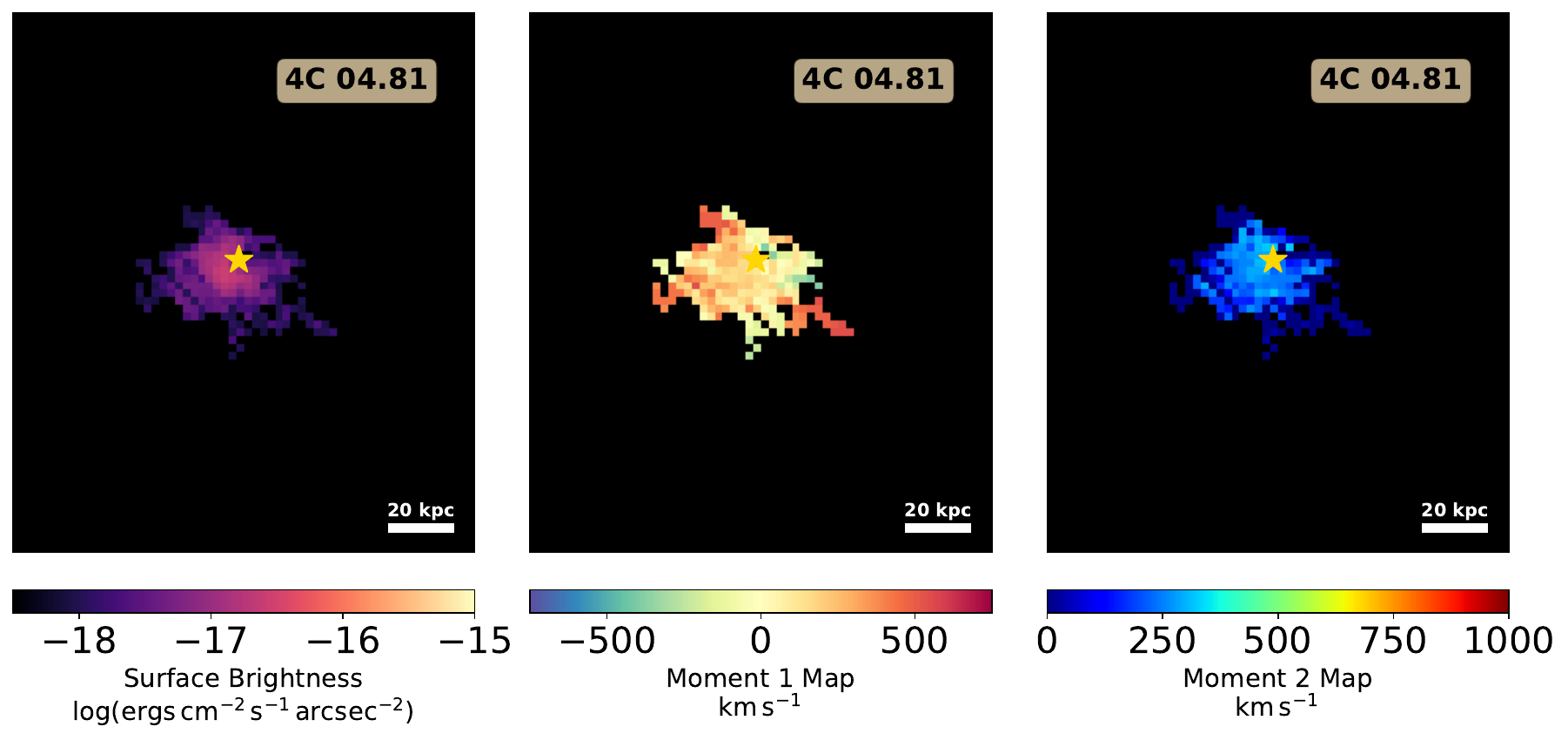}
    
    \caption{4C 04.81 \civ}
    \label{fig:4c0481_CIV}
\end{figure*}

\end{document}